\documentclass[12pt]{article}

\pdfoutput=1

\usepackage{graphics}
\usepackage{amssymb}
\usepackage{amsmath}
\usepackage{tikz}
\usepackage{cite}

\newcommand{\beq}{\begin{equation}}

\newcommand{\eeq}{\end{equation}}
\newcommand{\bea}{\begin{eqnarray}}
\newcommand{\eea}{\end{eqnarray}}

\begin{document}

\begin{center}
${}$\\
\vspace{60pt}
{ \Large \bf Quantum Gravity from Causal Dynamical \\
\vspace{0.3cm}
Triangulations: A Review
}

\vspace{46pt}

{\sl R. Loll}

\vspace{24pt}
{\footnotesize

Institute for Mathematics, Astrophysics and Particle Physics, Radboud University \\ 
Heyendaalseweg 135, 6525 AJ Nijmegen, The Netherlands.\\ 

\vspace{5pt}
{\it and}\\
\vspace{5pt}

Perimeter Institute for Theoretical Physics,\\
31 Caroline St N, Waterloo, Ontario N2L 2Y5, Canada.\\
}
\vspace{48pt}

\end{center}

\begin{center}
{\bf Abstract}
\end{center}

\noindent 
This topical review gives a comprehensive overview and assessment of recent results in Causal Dynamical
Triangulations (CDT), a modern formulation of lattice gravity, whose aim is to obtain a theory of quantum gravity nonperturbatively 
from a scaling limit of the lattice-regularized theory.
In this mani\-festly diffeomorphism-invariant approach one has direct, computational access to a Planckian spacetime regime,
which is explored with the help of invariant quantum observables. During the last few years, 
there have been numerous new and important developments and insights 
concerning the theory's phase structure, the roles of time, causality, diffeomorphisms and global topology, 
the application of renormalization group methods and new observables.
We will focus on these new results, primarily in four spacetime dimensions, and discuss some of their geometric and 
physical implications. 

\vspace{12pt}
\noindent

\newpage

\section{Introduction} 
\label{intro:sec} 

{\it CDT Quantum Gravity} or {\it CDT} for short denotes a nonperturbative path integral approach to quantum gravity, based on 
{\it Causal Dynamical Triangulations}. These represent curved, Lorentzian spacetimes and
play the role of regularized lattice confi\-gurations in this prominent version of modern lattice gravity.
CDT's key idea of how to introduce causal, Lorentzian features into gravitational path integrals  based on piecewise flat
spaces -- which up to that 
point had been primarily Euclidean in nature\footnote{``Euclidean spacetimes" have a Riemannian, positive-definite metric, 
instead of the physical, Lorentzian signature, and are used in the gravitational path integrals of 
Dynamical Triangulations (DT), the Euclidean precursor of CDT \cite{dt1,dt2}, and of Quantum Regge Calculus \cite{qregge1,qregge2}. } --
while rendering them amenable to computation, dates back to 1998.

From the initial formulation of this idea in two \cite{cdt2d} and subsequently higher spacetime dimensions \cite{cdt1,cdt2} 
it took several years of conceptual and technical development 
until the first breakthrough results were obtained for the full theory in four dimensions. 
They demonstrated the emergence of an extended four-dimensional   
universe with semiclassical features on large scales in one of the phases of CDT quantum gravity \cite{emergence,semi1}, 
something not seen previously (and since) in models of this type. In addition, the phenomenon of 
dynamical dimensional reduction was revealed as a nonperturbative quantum
signature of spacetime on Planckian scales \cite{spectral,reconstructing}, sparking numerous investigations of the spectral
dimension in other approaches to quantum gravity \cite{carlip}. 

The next important finding came in 2007, when it was shown that the expectation value of
the global shape of the dynamically generated universe behaves like the shape of a classical de Sitter universe \cite{desitter1,desitter2}. 
This further strengthened the evidence for the existence of a meaningful classical limit, something that nonperturbative approaches to 
quantum gravity generally struggle to demonstrate. Following this, strong evidence was found in 2011 for the presence of second-order phase
transitions in the phase space of the statistical model underlying CDT \cite{trans1,trans2}. Higher-order transitions are a crucial 
prerequisite for the existence of a well-defined continuum limit of a given lattice theory, and a feature that so far
has eluded similar models of four-dimensional quantum gravity. 

The last comprehensive review of CDT quantum gravity, published in 2012 \cite{physrep}, covered 
detailed derivations and descriptions of the geometric and technical set-up of the formalism, including the Monte Carlo simulations,
a description of the model's phase structure and all of the above-mentioned classic CDT results in four dimensions,
additional material on CDT systems with and without matter in two and three dimensions, many of which are interesting in their own right,
and an extensive bibliography. 

The main aim of the present article is to review the major new developments in CDT quantum gravity that
have occurred since then, focusing primarily on the four-dimensional theory, and the potential physical significance of the
results obtained. The latter include (i) the discovery of a new line of phase transitions, presumed to be of higher order, and an
associated new phase, (ii) the introduction of two new observables, the gravitational Wilson loops and the
quantum Ricci curvature, (iii) studies on the influence of global topology, and (iv) an implementation of the renormalization group.
In addition, considering everything that has been learned so far, we will also try to assess the specific features of CDT 
that underlie its success to date as a candidate theory of nonperturbative quantum gravity. 
As described in more detail in Sec.\ \ref{geom:sec} below, a key distinguishing feature appears to be the treatment of diffeomorphisms.

\bigskip
\section{Portrait of CDT as a young theory}
\label{portrait:sec}

In a nutshell, CDT aims to construct a theory of quantum gravity by taking a suitable scaling limit of a 
lattice theory, whose dynamics is given in terms of a regularized, nonperturbative path integral over geometric lattices 
that in a direct way approximate the curved spacetimes of classical general relativity. In complete analogy with lattice QCD, say,
the path integral depends on a number of bare coupling constants, which appear in a lattice version of the gravitational
action\footnote{After fine-tuning
the cosmological constant to its critical value, the phase diagram of CDT is two-dimensional, spanned by the inverse bare Newton constant
and the so-called asymmetry parameter, c.f. Sec.\ \ref{phase:sec} below.}, and on a 
UV lattice cutoff $a$ representing the shortest length unit on the lattice. 
As the regulator (or ``lattice spacing") $a$ is removed, $a\rightarrow 0$, one searches for continuum limits in the vicinity of a critical point, 
characterised by the divergence of a suitable correlation length in terms of lattice units. This will usually involve a fine-tuning and
renormalization of the dimensionless bare couplings of the lattice model. 

Unlike quantum chromodynamics, quantum gravity is a theory we presently know rather little about.
As a classical field theory, gravity has completely different degrees of freedom, dynamics and symmetries than a
gauge theory. It is therefore
not clear {\it a priori} that the construction of a fundamental quantum theory of gravity
as the continuum limit of a lattice formulation is technically or even in principle feasible,
and leads to physically meaningful results. Remarkably, CDT provides a blueprint for how potentially troublesome conceptual issues like 
the implementation of a Wick rotation and of symmetries, unitarity, and the unboundedness of the gravitational action
can be dealt with in a lattice regularization. With these ingredients in place, progress to date in the construction and
analysis of the nonperturbative quantum theory has been very promising indeed. 

The properties of the CDT model have
been explored extensively but not exhaustively, primarily with the help of Monte Carlo simulations. 
Despite the fact that the CDT path integral is defined for Lorentzian signature, these powerful numerical methods
can be applied {\it because of the presence of a well-defined Wick rotation}, a rare occurrence in quantum gravity beyond
perturbation theory. 
More concretely, the quantitative evaluation of a number of invariant 
quantum observables has established several milestones in the form of concrete evidence for the existence of second-order phase
transitions, the emergence of classical geometry, and the applicability of renormalization group methods, none of
which was obvious at the outset. 
As transpires from the above, the means employed in CDT to search for the holy grail of quantum gravity are rather conventional:
they are anchored in quantum field theory, without using
fundamental higher-dimensional excitations like strings, loops or branes, postulating supersymmetry, invoking gauge-gravity dualities
or a dynamical principle unifying all of the fundamental interactions.

Whether the no-frills approach of CDT quantum gravity, or any alternative approaches \cite{handbook,foundations,approaches}
will turn out to be successful or dead ends, 
complementary or mutually exclusive, 
or whether they are collectively barking up the wrong tree(s), is impossible to decide in the absence of a reliable 
quantum gravity phenomenology and in view of the incompleteness of our current candidate theories. 
Also CDT, despite its achievements, is still incomplete, with both conceptual and computational challenges ahead.
In terms of physics, these broadly speaking concern the nature and properties of the theory and its vacuum state on Planckian 
scales. For example, can CDT provide independent, corroborating evidence for 
``asymptotic safety" \cite{ReuSaubook,percaccibook,niedermaier}, or indications of some other form of nonperturbative UV completion? 
Can its physics be described 
in terms of (generalized) geometry or some other effective degrees of freedom? 

While work is ongoing to improve the efficiency of the simulations and to speed up measurements in the vicinity of the phase transitions --
issues that are largely CDT-specific -- all questions regarding the physical content of the theory and the properties of its ground state
(``quantum spacetime") must be phrased in terms of {\it observables}. The search for observables transcends any particular approach to quantum gravity and is key to understanding the physical implications of any given theory. Stated briefly, observables in this context are
diffeomorphism-invariant quantities, which are operationally well-defined and computable in the nonperturbative sector of the theory.   
Just a few observables are currently known, see \cite{physrep} for further details and Sec.\ \ref{observ:sec} below for an update.

An essential feature of CDT quantum gravity is the fact that it comes with its own reality check in the form of  
a well-developed ``computational lab", allowing in principle the evaluation of arbitrary quantum-gravitational observables, 
subject to the usual limitations in terms of computing power, lattice size, finite-size effects, lattice artefacts and numerical errors. 
In the context of full-fledged quantum gravity, CDT has played its part in demonstrating 
that numerical methods can yield important, nonperturbative information, which currently cannot be obtained by 
any other means.  
Far from being ``just numerics", numerical simulations allow the systematic testing of theoretical conjectures, be it on the
analytical form of an effective action for the Friedmann scale factor in ``true cosmology" (c.f. Sec.\ \ref{lessons:sec} below), 
or on specific properties of the theory's ground state. In this way numerical results 
feed back into the construction of
a coherent and comprehensive theoretical framework for quantum gravity.

At the same time, CDT has helped to move the discussion of ``which is the correct theory?" away from 
comparing the properties of specific approaches and formalisms,  
which often appear irreconcilable in terms of their initial assumptions and ingredients, 
to a better defined and potentially more fruitful comparison of (the spectra of) observables.  
The discussions around the spectral dimension -- one example of such a quantum observable --
illustrate that there is scope for unexpected coincidences between different approaches, 
which in turn can motivate further research.

This particular example also shows that in nonperturbative quantum gravity no exotic ingredients 
are needed to produce highly surprising results, which could not have been anticipated from the (semi-)classical theory.
In the case at hand this is the phenomenon of continuous dynamical reduction of the dimension of 
spacetime from the classical value of 4 to (at or near) 2 as one approaches
the Planck scale. The computation and comparison of quantum observables of this kind can be a way forward
in quantum gravity, helping us to identify its universal properties,
until we manage to establish a stronger link with observation or experiment. Of course, any candidate
theory must be sufficiently developed to permit unambiguous calculations, preferably without freely adjustable parameters 
and ad hoc auxiliary assumptions. 

At age twenty, counting from the initial idea of making the gravitational path integral Lorentzian again \cite{cdt2d}, 
CDT quantum gravity has established itself as a serious contender 
for a nonperturbative theory of quantum gravity, on the basis of a significant body of work {\it and results}, several
of which are unique to this approach. 
Of course, it is up to the informed readers to weigh the achievements of CDT relative 
to those of other approaches,
and relative to the overall goals and past setbacks in quantum gravity. 
In our view, the outlook of CDT quantum gravity is bright: 
there are already beautiful results, the approach has not hit any major roadblocks or dead ends,
and -- as illustrated by the developments described in this review -- 
there is a steady pace of progress and 
a clear way forward, hopefully with equally interesting results awaiting us in the near future.

Since CDT is in essence a quantum field-theoretic approach, its results can most easily be
compared to continuum results obtained with the help of functional renormalization group methods 
in a setting with asymptotic safety, which are usually formulated
in terms of the (gauge-fixed) spacetime metric $g_{\mu\nu}(x)$. This has already
happened for two observables: the spectral dimension \cite{specfrg1,specfrg2} and the so-called volume profile, 
defined as the (expectation value of the) volume of space as a function of cosmological time \cite{knorr}. 
However, recalling an earlier remark, because
of the very different ways in which gauge and physical degrees of freedom are treated in the continuum and on the lattice, it is virtually 
impossible to compare such derivations meaningfully at an intermediate stage, say, at the level of some coarse-grained version of the metric,
and for quantities that are not observables.

CDT quantum gravity in its regularized, Wick-rotated form
can be thought of as a particular type of statistical mechanical system, whose degrees of 
freedom are not spins or matter fields on a given, fixed lattice, but instead the parameters 
describing the metric properties of the lattice, viewed as a piecewise flat space, which is itself subject to
dynamics. In dimension two, where both DT and CDT were studied first, they constitute
statistical systems of {\it random geometry}, which are soluble exactly with a variety of analytical methods, including 
combinatorial, matrix model and transfer matrix methods \cite{david,qgeobook,loreu}. 
This provides an additional perspective in the sense that
one can try to import two-dimensional structural analyses and solution strategies into four-dimensional CDT. 
More generally, as illustrated by the construction of the quantum Ricci curvature in Sec.\ \ref{sec:curv}, 
CDT has overlap with issues discussed in the fields of discrete geometry and computational geometry, and may 
look to these subjects for inspiration, for example, in its search for observables.

\section{CDT path integral -- the bare essentials}
\label{setup}

Since the details of how the CDT path integral over triangulations is set up have been covered in \cite{physrep}
and other previous reviews and overviews (see, for example, \cite{rev1,rev2,rev3,rev4,rev5}), we will confine ourselves to a summary of the essentials, 
while highlighting some issues which are of special importance.
The central object of interest in CDT quantum gravity is the nonperturbative gravitational path integral.
For pure gravity, the CDT path integral $Z^{\rm CDT}$ takes the form of a continuum limit of
a regularized lattice expression. Schematically, we have
\begin{equation}
Z^{\rm CDT}(G_{\rm N},\Lambda)=\lim_{{a\rightarrow 0}
}\,\sum_{T\in{\cal T}}\frac{1}{C(T)}
\, {\rm e}^{i S^{\rm CDT}[T]},
\label{cdtact}
\end{equation}
where the sum is taken over inequivalent Lorentzian triangulations $T$, assembled from two types of 
elementary four-simplices according to certain causal gluing rules.

Each term in the sum contributes with an amplitude depending on a lattice discretization $S^{\rm CDT}$ 
of the continuum Einstein-Hilbert action
\begin{equation}
S^{\rm EH}=\frac{1}{G_{\rm N}}\int d^4x\, \sqrt{-\det g}\, (R[g,\partial g,\partial^2g]-2\Lambda),
\label{ehact}
\end{equation}
where $G_{\rm N}$ and $\Lambda$ are Newton's constant and the cosmological constant respectively,
and $R$ denotes the Ricci scalar of the metric tensor $g$.
The quantity $C(T)$ in eq.\ (\ref{cdtact}) is the number of elements in the automorphism group of the
triangulation $T$. It is equal to 1 whenever $T$ does not possess any symmetries, as is almost always the case.
The lattice version of the action (\ref{ehact}) is subject to the usual discretization ambiguities. The standard choice for $S^{\rm CDT}$
is the so-called Regge version of the gravitational action \cite{regge,physrep}, but this choice is in no way 
fundamental or ``exact", in the same way as Regge calculus in four dimensions has no fundamental
physical status, but is merely a discrete approximation of general relativity.
Recall that lattice quantum gravity -- like lattice QCD -- requires a continuum limit, guaranteeing 
the existence of universal properties on all scales, including Planckian, which are independent of the arbitrary choices made
as part of the regularized lattice set-up (for example, the types of building blocks, 
the detailed form of the discretized action, the path integral
measure and the gluing rules).

\begin{figure}
\centering
\includegraphics[width=0.75\textwidth]{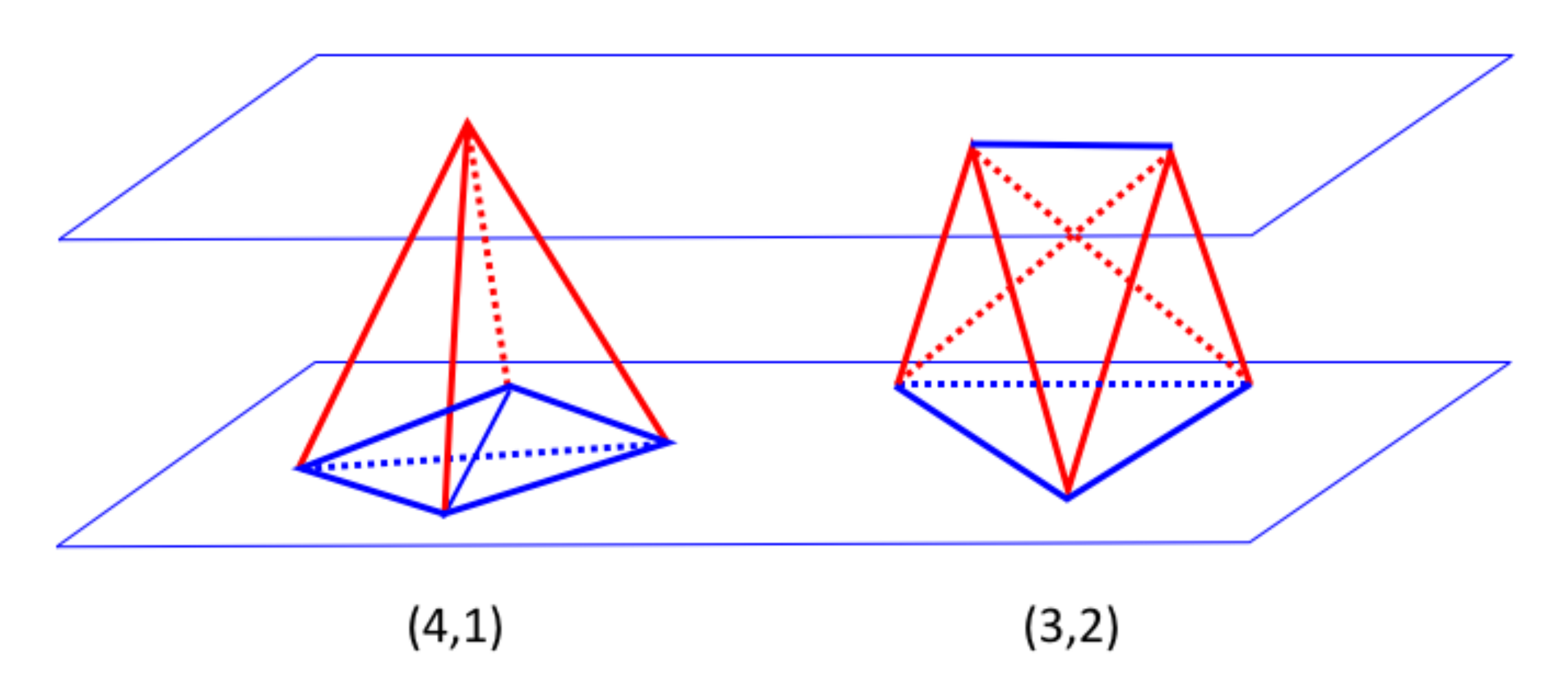}
\caption{\label{fig:simplices}Regularized spacetimes in CDT quantum gravity are built from 
two elementary four-simplices, the
(4,1)-simplex (left) and the (3,2)-simplex (right). They differ
in their assignments of space- and time-like edges, drawn in blue and red respectively, see \cite{physrep} for details.}
\end{figure}

The action $S^{\rm CDT}$, derived explicitly in \cite{cdt2,physrep}, has a simple form and is easy to compute for
a given simplicial manifold $T$. It is a linear combination of
several counting variables $\{ N_i^I\}$, where $N_i^I(T)$ denotes the number of simplices of dimension $0\leq i\leq 4$ 
contained in $T$. The index $I$ labels different subtypes. For instance, 
the four-simplices of CDT come in two different variants, (3,2) and (4,1), with corresponding counting variables
$N_4^{(3,2)}$ and $N_4^{(4,1)}$ (see Fig.\ \ref{fig:simplices}). 
Schematically, the action can be written as
\begin{equation}
S^{\rm CDT}[T]= k_b\pi \sqrt{4\alpha +1} N_0(T)+{\cal A}(\alpha,k_b,\lambda_b) N_4^{(4,1)}(T)+
{\cal B}(\alpha,k_b,\lambda_b)N_4^{(3,2)}(T),
\label{actlor}
\end{equation}
where $\cal A$ and $\cal B$ are specific linear combinations of the bare inverse Newton constant $k_b$ and the bare cosmological
constant $\lambda_b$, and $\alpha\! >\! 0$ is a finite, fixed para\-meter describing the geodesic length $\ell_t$ of a time-like edge in terms
of the geodesic length $\ell_s\!\equiv\! a$ of a space-like edge according to 
\begin{equation}
\ell_s^2=a^2,\;\;\;
\ell_t^2=-\alpha a^2,\;\;\; \alpha >0, 
\label{edges}
\end{equation}
see \cite{physrep} for further 
details.\footnote{Compared with formula (60) in \cite{physrep}, we have dropped a term proportional to the Euler characteristic
of the spacetime, because it plays no role in the path integral.}

A frequently asked question is whether using (a discretized version of) the Einstein-Hilbert action in the path integral (\ref{cdtact})
amounts to a truncation of the theory. This is not the case. Since $S^{\rm CDT}$ is only a bare lattice action, not 
including curvature terms of quadratic or higher order in it does {\it not}
imply a truncation, because -- to use a continuum expression -- such terms are generated during renormalization. 
More precisely, the {\it full} action governing the behaviour of the system at short distances receives contributions from 
both the bare action and the entropy of states (in a continuum language, the ``path integral measure").

The difference it would make to include an $R^2$-term already in the bare action would be that
the corresponding term in the full action would be associated with an additional, tuneable bare coupling parameter.    
One would be naturally led to consider such an extension of the parameter space of CDT if nothing interesting
(for example, no higher-order phase transition) had been found in the version studied presently. Fortunately, the opposite
is the case, which means that for the time being exploring CDT quantum gravity in its current form takes priority over
possible extensions of the phase space.   

Another way in which the action may be generalized is by including matter, accompanied by an additional sum over matter
configurations for each spacetime configuration in the path integral. There are no
specific technical difficulties in implementing this in dynamically triangulated quantum gravity, where gravity-matter models have been
studied extensively in Euclidean-signature DT (see, for example, \cite{dt2,physrep}). Systems of CDT geometry and
matter were investigated in the early days of CDT, and were centred on questions specific to two dimensions, like the existence
of a $c\! =\! 1$ barrier in CDT \cite{cdtmatter2,cdtmatter4} and whether matter critical exponents coincide with those on flat two-dimensional lattices \cite{cdtmatter1,cdtmatter3}.

There are several reasons for why matter-coupled models 
have not yet been researched extensively in CDT in four dimensions.
A key motivation for considering matter coupling in four-dimensional Euclidean DT quantum gravity in the 1990s was the 
hope to find a continuous phase transition and an associated continuum theory after all \cite{bilke1}, 
a hope that up to now has not been realized\footnote{Note that integrating out matter fields in the path integral 
can lead to results 
equivalent to adding a higher-order curvature term in the gravitational action or including a nontrivial
measure term, see \cite{edt1} for a discussion in the context of four-dimensional Euclidean DT. 
All of these situations are associated with an extra coupling constant.}. By contrast, 
four-dimensional CDT currently has plenty of interesting and nontrivial results even before matter is coupled,
which are under active investigation.
However, the issue of matter coupling is definitely interesting, and also presents new challenges. The primary 
one is to come up with physically interesting diffeomorphism-invariant 
matter observables, which are well-defined and finite {\it in a nonperturbative 
regime where the underlying local spacetime geometry is very far from classical}, as is the case in CDT.
Technically, they need to be implementable and measurable on CDT lattices.
Few such observables are currently known. On the conceptual side, as illustrated by studies of
CDT coupled to a single massive particle \cite{point,wilson}, there are subtleties in 
relating Euclidean and Lorentzian results, which still need to be understood better. 
  
Other important constructional aspects of the CDT path integral, discussed at length in \cite{physrep}, 
are its reflection positivity, virtually guaranteeing unitarity in the continuum limit, the existence of a rigorous {\it Wick
rotation} which analytically continues the complex path integral (\ref{cdtact}) to a sum of real Boltzmann weights, and the 
implementation of Monte Carlo 
simulations. Apart from some remarks on the Wick rotation in Sec.\ \ref{time:sec} below, these will not be reviewed again here. 

One can hardly overemphasize the importance of having well-developed quantitative tools available.
CDT quantum gravity enables an interplay between theoretical considerations and numerical simulations:
nonperturbative observables, suggested on theoretical grounds, can be implemented, tested and measured,
and results from simulations can feed back into the theoretical modelling of Planckian physics, allowing 
an ever better understanding of what nonperturbative quantum gravity is about. 
This requires nontrivial expertise in the theoretical and practical aspects of Monte Carlo and lattice methods,
and individuals trained in them. 
Many teams and researchers who worked or work with higher-dimensional
dynamical triangulations have links with the
lattice quantum field theory community, where these tools are commonplace, but this is by no means
essential.  
Neither is there a need for investing in dedicated hardware, since the 
simulations are typically run on laptops or local computer clusters.
In recent years, Markov chain Monte Carlo and other computer algorithms have
gradually been making inroads, beyond lattice QCD and lattice gravity, into other quantum gravity
approaches and more generally theoretical high-energy physics beyond the standard model
(see \cite{hanada,jha,cunningham} for some recent examples), and their appreciation appears to be on the increase.
Like in CDT, the emphasis is often on exploring the properties of given models and candidate theories,
and performing ``reality checks" of their viability, rather than improving the measuring accuracy of any given   
quantity by so many decimal places.

Broadly speaking, DT and CDT represent similar computational challenges in terms of algorithms, data storage and addressing. 
Computer code used in four-dimensional CDT in the early 2000s
was adapted from previous DT versions, heavily used and tested over the previous decade.
Computational efficiency of CDT has been improving steadily since. 
At the same time, a number of individuals and groups have independently written CDT code from scratch and
run simulations \cite{laiho,kommu,clemente1}, finding consistent results.\footnote{A publicly available version of the 
code, called {\it CDT-plusplus}, suitable for running on a variety of operating systems and cluster computing 
environments, will be released in the near future \cite{getchell}.}

\section{CDT geometry and diffeomorphism invariance} 
\label{geom:sec}

CDT quantum gravity combines several key features in a unique way to produce its results.
The two most important ones are 
\begin{itemize}
\item[(i)] the encoding of the geometric degrees of freedom of gravity, 
without introducing coordinates and their associated redundancy, and
\item[(ii)] the incorporation of Lorentzian features in the gravitational path integral, eq.\ (\ref{cdtact}), 
in a way that allows for its analytic continuation to a real partition function and the application of
Monte Carlo methods.
\end{itemize}
This section spells out some details regarding point (i), while the next section summarizes the
state of affairs with respect to point (ii). --
The starting point for encoding the geometric degrees of freedom of gravity
in a regularized way is Regge's proposal to describe
``general relativity without coordinates" \cite{regge}, where curved spacetimes are approximated
by piecewise flat, simplicial manifolds, and a discrete analogue of the Einstein equations is derived.
In this setting, a curved four-dimensional spacetime is described
by its connectivity data (specifying which pairs of four-simplices share a three-dimensional face)
and by the invariant lengths of the one-dimensional edges of the triangulation. 
All information about the intrinsic metric and curvature properties of a 
piecewise flat spacetime is expressible in terms of these data.

Note in passing that the space of piecewise flat spacetime geometries in CDT -- serving as carrier space
for the regularized gravitational path integral -- is not identical to the one on which classical evolution is defined in
\cite{regge}. The latter allows for continuous changes in the edge lengths of a given (topological) 
triangulation. This is by construction not possible in (C)DT, where all edge lengths are ``frozen in" to 
one common length in DT, or two lengths (one for time-like, one for space-like edges) in CDT. It
implies that CDT is not particularly well-suited for describing classical continuous evolution, a purpose it also was
never invented for. However, this is not an obstacle 
to formulating and computing the nonperturbative path integral, quite the contrary, as we will see. 
To summarize, (C)DT quantum gravity operates with a variable lattice topology\footnote{By this we mean different,
inequivalent gluings of the flat building blocks, not the overall spacetime topology of the lattice, which in CDT is
usually kept fixed.} and fixed edge lengths, while classical and quantum Regge calculus
operate with fixed lattices and variable edge lengths, subject to triangle inequalities.

In concrete applications of CDT one often introduces discrete labels to refer to parti\-cular (sub-)simplices
of a given triangulation, but this does not involve any coordinate system, and the ensuing relabeling redundancy
is easily taken into account.\footnote{Basically, by dividing by $N!$ for the case of $N$ labelled simplicial building blocks,
leading to the path integral measure exhibited in (\ref{cdtact}) for unlabelled triangulations;
see \cite{physrep} for a more detailed discussion.}
One is of course free to introduce local coordinates in individual simplices or in a simply connected
neighbourhood of a triangulation, as long as it does not contain any curvature singularities. However, this is redundant 
from a geometric point of view, and associated with the usual ambiguity of such a choice.  
{\it The true strength of Regge's description is the absence of
coordinates and their associated gauge freedom, together with a direct link of the edge length variables to intuitive geometric
notions.} The former is particularly important in quantum gravity beyond perturbation theory,
where this freedom turns into the formidable difficulty of either having to gauge-fix or otherwise having to implement 
the four-dimensional diffeomorphism symmetry at the quantum level. 
CDT arguably makes optimal use of this strength in its construction of a nonperturbative gravitational path integral 
without any gauge redundancies.\footnote{The situation in quantum Regge calculus is less clear-cut, see \cite{dt2} for a discussion
and further references.} 

The pedagogical value of explaining curvature in terms of piecewise flat triangulations is illustrated nicely in \cite{zk}. 
More than that, when working in CDT and defining observables, the formulation forces one to think in terms of pure geo\-metry
and therefore true physics, 
very much in the spirit of Einstein's ``rods and clocks", as opposed to resorting to some coordinate system. 
This latter feature is an {\it asset} of the formulation, even though it may appear unusual
from the point of view of the classical theory, where differentiable manifolds and smooth metric fields
provide powerful and convenient models of spacetime, and where choosing local coordinates is essential 
when performing concrete calculations. By contrast, in (C)DT quantum gravity the assignment of metric degrees of freedom is
not smooth, there are no coordinates, and the four-dimensional diffeomorphism group simply does not act. 
This holds for the regularized path integral with finite UV-regulator $a$ and any continuum limit where $a\rightarrow 0$.
In this sense, the folklore that ``putting gravity on the lattice necessarily breaks diffeomorphism invariance" does {\it not} apply to 
CDT quantum gravity. As emphasized before, the treatment of diffeomorphisms in nonperturbative quantum gravity
is in general highly nontrivial; by working directly on the quotient space of spacetime metrics modulo diffeomorphisms (or a suitable
regularization thereof), this issue does not arise in CDT. Of course, any comparison of results and predictions of CDT, 
for example concerning
its classical limit, with those coming from a more conventional smooth continuum formulation should be made in terms
of appropriate geometric, coordinate-invariant observables. 
The fact that the DT approach can reproduce the results of a diffeomorphism-invariant continuum quantum field theory 
is well documented in the case of Liouville quantum gravity in two dimensions \cite{dtliou1,dtliou2}.

\section{Causal structure, Wick rotation, time and all that}
\label{time:sec}

As highlighted in the introduction, the main motivation behind CDT was the search for a nonperturbative path integral formulation
of gravity for physical spacetime geometries $g^{\rm lor}$ of Lorentzian signature, schematically,
\begin{equation}
\int\limits_{{\rm Lorentzian}\atop{\rm spacetimes}}
\!\!\! Dg^{\rm lor}\, {\rm e}^{iS[g^{\rm lor}]},\;\;\;\;\; {\rm signature}(g^{\rm lor})=(-\! +\! ++).
\end{equation}
This is different from so-called Euclidean quantum gravity, which uses 
a different starting point, namely, the path integral 
\begin{equation}
\int\limits_{{\rm Riemannian}\atop{\rm ``spacetimes"}} 
\!\!\! Dg^{\rm eu}\, {\rm e}^{-S[g^{\rm eu}]},\;\;\;\;\;  {\rm signature}(g^{\rm eu})=(+\! +\! ++),
\end{equation}
where Lorentzian spacetimes have been replaced in
an ad hoc manner by Riemannian or (in physicists' parlance) Euclidean metric spaces $g^{\rm eu}$.
Since no Wick rotation\footnote{By Wick rotation we mean a suitable curved-space analogue of the flat-space prescription of
substituting real time $t$ by imaginary time $\tau =it$, for example through analytic continuation in
the complex $t$-plane (see \cite{wick} for a discussion of why this is difficult to achieve).} 
is known that would map general curved, smooth Lorentzian metrics to Euclidean ones, or diffeomorphism equivalence
classes of metrics of either signature into each other, there is no {\it a priori} physical justification for working with Euclidean 
geometries, and no immediate reason why an associated nonperturbative path integral -- if it exists -- should agree with (or 
in some sense be equivalent to) any Lorentzian counterpart, or indeed have any physical interpretation at all.

The difficulty prior to the introduction of CDT was to find examples of Lorentzian nonperturbative path integrals that could be 
evaluated at all, either analytically or numerically, given that they are of the form of (weighted) sums or integrals 
over complex phase factors. One of the beautiful features of CDT quantum gravity
is the existence of a well-defined Wick rotation that maps the piecewise flat spacetime configurations in the ``sum over histories",
eq.\ (\ref{cdtact}), to unique piecewise flat Riemannian spaces, while transforming the sum itself into a sum over
real Boltzmann weights \cite{cdt1,cdt2,physrep}. It turns out that in two spacetime dimensions 
the analytically continued, real partition function $Z^{\rm CDT}$ can be 
evaluated analytically, which led to the first explicit demonstration that Lorentzian and Euclidean
nonperturbative gravitational path integrals in general yield inequivalent results \cite{cdt2d,loreu}.
The same seems to be true in the physically relevant case of gravity in four dimensions, where the presence of the Wick rotation
enables the evaluation of the CDT path integral and associated observables by Monte Carlo simulations, and thus an
explicit comparison with corresponding DT results for Euclidean quantum gravity. Despite renewed efforts \cite{edt1,edt2,edt3}, 
there is still no evidence of higher-order phase transitions or semiclassical behaviour \`a la CDT in the latter. 

\begin{figure}
\centering
\includegraphics[width=0.55\textwidth]{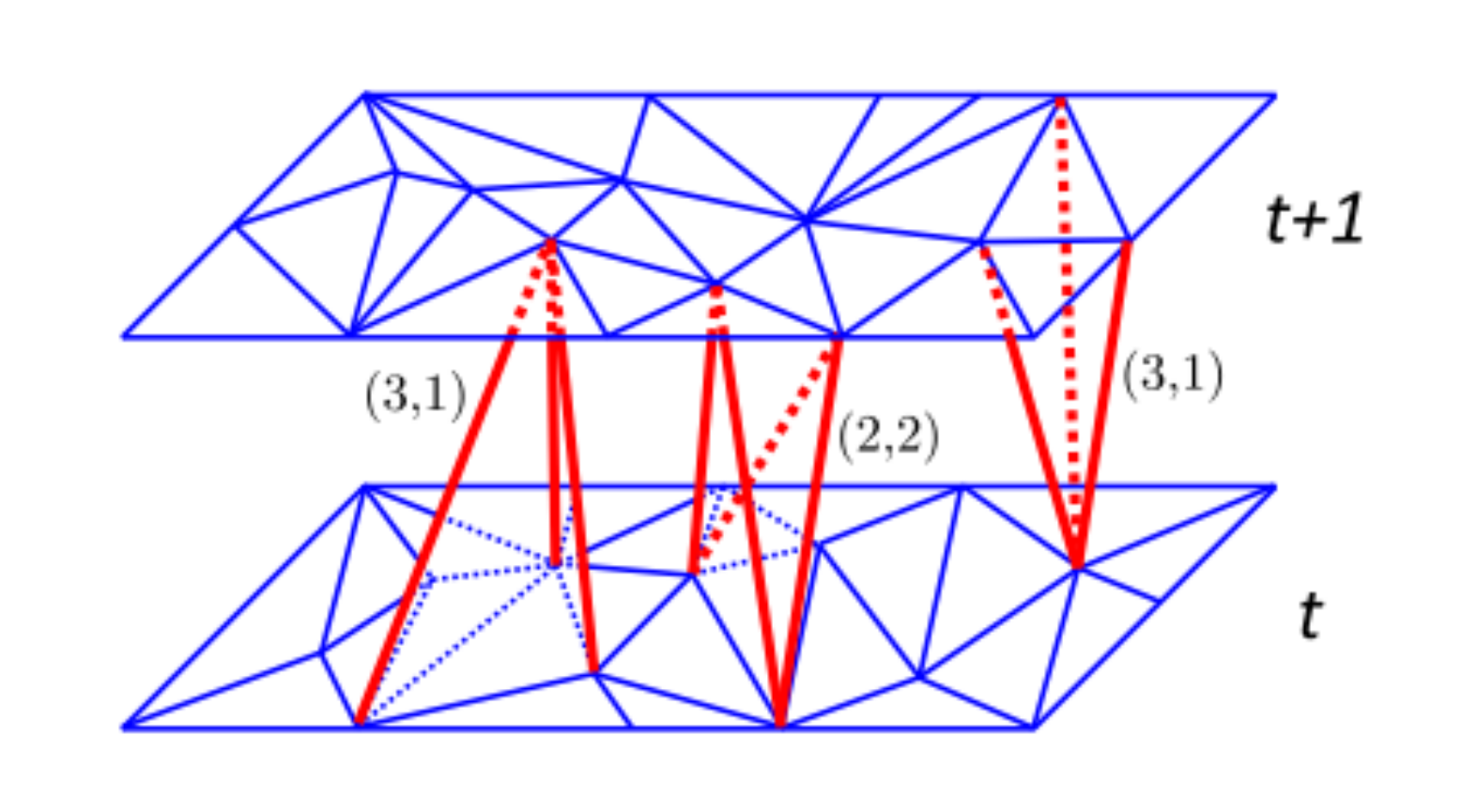}
\caption{\label{fig:3dtriang}A ``sandwich geometry" of topology $I\!\times\! {}^{(2)}\Sigma$ for
CDT in $2\! +\! 1$ spacetime dimensions.
The sandwich consists of a layer of three-simplices (tetrahedra) extrapolating between two adjacent spatial slices at times $t$
and $t\! +\! 1$ 
made up of two-dimensional triangles. The space between the lower and upper triangulation is completely filled with tetrahedra, 
but for simplicity only three are shown here. (Space-like edges depicted in blue, time-like ones in red.) }
\end{figure}
The spacetimes summed over in the CDT path integral are assembled from two types of
four-dimensional simplicial building blocks, as illustrated in Fig.\ \ref{fig:simplices}. Each simplex can be thought of as a piece of flat Minkowski
space, whose geometric properties are fixed uniquely by the assignment of (squared) geodesic lengths to its 10 edges, given by
$\ell_s^2\! =\! a^2$ for all space-like edges, and $\ell_t^2\! =\! -\alpha a^2$ for all time-like edges, for some $\alpha >0$. 
The gluing rules for these
building blocks are such that the resulting spacetime configuration satisfies a lattice version of global hyperbolicity,
in the sense that it has the form of a stack of ``sandwich geometries". Each sandwich has topology $I\times\! {}^{(3)}\Sigma$, where 
$\! {}^{(3)}\Sigma$ denotes the fixed topology of a spatial slice, which in the situations studied so far is either a three-sphere $S^3$ or
a three-torus $T^3$. Geometrically, a sandwich is a layer of thickness one in terms of the simplicial building blocks, with
two purely space-like three-dimensional triangulations as initial and final boundaries (see Fig.\ \ref{fig:3dtriang} for
an illustration of the analogous situation in one dimension less). One can attach a {\it discrete
proper time} label $t$ to these sandwiches or, equivalently, to the spatial triangulations. In other words, each CDT spacetime can be seen
as a sequence $t=0,1,2,\dots, t_{tot}$ of spatial triangulations of topology $\! {}^{(3)}\Sigma$, with $t_{tot}$ four-dimensional
sandwich geometries extrapolating between pairs of adjacent spatial slices, to produce a simplicial Lorentzian manifold of
total time extension $t_{tot}$ with an initial and a final spatial boundary triangulation.

\begin{figure}[t]
\centering
\includegraphics[width=0.8\textwidth]{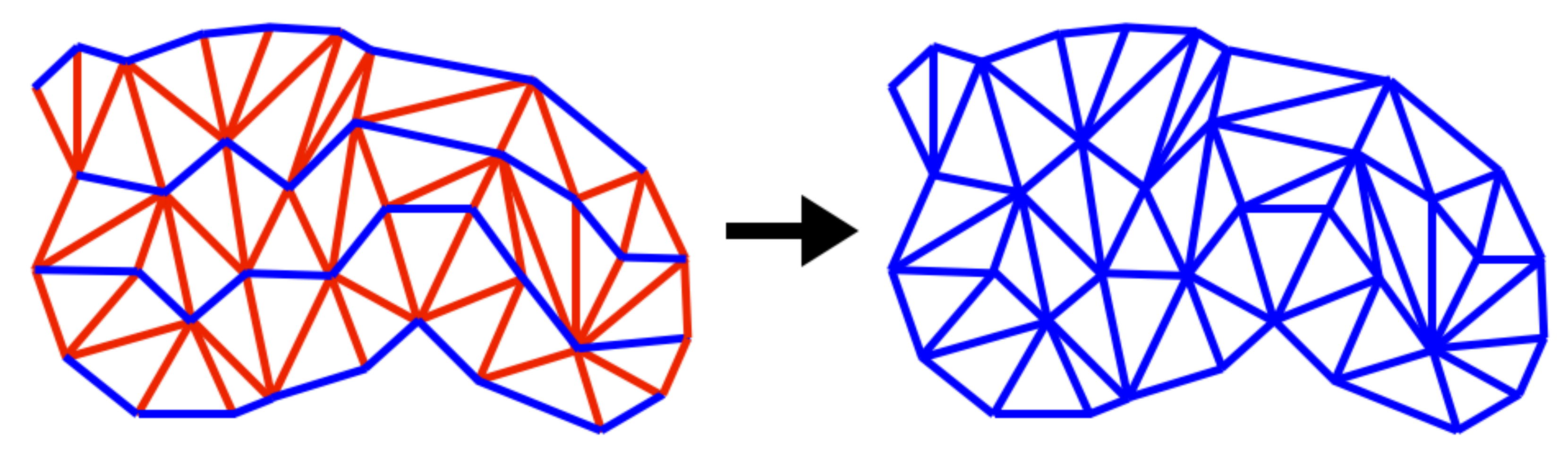}
\caption{\label{fig:2dtriang}The layered structure of a two-dimensional CDT configuration, 
with space-like edges drawn in blue and time-like ones in red.  
Left: CDT geometry consisting of a stack of three sandwiches. 
The spatial geometries at discrete times $t$ are open 
chains consisting of a variable number $l_t >0$ of space-like edges. Right: the corresponding Euclidean geometry after Wick rotation. 
The underlying lattice structure is unchanged, but all time-like edges have been replaced by space-like ones. 
}
\end{figure}
For the analogous but simpler case of CDT in two spacetime dimensions, 
Fig.\ \ref{fig:2dtriang} depicts a sequence of three sandwich layers. 
The elementary CDT building block in two dimensions is a Minkowskian triangle with 
one space-like and two time-like edges, and without loss of generality one can set 
$\alpha\! =\! 1$. Wick-rotating in CDT amounts to analytically continuing $\alpha\mapsto -\alpha$ in the lower-half
complex $\alpha$-plane \cite{cdt2,physrep}, which in the case at hand implies that all edges after the Wick rotation have 
equal length $\ell^2\! =\! a^2$.
The resulting Euclidean triangulated curved space is shown in Fig.\ \ref{fig:2dtriang} on the right. 
Note that the triangulations cannot be represented isometrically in the plane because of their nonvanishing curvature
assignments, associated with nonvanishing deficit angles at the vertices. Unlike what is usually done 
(including in Fig.\ \ref{fig:3dtriang} above), we have chosen 
deliberately to not depict the spatial universes by straight horizontal lines. The latter is a convenient, but ultimately arbitrary and
potentially misleading choice, because it (wrongly) suggests an absence of extrinsic curvature of the spatial slices.

The layered structure is an important element of CDT and allows the straightforward implementation of a key feature,
which distinguishes 
CDT geometries (before and after Wick rotation) from Euclidean DT geometries, namely, the
absence of so-called ``baby universes". These are violations of the local light cone (i.e.\ causal) structure, like
those associated with branching points, where the spatial universe changes topology by branching out into two or 
more disconnected components. Such branchings are by definition disallowed in CDT, but they are generically present in DT, 
no matter what notion of time one superimposes on the Euclidean triangulations 
(for illustration, see for example \cite{lollnpps,lollcqg}). 
Although such violations of the causal structure are not allowed in the classical theory, there is no {\it a priori} reason to not
permit them in path integral configurations. A major insight gained from CDT quantum gravity is that 
insisting on a well-behaved causal structure at this level appears to be essential for obtaining a nonperturbative quantum theory with a 
good classical limit.
Note that analogous branchings of the geometry in the {\it spatial} directions are still perfectly allowed, are not in conflict
with the causal structure and will generically
occur. It means that in CDT -- just like in the classical theory -- space and time directions are not equivalent or 
interchangeable by symmetry.\footnote{Note also that current investigations are far from a regime where 
one could make meaningful statements about the presence of local Lorentz invariance. 
This would require an (at least approximate) notion of local frames and the construction of a diffeomorphism-invariant 
quantum observable testing for Lorentz symmetry, objectives which at this stage seem out of reach.} 
The fact that this distinction persists after the Wick rotation is at the heart of why the Lorentzian path integral can be
inequivalent to (and lie in a different universality class than) its Euclidean counterpart.

Let us 
take the opportunity to remark on some properties of CDT quantum gravity associated with causal properties. 
Firstly, as just noted, the regularized CDT configurations come with a well-defined causal structure on all scales, 
which appears to be necessary for the model to have a four-dimensional behaviour compatible with classicality.
It does not imply that a spacetime that emerges
in a suitable scaling limit from a superposition of such configurations will necessarily inherit this structure.
This is a property that will have to be checked in each case.
It should also be noted that the causal structure of CDT configurations
is part of their Lorentzian metric structure and as such is of course not fixed,
but subject to quantum fluctuations, along with all other modes of geometry.    
Secondly, {\it causal structure} should not be confused with {\it causality}. The latter usually refers 
to the behaviour of matter or fields on a spacetime that is {\it already} endowed with a causal structure. 
The presence of the latter in the classical or quantum theory is a prerequisite for deciding whether or not
excitations propagate causally.

The global, discrete time label $t$ and the associated family of three-dimensional spatial triangulations labelled 
by $t$ form a lattice substructure that is present in each CDT configuration. Note that $t$ is a parameter which
can be used to (partly) specify the location of vertices, say, but is {\it not} related to any choice of gauge, 
because (i) there is no residual
gauge invariance, as explained in Sec.\ \ref{geom:sec}, and (ii) there are no coordinates in the first place,
nor is it in general possible to introduce coordinate patches that extend
beyond small neighbourhoods of pairs of adjacent simplices. 
Moreover, $t$ has an invariant geometric meaning on the piecewise flat CDT geometries.
For the case of open spatial boundaries, it measures the geodesic (link) distance to the initial spatial boundary of the spacetime, 
while for the case of cyclically identified boundaries -- at least in a phase with de Sitter behaviour -- it measures the geodesic 
distance to the beginning of the universe, reminiscent of the diffeomorphism-invariant 
notion of cosmological time in classical general relativity \cite{cosmic}. 

Whether or not the label $t$ can be related to some physical notion of time
in a suitable continuum limit cannot be determined a priori. What has been established is that in the
phases of CDT where a de Sitter behaviour for the spatial volume of the universe is found, a
renormalized version of $t$ assumes the role of global, cosmological proper time \cite{desitter1,desitter2}. Whether there exist
observables, for example, suitable two-point functions, whose behaviour allows for an interpretation of $t$ in terms of 
a specific, more local notion of time is currently not known.

While it is true that the regularized CDT geometries possess a layered or stacked structure, as we have just explained, 
there are several reasons to believe that this is just a lattice artefact and has no implications for the continuum limit.
Firstly, although the situation resembles superficially the preferred foliation present in Ho\v rava-Lifshitz gravity \cite{horava}, 
which is associated with a breaking of full diffeomorphism invariance to an invariance under foliation-preserving
diffeomorphisms, the discrete layers of CDT do 
{\it not} break any diffeomorphism symmetry, because the formulation does not have this symmetry to start with. 
Secondly, the action used in CDT quantum gravity is just the standard Einstein action, although one
could in principle try to implement an explicitly noncovariant action of Ho\v rava-Lifshitz type, as has been done
in three dimensions \cite{3dHL}. Thirdly, there is a generalized version of CDT quantum gravity, where a 
well-behaved causal structure is implemented in terms of suitable local gluing rules, without any preferred lattice foliation. 
It has been investigated extensively in three dimensions, 
reproducing several key results of standard CDT \cite{jordanloll1,jordanloll2,Jordan}, including the
emergence of a de Sitter-shaped universe, thereby supporting the conjecture that it lies in the same universality 
class. Since it touches on the fundamental issue of time in CDT, we will describe the construction and results of this model in some detail.

\begin{figure}
\centering
\includegraphics[width=0.8\textwidth]{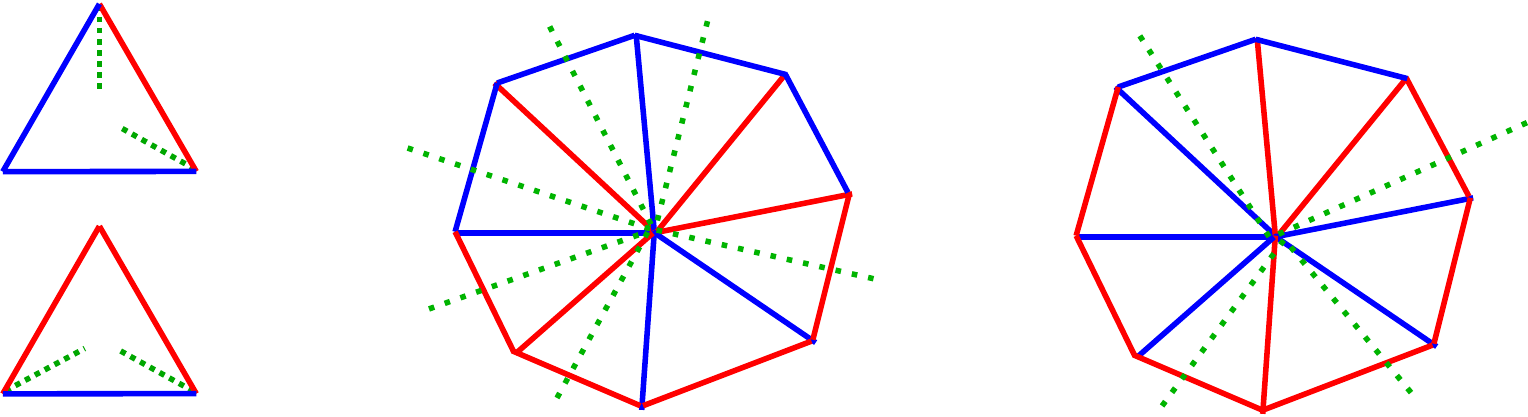}
\caption{\label{fig:caus2d} Left: the two elementary Minkowskian building blocks of CDT without preferred foliation in two dimensions; in standard CDT only the lower one is present. Dotted lines indicate light rays through the corner points. 
Centre: gluing these triangles together can 
result in local causality violations, like too many light cones meeting at a vertex, as shown. Right: at a causally 
well-behaved vertex, one crosses exactly four light-like directions when going around the 
vertex.
}
\end{figure}

The primary motivation for studying ``CDT without preferred foliation" or ``locally causal DT" 
was to dissociate the causal structure of CDT from the preferred 
discrete foliation of the individual path integral histories, 
while staying in the purely geometric set-up of dynamical triangulations. 
The formulation introduced in \cite{jordanloll1} achieves this by allowing for additional simplicial building blocks 
without enlarging the set of allowed edge
lengths, which are still given in terms of the relations (\ref{edges}). The gluing rules for these generalized elementary building blocks
have to obey {\it local causality conditions}, to ensure that the local light cone structure is regular. (Fig.\ \ref{fig:caus2d} illustrates
the analogous situation in two dimensions, where the local causality conditions forbid the occurrence of a causality-violating
configuration like the one depicted in the middle.)
Unlike in standard CDT, a well-defined local causal structure is not
guaranteed automatically, because the sandwich structure described earlier in this section is no longer present. 
Consequently, there is no preferred discrete lattice foliation and no preferred notion of time, although the CDT Wick rotation
$\alpha\!\mapsto\! -\alpha$ is still applicable. Note that this also implies that
there is no notion of performing the simulations for a fixed time extension $t_{tot}$.

For simulation-technical reasons the most convenient choice of the overall topology is not $S^1\!\times\! S^2$ with
a cyclically identified time direction, but open boundary conditions where the two spatial two-spheres at the boundaries are
compactified to points, resulting in an effective $S^3$-topology. 
Because the set of building blocks is enlarged and the Monte Carlo moves are more numerous and geometrically involved,
the development of the simulation software turned out to be extremely complex. The system was simulated in a 
region of
phase space spanned by the bare inverse Newton constant $k$ in the range $k\! \in\! [-1,1]$ and the asymmetry parameter
$\alpha$ in the range $|\alpha |\! \in\! [0.5,1]$, and for fixed volumes
$N_3\!\leq\! 160k$. Just like standard CDT in three dimensions \cite{3dcdt}, the system exhibits a phase transition as a function of $k$,
with a physical, ``de Sitter-like" phase at low vertex density $N_0/N_3$ for $k\! < \! k^{crit}\!\approx\! 0.25$, and one at 
high vertex density above.

In order to compare with the results of standard CDT, one needs to re-introduce a notion of time, by referring only to intrinsic, 
geometric properties of the configurations. A natural choice, which can be used to associate a time
variable with individual simplices and subsequently to define the notion of a spatial slice of constant time,
is an average version of ``link distance to the poles" of the three-sphere (see \cite{jordanloll2,Jordan} for details).
Volume profiles as a function of this intrinsic time $\bar{t}$ were analyzed throughout phase space for $k\! <k^{crit}$ and found to
both exhibit finite-size scaling (leading to a global Hausdorff dimension compatible with 3) and a high-precision
matching with a cos$^2(\bar{t}/const)$-shape expected for Euclidean three-dimensional de Sitter space. In addition, when
approaching the upper kinematical range of $\alpha\! =\! 0.5$ inside the physical phase, spacetime shows an increasing
tendency to foliate dynamically, in the sense of resembling more and more that of a standard CDT configuration. 

This provides additional nontrivial evidence that the 
preferred time slicing of Causal Dynamical Triangulations is not an essential element of its background structure, 
which is in line with the expectation formulated earlier. 
It also confirms the experience so far with dynamical geometric systems of DT-type, which suggests that universality is strong and 
there are very few universality classes.
Because of the formidable difficulties that are already present
in three dimensions, a similar analysis in four dimensions seems neither to be
in reach nor to have high priority at this moment.\footnote{Perhaps more feasible, although less relevant physically, is the
analytic solution of CDT without preferred slicing in two dimensions, to establish its equivalence or otherwise with
standard CDT, which so far has not been achieved (see \cite{lollruijl} for some numerical hints).} 
Note that the absence of the layered structure and preferred time of standard CDT 
implies that reflection positivity at the regularized level is not manifest in this generalized variant of CDT,
which means that the status of unitarity is less clear-cut. 

\section{Phase structure and phase transitions}
\label{phase:sec}

The regularized implementation in CDT of the Wick-rotated path integral for pure gravity has the form
\begin{equation}
Z^{\rm CDT}_{eu}=\sum_{T\in {\cal T}} \frac{1}{C(T)}\ {\rm e}^{-S^{\rm CDT}_{eu}[T]}.
\label{partition}
\end{equation}
The Euclidean action $S^{\rm CDT}_{eu}$ in the exponent is the result of analytically continuing the 
parameter $\alpha$ in the Lorentzian action (\ref{actlor}) to $-\alpha$, according to the prescription outlined 
in Sec.\ \ref{time:sec}
above\footnote{In order to satisfy the triangle inequalities, one needs $\alpha\! >\! 7/2$ after the Wick rotation.}.
The analytically continued action is purely imaginary, and combines with the imaginary $i$ in the
exponent of the path integral (\ref{cdtact}) to yield the corresponding real Boltzmann weight in the
partition function (\ref{partition}). A form of the Wick-rotated action that is convenient in simulations 
is obtained by substituting $\alpha$ by a new parameter $\Delta$, such that
\begin{equation}
S^{\rm CDT}_{eu}[T]\!=\! - (\kappa_0 +6\Delta) N_0(T) + \kappa_4 (N_{41}(T)+N_{32}(T)) +
\Delta (2 N_{41}(T)+N_{32}(T)),
\label{eqS1a}
\end{equation}
where for ease of notation we have introduced $N_{41}\! :=\! N_4^{(4,1)}$ and $N_{32}\! :=\! N_4^{(3,2)}$,
which we will use from now on.\footnote{In some of the recent literature, the coupling $\kappa_4$ is redefined
to $K_4\! := \! \kappa_4\! +\!\Delta$, resulting in a rewriting of the action (\ref{eqS1a}) to
$S^{\rm CDT}_{eu}[T]\!=\! - (\kappa_0 +6\Delta) N_0(T) + K_4 (N_{41}(T)+N_{32}(T)) +
\Delta N_{41}(T)$. }  
To arrive at expression (\ref{eqS1a}) from (the analytic continuation of) the action (\ref{actlor}), one also has performed a
linear redefinition of the bare couplings from $(k_b,\lambda_b)$ to $(\kappa_0,\kappa_4)$, see also ref.\ \cite{physrep}. 
The so-called {\it asymmetry parameter} $\Delta$ depends on the parameter $\alpha$ -- describing the finite, relative scaling between
the chosen geodesic lengths of time- and space-like links -- in such a way that for equilateral triangulations (after Wick rotation) 
we have $\Delta\! =\! 0$.
However, it should be emphasized that neither this nor any other value of $\Delta$ is in any way distinguished 
on physical grounds {\it a priori}. At the level of the discretized action, different values correspond to building blocks which are
more or less elongated in the time direction, but equally well suited to describing the (discretized) Einstein-Hilbert action.
However, it has turned out that in the region of phase space associated with interesting physical behaviour, where
contributions from the action and the entropy of states combine nonperturbatively, $\Delta$ {\it is} an independent coupling
constant, which may need to be fine-tuned in a scaling limit near one of CDT's critical points. 

The coupling $\kappa_0$ in the action (\ref{eqS1a}) is, up to a numerical constant, the inverse bare gravitational coupling,
while $\kappa_4$ plays the role of bare cosmological constant. In CDT computer simulations, the lattice volume is (approximately)
held fixed, to maximize efficiency. Since lattice volumes are necessarily finite, this is equivalent to fixing $\kappa_4$ to its (pseudo-)critical value, see \cite{physrep} for an extended discussion. As usual in lattice field theory, one extrapolates to the
continuum theory by studying the scaling properties of observables for a sequence of large and growing discrete lattice volumes,
using so-called finite-size scaling techniques. Results reported for quantum observables, including in this review, 
usually refer to an extrapolation of 
measurements to infinite lattice volume, also known as the thermodynamic limit.

To summarize, the phase diagram of CDT quantum gravity  
is spanned by two 
bare coupling constants that can be varied independently, $\kappa_0$ and $\Delta$. 
In standard simulations of CDT, two different prescriptions have been used to enforce an approximate lattice volume fixing, in terms of 
either the total number of four-simplices, $N_4\! =\! N_{41}\! +\! N_{32}$, or the number $N_{41}$ of (4,1)-simplices only.
In physically interesting regions of phase space these lead to equivalent results,
since at a given point $(\kappa_0,\Delta)$ the two four-simplex types occur approximately at a fixed ratio \cite{physrep}, where the
ratio depends on the location in phase space. However, as will be mentioned
below, the different prescriptions can influence the way in which phase transitions manifest themselves.
To allow the volume to fluctuate in a small interval around a chosen target value
$\bar{N}_4$ or $\bar{N}_{41}$, one includes a corresponding quadratic volume-fixing term
of the form 
\begin{equation}
S_{\it fix}^{\bar N_4}(N_4)=\varepsilon (N_4-\bar N_4)^2\label{sfix1}
\end{equation}
or
\begin{equation}
S_{\it fix}^{\bar N_{41}}(N_{41})=\varepsilon (N_{41}-\bar N_{41})^2
\label{sfix2}
\end{equation}
in the bare action, where in either case $\varepsilon$ is an appropriately chosen small, positive parameter. 
This is needed because the update moves in the simulations are not volume-pre\-serving. 

\begin{figure}[t]
\centering
\scalebox{0.85}{\includegraphics{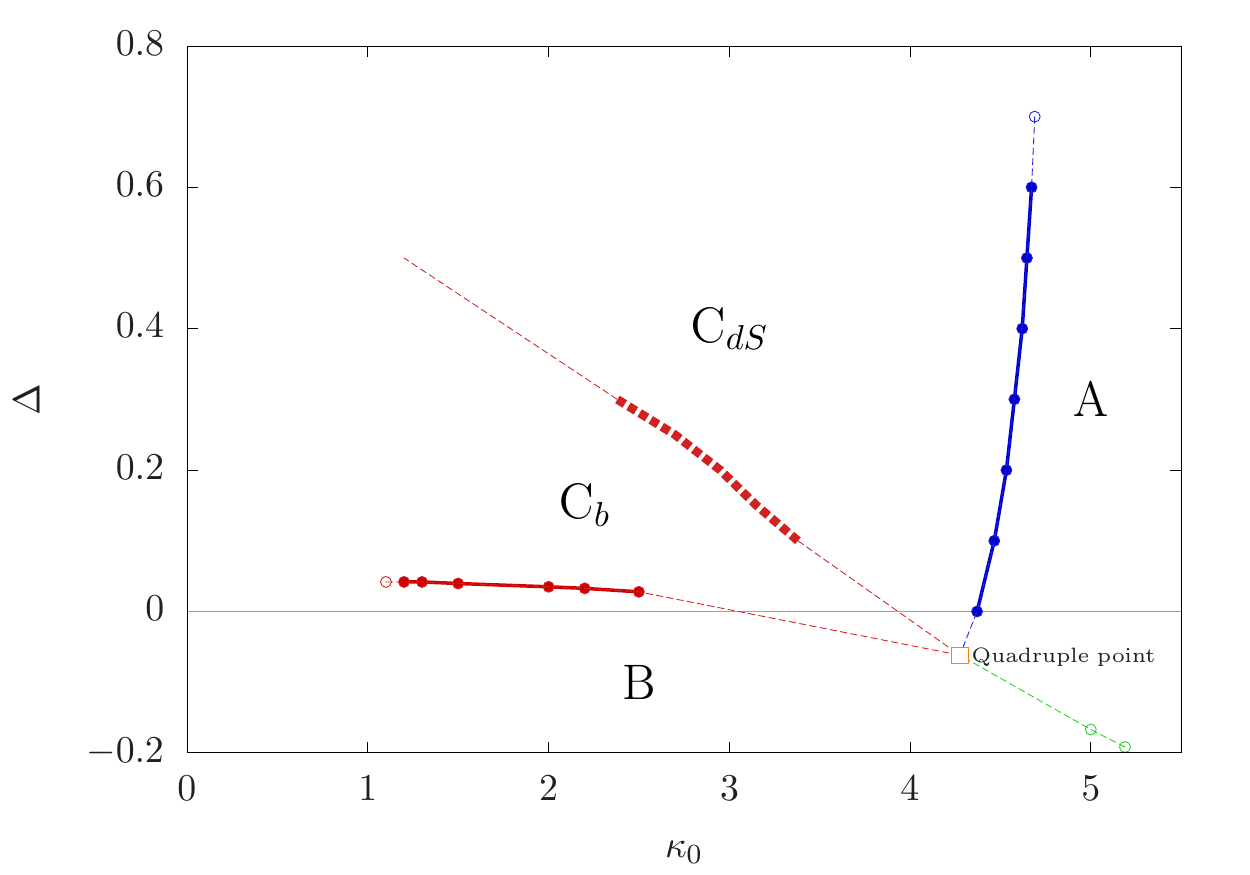}}
\caption{CDT phase diagram spanned by the bare couplings $\kappa_0$ and $\Delta$, consisting of 
the de Sitter phase $C_{dS}$, the bifurcation phase $C_{b}$, and two unphysical phases $A$ and $B$. 
(Fat dots and squares refer to actual measurements.)
}
\label{phasedia}
\end{figure}

The current understanding of the phase structure of CDT quantum gravity is illustrated by its phase diagram,
depicted in Fig.\ \ref{phasedia}.\footnote{The presence of a quadruple point in Fig.\ \ref{phasedia} is speculative. New evidence
from measurements for spatial tori, as described in more detail in Sec.\ \ref{sec:torus},
suggest there are two triple points instead.} A new feature compared to earlier discussions \cite{physrep} is the
phase transition line inside the physically interesting phase $C$, dividing it into a {\it de Sitter phase} $C_{dS}$
and a {\it bifurcation phase} $C_b$. 
It was discovered during investigations of the system with the help of the effective transfer matrix, 
to be described in more detail below. It had previously evaded attention because for the system sizes studied the global 
volume profile observable used to distinguish the different physical behaviours inside phases $A$, $B$ and $C$
is not sensitive to the $C_{b}$-$C_{dS}$ phase transition. 
The likely explanation is the nature of the bifurcation phase, whose distinguishing feature is 
a particular localized structure \cite{bifurcation2,newphase} that does not affect
the overall shape of the universe, at least not near the transition to the de Sitter phase. 
A similar observation holds for the order parameters
that were used to locate the previously known transition lines $A$-$C$ and $B$-$C$ and determine their 
order \cite{CDTHL,trans1,trans2,practitioner,Jordan}.
Although they can also be used for the $C_{b}$-$C_{dS}$ transition, the corresponding signals are much weaker
than those of alternative order parameters that are attuned to more local features of the 
geometry \cite{bifurcation2,cgj}.  

In line with the usual logic of lattice quantum field theory, results on the order of CDT phase transitions are highly significant,
because any transition point of second or higher order is a natural candidate for taking a scaling limit, and thus for the 
potential existence of a well-defined continuum limit independent of regularization details. 
CDT quantum gravity is currently unique in combining strong evidence for higher-order transition points and the applicability of
Wilsonian renormalization group methods (see also Sec.\ \ref{sec:rengroup} below) with strong evidence of the existence of a 
classical limit and the emergence of four-dimensional spacetime.  

\begin{figure}[t]
\centering
\includegraphics[width=0.55\textwidth]{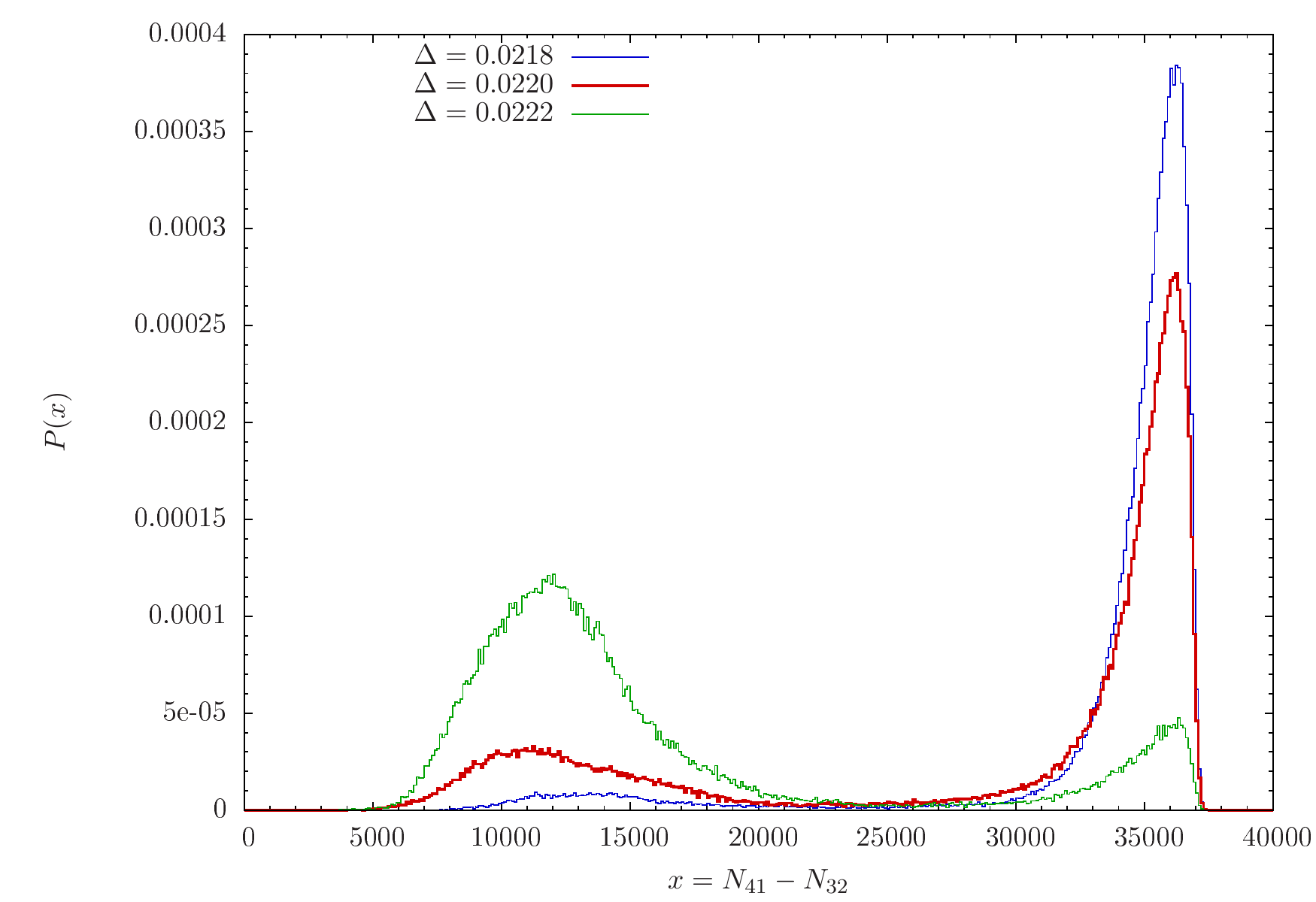}
\caption{
Probability distribution $P(x)$ of the order parameter $x\! =\! N_{41}\! -\! N_{32}$, closely related to (\ref{conj}), 
for fixed volume $N_4\! =\! 40k$, $t_{tot}\! =\! 80$ and $\kappa_0\! =\! 2.2$. The measurements at three 
different $\Delta$-values near the critical point $\Delta_c\!\approx\! 0.0220$ on the $B$-$C$ transition line
exhibit the double-peak structure characteristic for this volume fixing. 
}
\label{fig:distN4}
\end{figure}

A first analysis of the order of CDT phase transitions
was published in \cite{CDTHL} on the basis of measurements for geometries of topology $S^1\times S^3$, 
time extension $t_{tot}\! =\! 80$ (cyclically identified) and fixed
four-volume $N_4$. From the evolution in Monte Carlo time of the order parameter $N_0/N_4$ 
and its histogram at the $A$-$C_{dS}$ (former $A$-$C$) transition,
it was concluded that this transition was most likely of first order. Repeating a similar analysis for the order parameter 
\begin{equation}
conj(\Delta):=(-6 N_0+\! 2 N_{41}\! +\! N_{32}), 
\label{conj}
\end{equation}
the quantity conjugate to $\Delta$ 
in the action (\ref{eqS1a}), at the $B$-$C_b$ (former $B$-$C$) transition yielded an inconclusive result. 
While the probability distribution for $conj(\Delta)/N_4$ exhibited a double-peak structure, 
typically associated with two metastable states on either side of a first-order transition, 
the two peaks showed a tendency to approach each other for
growing four-volume $N_4$, pointing to a second-order transition instead. 

This motivated a careful study of the same system for larger volumes of up to $N_4\! =\! 160k$ \cite{trans1,trans2},
repeating the histogram analysis, and supplementing it with independent criteria to establish the order of
the phase transitions, namely, a measurement of the shift exponent (determined from the volume-dependence of 
the location of the maximum of the susceptibility) and an analysis of Binder cumulants. The results were mutually consistent
and within measuring accuracy established the second-order nature of the $B$-$C$ transition\footnote{The same
behaviour and a compatible numerical value for the shift exponent are found when using $t_{tot}\! =\! 2$ in an
effective transfer matrix treatment \cite{newphase}.\label{foot1}}, while confirming the
first-order character of the $A$-$C$ transition. 

\begin{figure}
\centering
\scalebox{0.65}{\includegraphics{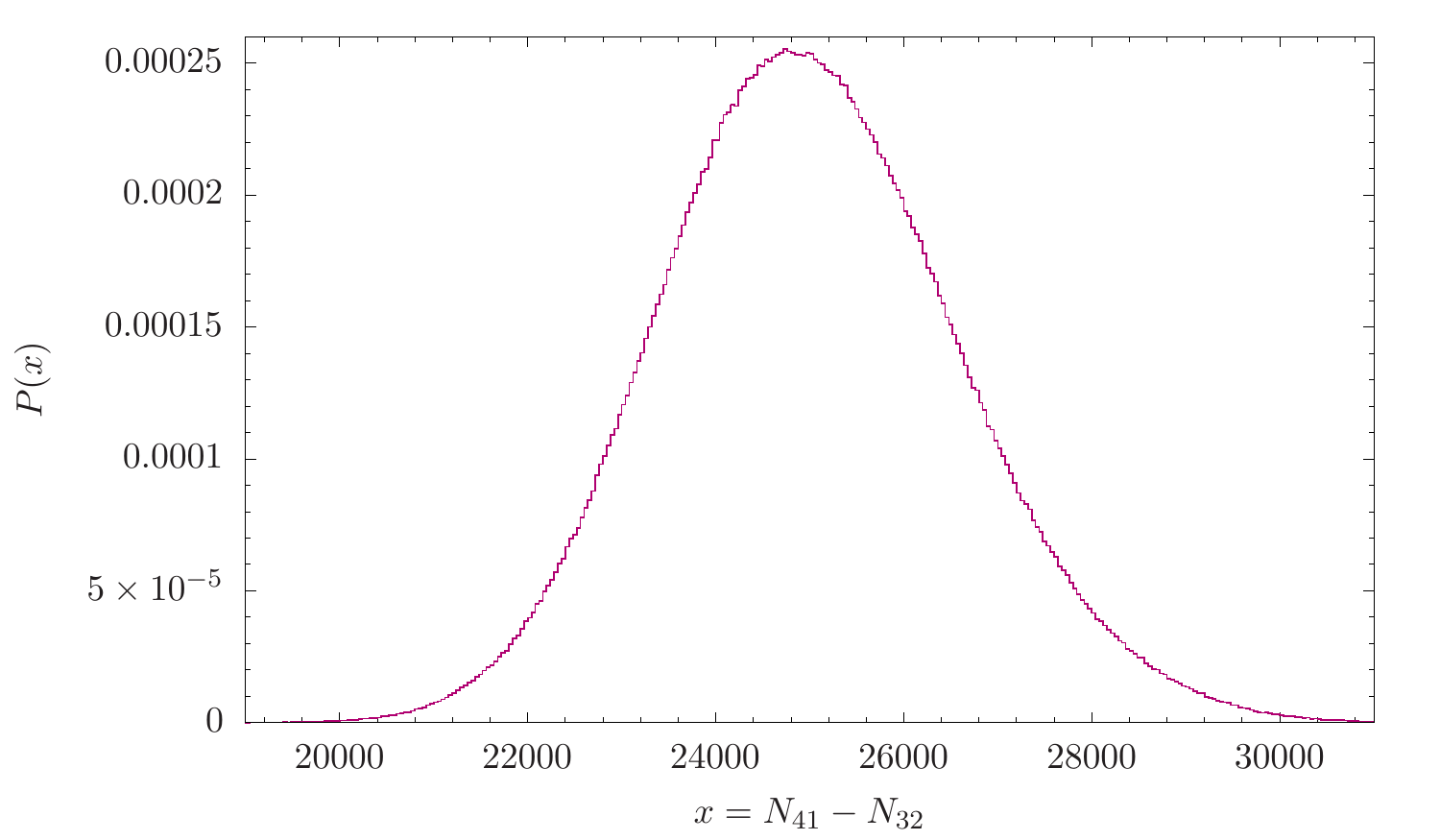}}
\caption{
For fixed volume $N_{41}\! =\! 33k$, $t_{tot}\! =\! 80$ and $\kappa_0\! =\! 2.2$, the measured 
probability distribution $P(x)$ of the order parameter $x\! =\! N_{41}\! -\! N_{32}$ 
close to the critical point $\Delta_c\!\approx\! 0.0220$ on the $B$-$C$ transition line has only a single peak.
}
\label{fig:distN41}
\end{figure}

A more recent study of the $B$-$C$ transition \cite{newphase} has confirmed its second-order status, and
somewhat surprisingly found that the (potentially misleading) double peak in histograms simply
disappears when $N_{41}$ instead of $N_4$ is kept fixed (see Figs.\ \ref{fig:distN4} and \ref{fig:distN41}). 
A detailed quantitative analysis in the 
same work revealed that the origin of the two different behaviours has to do with the ``entropy factor"
${\cal N}(N_{41}, N_{32},N_0)$, counting the number of configurations for given values of the bulk
variables $N_{41}$, $N_{32}$ and $N_0$, and the way in which it functionally depends on these
variables. This confirms that the double-peak structure at the $B$-$C$ transition first reported in \cite{CDTHL}
in no way contradicts the conclusion of \cite{trans1,trans2} that this phase transition is of second order. 

\subsection{Effective transfer matrix and new phase transition}
\label{subsec:eff}

Along with the layered structure of standard CDT configurations and the associated discrete proper time label $t$ described in
Sec.\ \ref{time:sec} above, CDT after the Wick rotation possesses a well-defined transfer matrix $\cal M$ \cite{cdt1,cdt2,physrep}. 
Its matrix elements describe the amplitude of going from one three-geometry $T_3$ (a spatial three-dimensional simplicial
manifold, usually of topology $S^3$) at time $t$ to another such geometry in the next time slice $t+1$,
\begin{equation}
\langle T_3(t+1)|{\cal M}|T_3(t)\rangle=\sum_T \frac{1}{C(T)}\, {\rm e}^{-S^{\rm CDT}_{eu}[T]},
\end{equation}
where the sum is taken over all four-dimensional sandwich geometries $T$ with $\Delta t\! =\! 1$ that extrapolate between the initial
geometry $T_3(t)$ and the final geometry $T_3(t+1)$. The transfer matrix acts on the infinite-dimensional vector space
labelled by distinct three-dimensional triangulations $T_3\!\in\!{\cal T}_3$. Its elements $|T_3\rangle$ are equipped 
with the natural scalar product
\begin{equation}
\langle T_3 | T_3'\rangle=\frac{1}{C(T_3)}\, \delta_{T_3,T_3'},\;\;\;\; 
\sum_{T_3} |T_3\rangle C(T_3)\langle T_3| =\hat{\bf 1}, \;\;\; T_3,T_3'\in {\cal T}_3,
\end{equation}
where as before $C(T)$ denotes the size of the automorphism group of the triangulation $T$. Transition amplitudes
between spatial three-geometries that are $t_{tot}$ time steps apart can be obtained by iterating the transfer matrix 
$t_{tot}$ times, resulting in ${\cal M}^{t_{tot}}$. Note that the Hilbert space decomposes into subspaces spanned by
three-geometries sharing the same discrete three-volume, namely, the number $n_T\! :=\! N_3(T)$ of tetrahedra of a given
$T\in {\cal T}_3$. The number of three-dimensional simplicial manifolds with volume $n$ is finite and for large volume
grows exponentially with $n$.\footnote{The notation $n$ is used in the literature 
both for the discrete three-volume of spatial
slices and for {\it twice} the discrete three-volume, sometimes even in the same paper \cite{TMfirst}. 
Note that in terms of the former interpretation one has $2n_T\! =\! N_{41}(T)$ for a configuration $T$ with compactified time, because
each space-like tetrahedron in a slice of constant integer time is shared by two four-simplices of type (4,1). 
Wherever the difference matters, for example, in some explicit fitting formulas for the effective action, readers are advised to
consult the original papers.}

The idea of the {\it effective transfer matrix} $M$, introduced in \cite{TMfirst}, 
is the construction of a simpler quantity than $\cal M$ itself, which can nevertheless capture important information of 
the full system.\footnote{See \cite{bogacz} for an attempt to
model the phase structure of CDT in terms of a one-dimensional lattice model whose transfer matrix resembles
the effective transfer matrix discussed here.} 
As will become clear below, this has proven a
valuable strategy for describing the dynamics of the three-volume and its associated correlators, as well as for
the analysis of CDT's phase structure. As mentioned earlier, it also led to the discovery of the $C_b$-$C_{dS}$
phase transition. The states $|n\rangle$ associated with the effective transfer matrix are labelled by
the three-volume $n$ only, and can be thought of as arising from uniform probability distributions of states $|T_3\rangle$
sharing the same three-volume $n$. Recall that the three-volume as a function of proper time plays an important role in CDT, since
its expectation value, the {\it volume profile} $\langle n_t \rangle$, in phase $C$ can be matched to that of a
(Euclidean) de Sitter space, and the behaviour of the quantum fluctuations of the three-volume
to those of a semiclassical minisuperspace treatment \cite{desitter1,desitter2,semi2,gizbert1}. 

It is entirely nontrivial that a {\it reduced transfer matrix} with matrix elements $\langle n_{t+1}|M|n_t \rangle$ should 
exist and can reproduce these results, but this is what has been found \cite{TMfirst,proceedings1,bifurcation1,gizbert1}. A
big computational advantage of this formalism, wherever it applies, is the direct accessibility of its matrix elements
in simulations, and the fact that it avoids time-consuming simulations of full-size CDT confi\-gurations with $t_{tot}\!\approx\! 80$, 
in favour of measurements involving a small number of time slices.
In \cite{TMfirst}, transfer matrix elements $\langle n_{t+1}|M|n_t \rangle$ were obtained from combining measurements
of various probability distributions of pairs $(n_1,n_2)$ of three-volumes, for (cyclically identified) total times in
the range $t_{tot}\in[2,6]$, at $(\kappa_0,\Delta)\! =\! (2.2,0.6)$, and using a novel, local volume fixing. Independent of
the specific choices made and for a wide range of three-volumes 
the measured matrix elements turns out to be well described by the ansatz
\begin{equation}
\langle n_{t+1}|M|n_t \rangle\propto {\rm e}^{-L_{\rm eff}[n_t,n_{t+1}]},
\label{mael}
\end{equation}
with the effective Lagrangian 
\begin{equation}
L_{\rm eff}[n,m]=\frac{1}{\Gamma}\frac{(n-m)^2}{(n+m)}+\mu \Big(\frac{n+m}{2}\Big)^{1/3}\!\! -\lambda\Big(\frac{n+m}{2}\Big).
\label{lagra}
\end{equation}
One recognizes the functional form of $L_{\rm eff}$ from the effective action for the three-volume reconstructed from 
the covariance matrix of spatial volume fluctuations around the semiclassical expectation value $\langle n_t \rangle$
in the full CDT simulations \cite{desitter1,desitter2,semi2}. It consists of a kinetic term $\propto\! 1/\Gamma$, 
a three-dimensional scalar curvature term
$\propto\! \mu$ and a cosmological constant term $\propto\! \lambda$, where the
parameters $\Gamma$, $\mu$ and $\lambda$ depend on the couplings $\kappa_0$ and $\Delta$.
Correction terms $\propto\! {\cal O}((n+m)^{-1/3})$
to the essentially classical form of the Lagrangian (\ref{lagra}), indicative of higher powers of the scalar curvature,
could not be extracted reliably from small-volume measurements, which are subject to both finite-size effects
and lattice artefacts.

In a follow-up investigation \cite{bifurcation1} the matrix elements (\ref{mael}) were extracted from measurements
of a periodically identified two-slice system with $t_{tot}\! =\! 2$. It established that in phase $C$ one can to good accuracy
{\it reconstruct} from these data both the average volume profile $\langle n_t \rangle\propto \cos^3 (t/const) $ and the covariance
matrix 
\begin{equation}
C_{tt'}=\langle (n_t-\langle n_t\rangle )(n_{t'}-\langle n_{t'}\rangle )\rangle
\label{coma}
\end{equation}
previously measured in full CDT simulations\footnote{Strictly speaking, this involved an extrapolation for the
matrix elements with large arguments $n_t>700$, which appears well justified, see \cite{bifurcation1} for details.}.
The passing of this important consistency check
raised the interesting question of whether the effective transfer matrix can be used as a new tool to learn 
more about the rest of the phase space of CDT quantum gravity. Trying to fit the measured matrix elements at
the point $(\kappa_0,\Delta)\! =\! (5.0,0.4)$ inside phase $A$ to an effective Lagrangian similar to (\ref{lagra}) 
produced evidence for a Lagrangian of the form
\begin{equation}
L_{\rm eff}^{(A)}[n,m]=\mu (n^\alpha+m^\alpha)-\lambda (n+m),
\end{equation}
apparently without a kinetic term, and with a best fit yielding $\alpha\! \approx\! 0.5$ \cite{bifurcation1}. However, one should keep in
mind that phase $A$ most likely does not have any physical interpretation in terms of either gravity or (higher-dimensional) 
geometry, which means that an ansatz inspired by such an assumption may lead to misleading conclusions.
Another finding of \cite{bifurcation1} has a more straightforward interpretation:
the kinetic term in the Lagrangian (\ref{lagra}) vanishes gradually as one approaches the $A$-$C$ transition at fixed $\Delta$ 
{\it from inside phase} $C$ and is zero at the transition. This is consistent with the idea
that inside phase $C$ the entropy of the microscopic configurations prevents the system from developing a conformal 
divergence, but that for increasing $\kappa_0$ the negative kinetic term of the conformal mode in the bare action will
eventually become dominant \cite{wick,entropic}.  

\begin{figure}
\centering
\hspace{1.5cm}\includegraphics[width=0.75\textwidth]{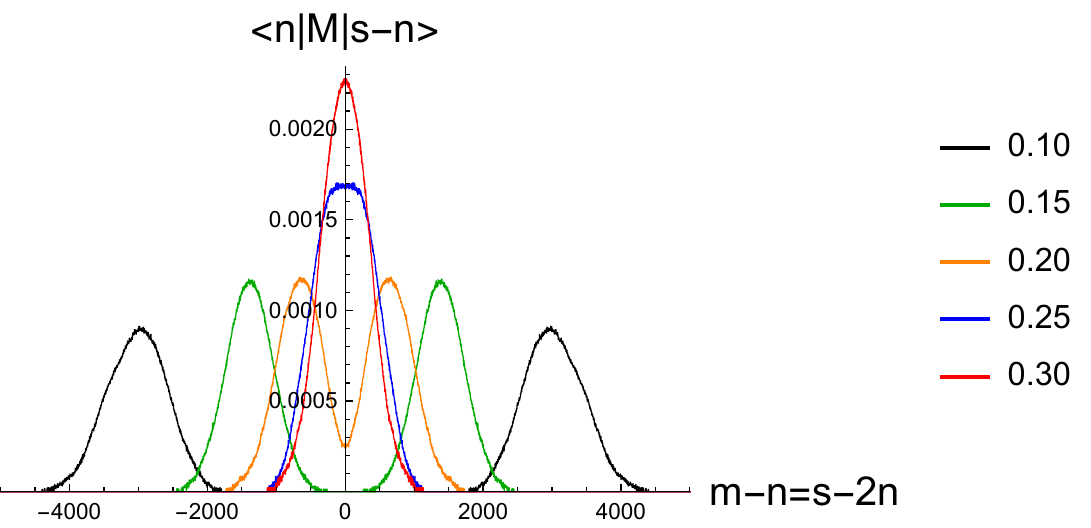}
\caption{\label{fig:cross} 
Measured matrix elements $\langle n|M|s-n\rangle$
for $s\! =\! 30k$, at fixed $\kappa_0\! =\! 2.2$ and for five different $\Delta$-values. Since $s\! =\! m+n$ stays constant,
these measurements illustrate the behaviour of the kinetic part of (\ref{shi}): the smaller $\Delta$, the more pronounced 
is the ``bifurcation" into two Gaussians. The $C_b$-$C_{dS}$ phase transition (merger to a single Gaussian) appears to lie
in the interval $\Delta\in[0.25,0.3]$. [Figure from \cite{bifurcation2}.]}
\end{figure}

While the $A$-$C$ transition is thus associated with a clear signal in the effective Lagrangian for the spatial
volume, this does not seem to be true for the $B$-$C$ transition, although the latter {\it is} visible in the effective 
transfer matrix approach when looking at other order parameters (c.f. footnote \ref{foot1}).

Instead, monitoring the effective transfer matrix throughout phase $C$ as phase $B$ is approached
has revealed the existence of a new phase transition {\it inside} phase $C$. This implies that the previous phase $C$ 
is now divided
into the de Sitter phase $C_{dS}$ and the bifurcation phase $C_b$, as indicated in the phase diagram of Fig.\ \ref{phasedia}.
The name of the latter derives from a property of the effective transfer matrix, whose kinetic part exp$[-(n-m)^2/(\Gamma (n+m))]$
in the bifurcation phase 
splits into a sum of two Gaussians with a relative shift, giving rise to matrix elements of the form
\begin{eqnarray}
\langle n|M|m\rangle \!\!\!\!&=&\!\!\!\!\Bigg[\exp\Big(-\frac{1}{\Gamma}\frac{((n-m)-c[n+m])^2}{n+m}\Big)\nonumber\\
&&\hspace{1.5cm} +\exp\Big(-\frac{1}{\Gamma}\frac{((n-m)+c[n+m])^2}{n+m}\Big) \Bigg]\, {\cal V}[n+m],
\label{shi}
\end{eqnarray}
where ${\cal V}[n+m]$ is the exponential of the potential part of eq.\ (\ref{lagra}) or some variant thereof \cite{bifurcation1,bifurcation2}. 
The function $c[n+m]$ in (\ref{shi}) goes to $c_0 (n+m+s_b)$ for large spatial volumes $n+m\!\gg\! s_b$ and to zero for
small volumes $n+m\!\ll\! s_b$. The two parameters $c_0$ and $s_b$ depend on the bare coupling constants $\kappa_0$ and
$\Delta$. Coming from inside $C_b$, 
the $C_b$-$C_{dS}$ transition is associated with the limits $c_0\!\rightarrow\! 0$ and $s_b\!\rightarrow\!\infty$, for which
the two Gaussians merge into one, resulting in the previous form (\ref{mael}) for the matrix elements. Going the other way 
in phase $C_b$, by lowering $\Delta$ for fixed $\kappa_0$, one eventually enters phase $B$. In the process, the two Gaussians
of (\ref{shi}) move further apart in a smooth way \cite{bifurcation2}. This is consistent with the overall geometry in phase $B$, where the
universe is known to collapse to a single time slice. Expanding the ansatz (\ref{shi}) for small $c_0$ and large $s_b$ and
interpreting the result as an effective Lagrangian like (\ref{lagra}), one finds for
sufficiently large spatial volumes an ``effective" kinetic term with a negative sign, which the authors of \cite{bifurcation2,recent} 
suggest to interpret as a scale-dependent effective signature change of the metric. 
Since the total volume is just one of an infinity of modes of the spatial metric (in the continuum),
this is a rather far-reaching conjecture. It may be worth re-examining once the physics of the bifurcation phase
is better understood.

\begin{figure}
\centering
\includegraphics[width=0.65\textwidth]{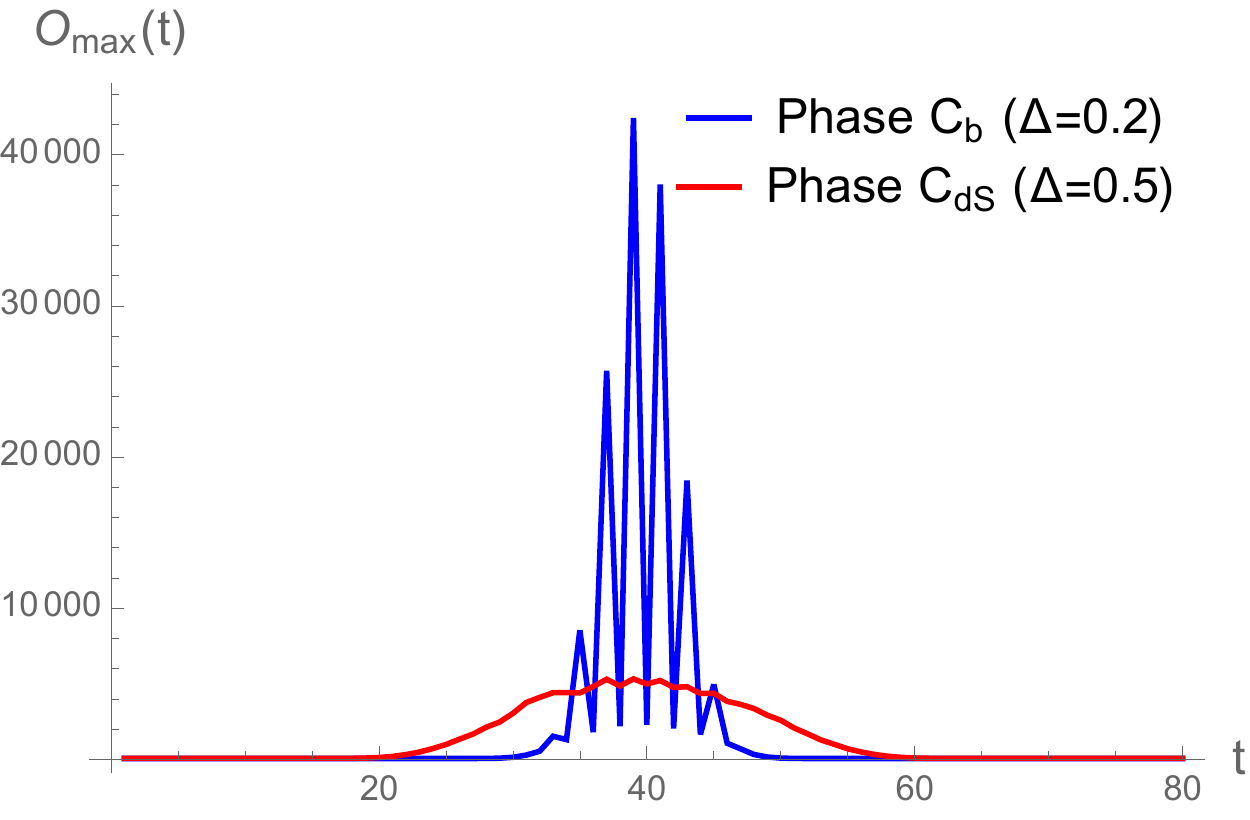}
\caption{\label{fig:alter} 
In the bifurcation phase $C_b$ ($\Delta\! =\! 0.2$), the maximal coordination 
number $O_{\rm max}(t)\! :=$max$[O(v(t))]$ jumps to a very large value on every second spatial slice, compared with the situation in
the de Sitter phase $C_{dS}$ ($\Delta\! =\! 0.5$). Measurements taken for $t_{tot}\! =\! 80$,
$N_{41}\! =\! 160k$, 
$\kappa_0\! = \! 2.2$, and averaged over many configurations. [Figure from \cite{higher}.]}
\end{figure}

The location of the new phase transition was first estimated from the vanishing of the parameter 
$c[s]\equiv c[m+n]$ appearing in the exponentials on the right-hand side of eq.\ (\ref{shi}), in a two-slice
system with volumes of up to $s\! =\! 60k$ \cite{bifurcation2}. The existence of the new $C_b$-$C_{dS}$ transition line
was confirmed in simulations at fixed volumes $N_{41}\! =\! 80k$ and $160k$ for a standard system of CDT configurations
with $t_{tot}\! =\! 80$ \cite{cgj}. Its authors studied the behaviour of three order parameters
near the transition, $conj(\Delta)$ of eq.\ (\ref{conj}), and two new quantities introduced in \cite{bifurcation2}, which both
refer to a particular spatial slice at time $t_0$.\footnote{the slice closest to the maximum of the
volume profile $n_t$ of a given configuration, or some variant involving the slice with the vertex of highest order, see 
\cite{bifurcation2,cgj,higher} for details} They are 
\begin{eqnarray}
&&OP_1:=|\bar{R}(t_0)-\bar{R}(t_0+1)|,\label{op1} \\
&&OP_2:=|\max [O(v(t_0))]-\max[ O(v(t_0+1)]|,\label{op2}
\end{eqnarray}
where up to an irrelevant constant $\bar{R}(t)\! =\! 2\pi N_0(t)/N_3(t)$ is the average scalar 
curvature of the spatial slice at integer time $t$, 
using the deficit angle prescription of Regge calculus \cite{regge}, and max$[O(v(t))]$ denotes the maximal coordination 
number (the number of four-simplices meeting at $v$) of any vertex
$v$ contained in the slice $t$. 
The choice of the new order parameters is motivated by a distinct ``modulation" of several geometric properties of 
the CDT configurations in phase $C_b$, with characteristic time period $\Delta t\! =\! 2$. For example, 
the maximal coordination number of any vertex in a given spatial slice oscillates strongly between alternating slices
(Fig.\ \ref{fig:alter}).

Varying $\Delta$ at fixed $\kappa_0\! =\! 2.2$ and for $N_{41}\! =\! 160k$, 
its (pseudo-)critical value 
at the new phase transition was estimated as $\Delta^{crit}\! =\! 0.35\pm 0.01$
from the peaks of the susceptibilities of the parameters (\ref{op1}) and (\ref{op2}) \cite{cgj}.
A preliminary investigation into the order of the transition in the same work did not yield conclusive results, 
but a subsequent, comprehensive analysis of several indicators established strong evidence that the
$C_b$-$C_{dS}$ transition is of second or higher order \cite{higher}. The indicators measured were the
dependence of the pseudo-critical point $\Delta^{crit}$ on the lattice volume (the shift exponent used
earlier for the other transitions \cite{trans1,trans2}), the peak separation in
the histogram of the normalized order parameter $OP_2/N_{41}$ as a function of the volume, and the frequency
of parameter jumps of $OP_2$ as a function of Monte Carlo time, normalized by the autocorrelation time.
Notable for the simulations near the transition is a severe critical slowing-down, associated with very long
autocorrelation times, as documented in \cite{cgj,higher}.

\subsection{The bifurcation phase}

Before the discovery of the bifurcation phase, much, though by far not all, of the analysis of CDT geometry
in the ``well-behaved" $C$-phase was done at the point $(\kappa_0,\Delta)\! =\! (2.2,0.6)$, thought
of as a generic point well inside phase $C$. This analysis, which includes the classic results of the de Sitter 
volume profile, with corresponding quantum fluctuations, and the dynamical dimensional reduction on short scales, 
remains valid inside the de Sitter phase $C_{dS}$, where this point is located. 
On the other hand, scaling behaviour compatible with a four-dimensional universe on large scales and a de Sitter-like
behaviour were also found in at least part of what is now identified as phase $C_b$ (this is exactly
why the $C_b$-$C_{dS}$ transition remained undetected for a long time). 
For the time being the question remains open of whether and to what extent -- in addition to the de Sitter phase -- also 
the bifurcation phase is interesting from a quantum gravity point of view, and has a good classical limit. 
More details need to be known on how the two phases differ in their geometric
properties, which will in turn help us understand the nature of the new phase transition.
\begin{figure}
\centering
\scalebox{.56}{\includegraphics{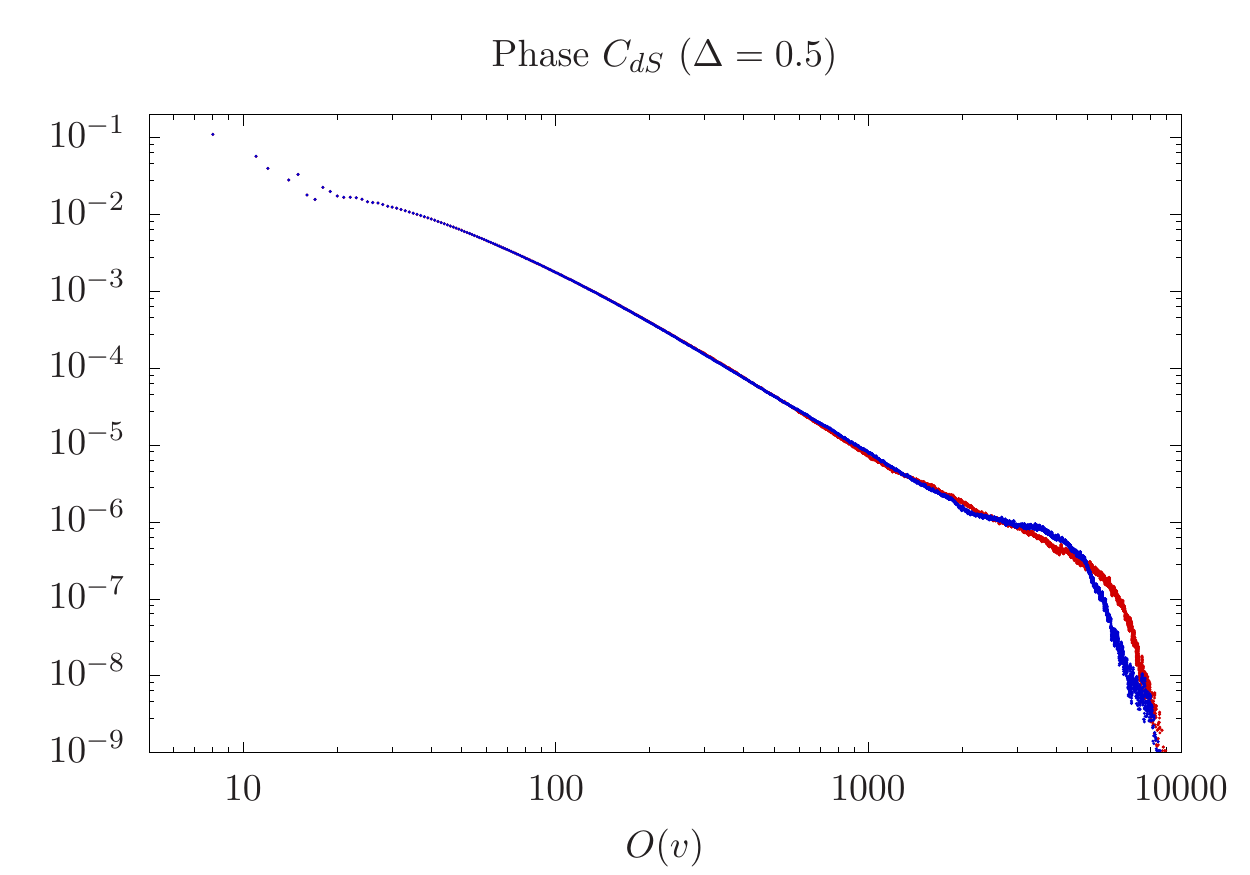}}
\scalebox{.56}{\includegraphics{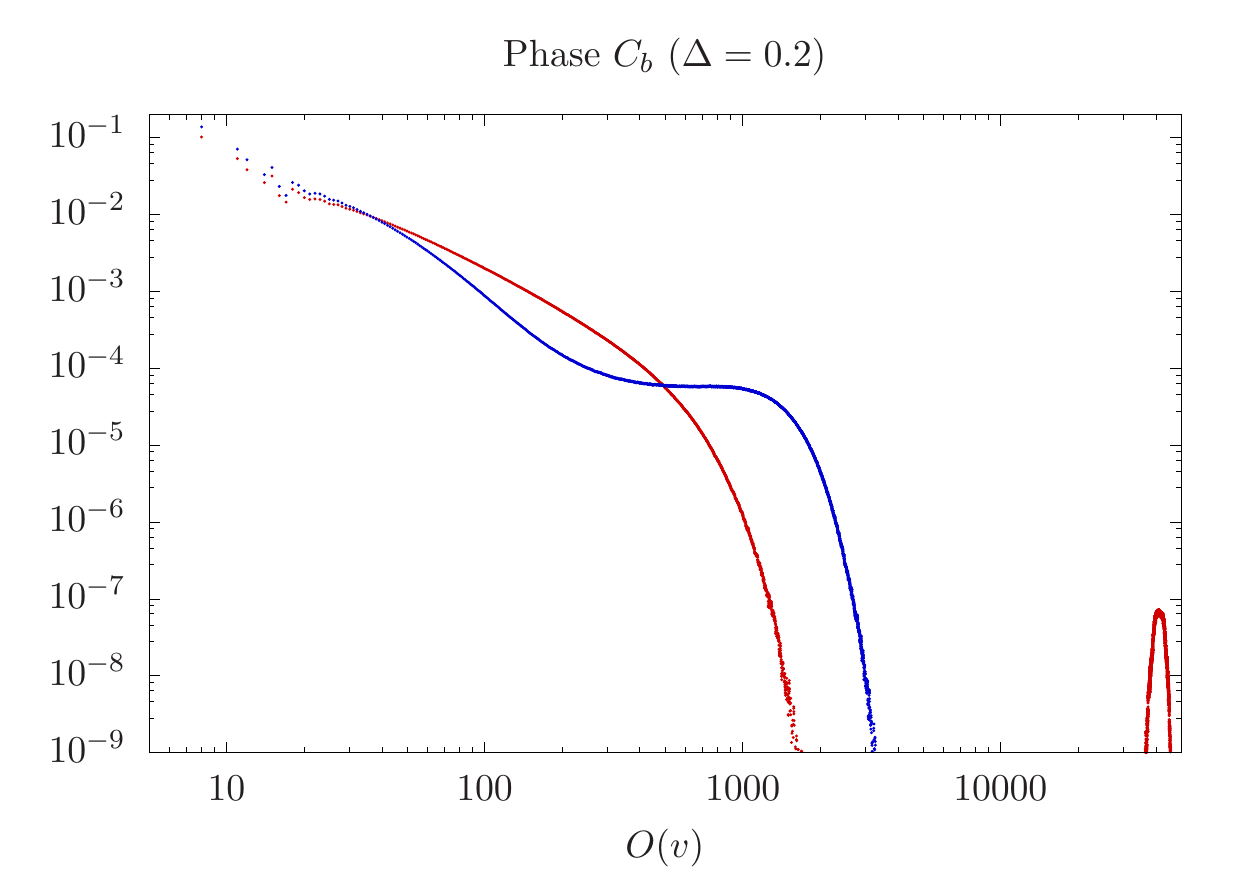}}
\caption{\label{fig:histocoord} 
Histograms of the coordination number
$O(v)$ in the spatial slice containing the vertex with maximal coordination number (red curves) and in a
neighbouring spatial slice (blue curves). In the de Sitter phase ($\Delta\! =\! 0.5$, left) there is hardly any difference.
In the bifurcation phase ($\Delta\! =\! 0.2$, right), the singular vertices form a peak in $O(v)$ isolated from the rest 
of the distribution, which is not present in slices without singular vertices.
Measurements taken for $t_{tot}\! =\! 80$, $N_{41}\! =\! 160k$ and $\kappa_0\! = \! 2.2$. [Figures courtesy of A.\ G\"orlich.]}
\end{figure}

There is a particular large-scale property that has already been identified as characteristic for the bifurcation phase,
and has motivated the introduction of the order parameters (\ref{op1}) and (\ref{op2}). Both measure a difference in
geometry of adjacent spatial slices, which {\it causes} the effective bifurcation behaviour expressed by 
the functional form of the matrix elements (\ref{shi}).
As first noted in \cite{bifurcation2}, inside phase $C_b$ each second slice appears to have a single ``singular" 
vertex\footnote{``Exceptional" instead of ``singular" would be 
more appropriate, since these vertices are perfectly regular from the point of view of finite, piecewise flat geometries. 
We will nevertheless use this notion, first coined in the context of Euclidean Dynamical Triangulations \cite{singular}.} with an
exceptionally high coordination number, both in a three- and a four-dimensional sense, reflected by the parameters (\ref{op1})
and (\ref{op2}) respectively, and illustrated by Fig.\ \ref{fig:alter} above.
Coming from the de Sitter phase $C_{dS}$ and moving into $C_b$ by decreasing $\Delta$, 
within a spatial slice that contains a singular vertex
a gap opens between the coordination number $O(v)$ of this vertex and the coordination number of the vertex
with the second-largest $O(v)$. 
The histograms of Fig.\ \ref{fig:histocoord} illustrate this phenomenon, which is absent in the de Sitter phase.
Well inside the bifurcation phase, in a slice containing such an exceptional vertex, its coordination number is 
a couple of orders of magnitude larger than the average coordination number of the slice.

The relation between maximal vertex order and spatial volume was made quantitative in \cite{newphase}, which for a system with
$t_{tot}\! =\! 2$, $\kappa_0\! =\! 2.2$, $\Delta\! =\! 0$ and $N_4\! =\! 10k$ found an approximately linear relation
between the expectation value of the highest vertex order, located in slice $t_1$, say, and the volume difference
$n_{t_2}-\! n_{t_1}$ with the neighbouring slice $t_2$. In other words, a high coordination number of a singular vertex in one slice
is associated with a large volume in an adjacent slice (which does not have a singular vertex). 
\begin{figure}
\centering
\scalebox{.75}{\includegraphics{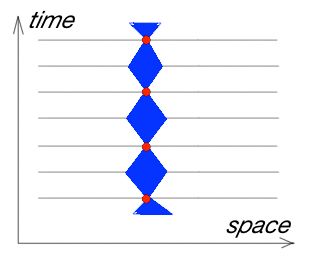}}
\scalebox{.54}{\includegraphics{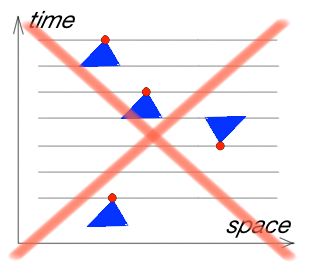}}
\caption{\label{fig:diamond} 
Schematic illustration of the spacetime geometry inside the bifurcation phase, 
where the singular vertices present on every second spatial slice align themselves into a chain of
diamond-like regions, as shown in the figure on the left. [Figures from \cite{bifurcation2}.]}
\end{figure}

The overall picture that emerges in the bifurcation phase for CDT configurations with large time extension $t_{tot}$ is 
that the singular vertices on alternating spatial slices are associated with a four-dimensional substructure of the 
triangulation, which consists of a chain along the time direction of roughly diamond-shaped regions. 
The tips of each ``diamond" are a pair of singular vertices a distance $\Delta t\! =\! 2$ apart, say, at times $t$ and $t\! +\! 2$,
and its body by definition consists of all four-simplices in the interval $[t,t\! +\! 2]$ that contain either of the two singular vertices.
They form a diamond because the two sets of four-simplices, lying in either $[t,t\! +\! 1]$ or $[t\! +\! 1,t\! +\! 2]$, turn out to 
have a large overlap at the intermediate time $t\! +\! 1$, in the form of shared (sub-)simplices. The situation is illustrated
schematically by Fig.\ \ref{fig:diamond}. The entire chain forms a substructure imbedded in the rest of the triangulation,
and contains a large, finite fraction of the total four-volume.
 
As already remarked in \cite{bifurcation2}, the singular vertices and their associated substructures in $C_b$ break
the spatial homogeneity and isotropy that appear to be present (in a statistical sense, and on sufficiently large scales) 
in the de Sitter phase and presumably are related to the fact that the large-scale properties of the
dynamically generated quantum universe in $C_{dS}$ are very well described by a minisuperspace model with built-in
homogeneity and isotropy \cite{emergence,semi1,reconstructing}. For this reason, the new 
$C_b$-$C_{dS}$ phase transition has been associated tentatively with the breaking of this 
symmetry \cite{bifurcation2,newphase,higher}.

\subsection{CDT on a spatial torus}
\label{sec:torus}

As already mentioned in Sec.\ \ref{time:sec}, CDT configurations satisfy a lattice version of global hyperbolicity,
which entails a fixed topology $ {}^{(3)}\Sigma$ for the three-dimensional spatial slices at fixed time $t$. 
All results discussed until now have used a spherical topology, $ {}^{(3)}\Sigma\! =\! S^3$, and -- for the convenience of
not having to deal with boundaries in the simulations -- a compactified time direction, tantamount to a global
topology $S^1\times S^3$. Since this amounts to a specific choice that is made {\it a priori}, the question arises of
whether and to what extent a different choice of spatial topology will affect any of the key findings of CDT quantum gravity.   
On the one hand, one would not expect the global topology to affect local physics, at least not for sufficiently large
pieces of spacetime. On the other hand the properties of quantum gravity are studied with the help of observables
that themselves are often very nonlocal.\footnote{Note that ``nonlocal" here is not meant 
in the sense of causality-violating, faster-than-light propagation of information, to the extent that this is an operationally 
well-defined concept at Planckian distance scales in the first place.}
Given the limited range of scales accessible in computer simulations, this means that
disentangling local and global features will in general be nontrivial. 

Several results for CDT with a spatial three-torus, $ {}^{(3)}\Sigma\! =\! T^3$, 
are now available \cite{impact,torus1,torus2}. 
The analysis proceeds largely along the lines of the spherical case and so far has included investigations of the volume profile, the 
associated effective transfer matrix and effective action, and mapping out the phase diagram.
As will be described in more detail below, the dynamics of the global scale factor (the three-volume) {\it is} sensitive
to the global topology, while the number, location and broad characteristics of the phases are not, in line with the
expectation articulated in the previous paragraph. Unlike what happens for spherical spatial slices, where
the chosen topology $S^1\times S^3$ appears to be driven dynamically to that of a four-sphere $S^4$, 
for toroidal spatial slices the
chosen topology $S^1\times T^3$ is unchanged, at least in the phase(s) with macroscopic semiclassical geometry.

One reason why the three-torus is not the first choice in simulations
is that the minimal size of a simplicial three-manifold with this topology 
(in terms of the number of equilateral tetrahedra it contains) is much bigger than that of a three-sphere,
namely, 90 as opposed to 5 \cite{impact}. This is bound to reduce the window between lattice artefacts and finite-size
effects where reliable measurements can be made and, generally speaking, will drive up the lattice volumes at
which specific phenomena are observed, compared to the spherical case. This effect has been confirmed in the
actual simulations.  

The simulations require configurations with spacetime topology $S^1\times T^3$,
consisting of $t_{tot}$ individual sandwich geometries $I\times\! T^3$. 
Reference \cite{impact} provides an explicit construction of an initial configuration of this kind, which is needed
as a starting point for the Monte Carlo simulations. It is obtained 
by gluing together regular hypercubes subdivided into four-simplices. 
Each sandwich geometry is a layer of thickness $\Delta t\! =\! 1$ of these cubes, 
assembled in a specific way to make sure that the triangulated three-geometries shared by adjacent layers match. 
Each sandwich contains a total of 384 four-simplices of type (3,2) and
640 of type (4,1), and its spatial boundaries consist of 320 equilateral tetrahedra each. 
Using the same set of ergodic Monte Carlo moves as for $S^1\times S^3$, this initial 
spacetime configuration is 
grown until a target volume $\bar{N}_{41}$ is reached at which the simulations are performed in the usual manner. 
\begin{figure}
\centering
\scalebox{.56}{\includegraphics{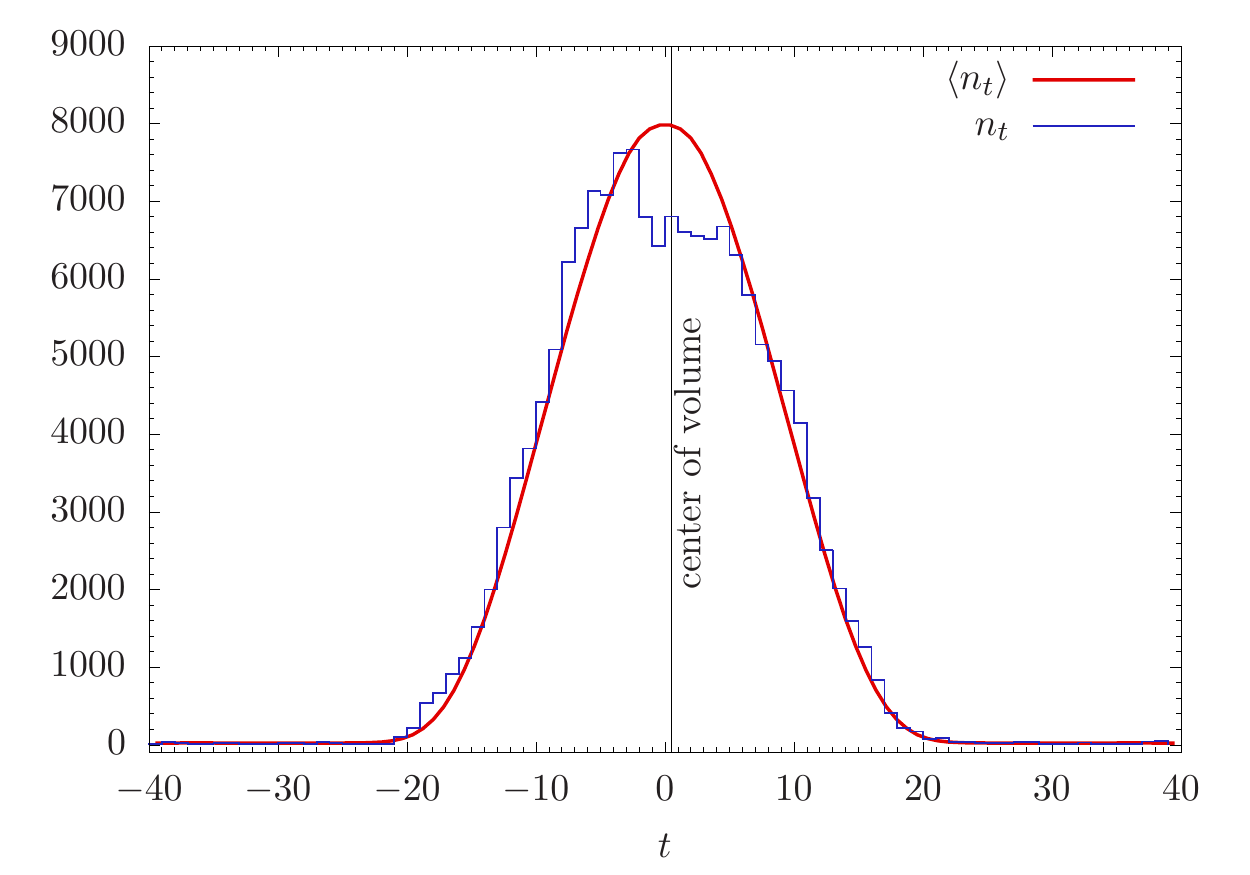}}
\scalebox{.56}{\includegraphics{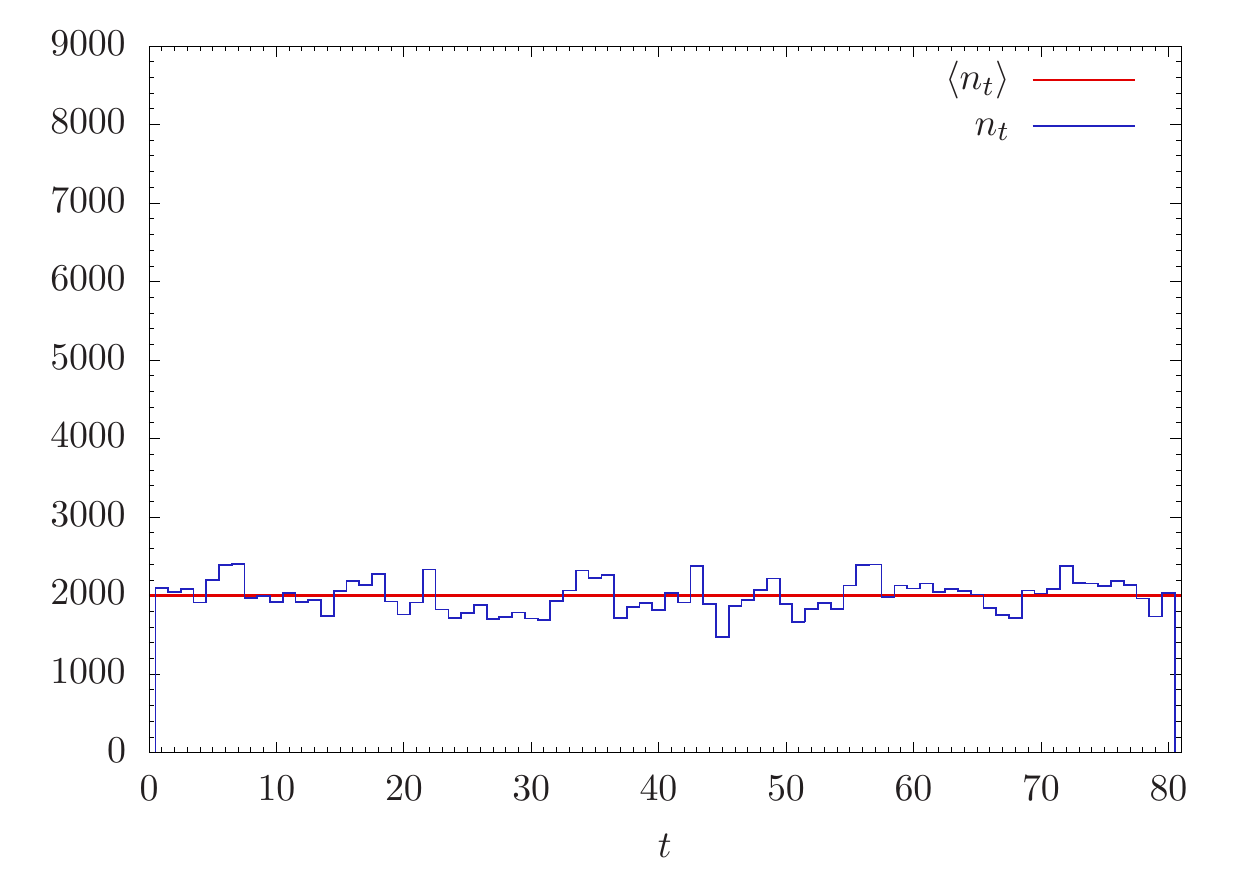}}
\caption{\label{fig:profiles} 
Snapshot of a typical volume configuration $n_t$ (blue) and the expectation value $\langle n_t\rangle$ (red), both for spherical
slices (left) and toroidal ones (right). All measurements taken at $(\kappa_0,\Delta)\! =\! (2.2,0.6)$, and for $N_{41}\! =\! 160k$
and $t_{tot}\! =\! 80$. [Figures from \cite{torus1}.]
}
\end{figure}

First simulation results of the toroidal system at couplings $(\kappa_0,\Delta)\! =\! (2.2,0.6)$, $t_{tot}\! =\! 80$
and with a quadratic volume fixing (\ref{sfix2}) were reported in \cite{impact}. Unlike for the spherical system,
for which this point is associated with a nontrivial expectation value $\langle n_t\rangle\!\propto\!
\cos^3 (t/const)$ for the volume profile, the corresponding quantity for the spatial torus universe appears to be time-independent,
$\langle n_t\rangle\!\propto\! const$. Fig.\ \ref{fig:profiles} illustrates both typical configurations $n_t$ for the two systems as
well as their mean values $\langle n_t\rangle$. In the spherical case, the effective time 
extension of the de Sitter universe (the time interval where the spatial volume differs appreciably from its kinematically 
allowed minimum $n\! =\! 5$) scales $\propto\! N_4^{1/4}$ 
as a function of the total four-volume $N_4$, while for the spatial
volumes one finds $n\! \propto \! N_4^{3/4}$ \cite{emergence,reconstructing,physrep}.    
The situation for the torus case is different: the four-volume is on average evenly distributed over the available
layers, such that $n\!\propto\! N_4/t_{tot}$. 

For small perturbations $\Delta n_t\! =\! n_t-\! \langle n_t\rangle$ around the constant average three-volume, one can again try to 
determine an associated effective action from measuring the volume-volume 
covariance matrix (\ref{coma}) and taking its inverse. 
The elements of neither the covariance matrices nor their inverses show any systematic time dependence. Taking
time averages of the diagonal and sub- and superdiagonal elements of the inverse covariance matrix, and studying their dependence
on the slice volume (by varying the four-volume in the range $N_{41}\in [80k,240k]$ and the time extension in the
range $t_{tot}\in [10,200]$) resulted in an effective action of the form
\begin{equation}
S_{\rm eff}=\sum_t \Big( \frac{1}{\Gamma}\,\frac{(n_t-n_{t+1})^2}{n_t+n_{t+1}}+\mu n_t^{-\gamma}+\lambda n_t   \Big),
\label{storus}
\end{equation}
with $\Gamma\!\approx\!26.3$, $\gamma\! =\!1.16\pm0.02$ and $\mu\! >\! 0$ \cite{impact}.\footnote{A cosmological constant
term is included here for generality; $\lambda$ does not play a role in the fixed-volume study of \cite{impact}, other
than possibly as a Lagrange multiplier. It has been determined 
in the effective transfer matrix treatment of \cite{torus1} as $\lambda\!\approx\! 3.5\times 10^{-4}$.}
Comparing this with the effective Lagrangian (\ref{lagra}) obtained for the spherical case, the kinetic term is identical, 
including the value of $\Gamma$ within measuring accuracy. However, the potential term proportional to $n^{1/3}$ is 
absent in the torus case. This latter finding is in agreement with the classical (Euclidean)
minisuperspace action for FLRW metrics with flat toroidal slices in proper-time form,
\begin{equation}
ds^2=dt^2 +a^2(t)(dx^2+dy^2+dz^2),
\label{flrw}
\end{equation}
given by 
\begin{equation}
S_{eu}^{\rm EH}\propto \frac{1}{G_{\rm N}}\int dt\,\Big( a \dot{a}^2-\frac{\Lambda}{3}\, a^3\Big),
\end{equation}
obtained by inserting the ansatz (\ref{flrw}) into the Euclidean counterpart of the continuum Einstein-Hilbert action
(\ref{ehact}). By contrast, the integrand of the corresponding minisuperspace action on spherical slices 
(of constant positive curvature) has an
additional potential term linear in the scale factor $a(t)$, which re-expressed in terms of the three-volume 
corresponds to a term $\propto\! n^{1/3}$.  
It is remarkable that the nonperturbative CDT quantum theory with both spherical and toroidal spatial topology
reproduces these detailed features of the potential terms of the corresponding classical minisuperspace cosmologies.
The term with negative power $n^{-\gamma}$ in eq.\ (\ref{storus}) does not appear in the classical action and
is interpreted as a genuine quantum correction. Such a term may in principle also be present in the spherical
case, but is out of reach with current measurement precision.
There is at this time no analytic argument for any particular value of the exponent $\gamma$. 
\begin{figure}
\centering
\includegraphics[width=0.6\textwidth]{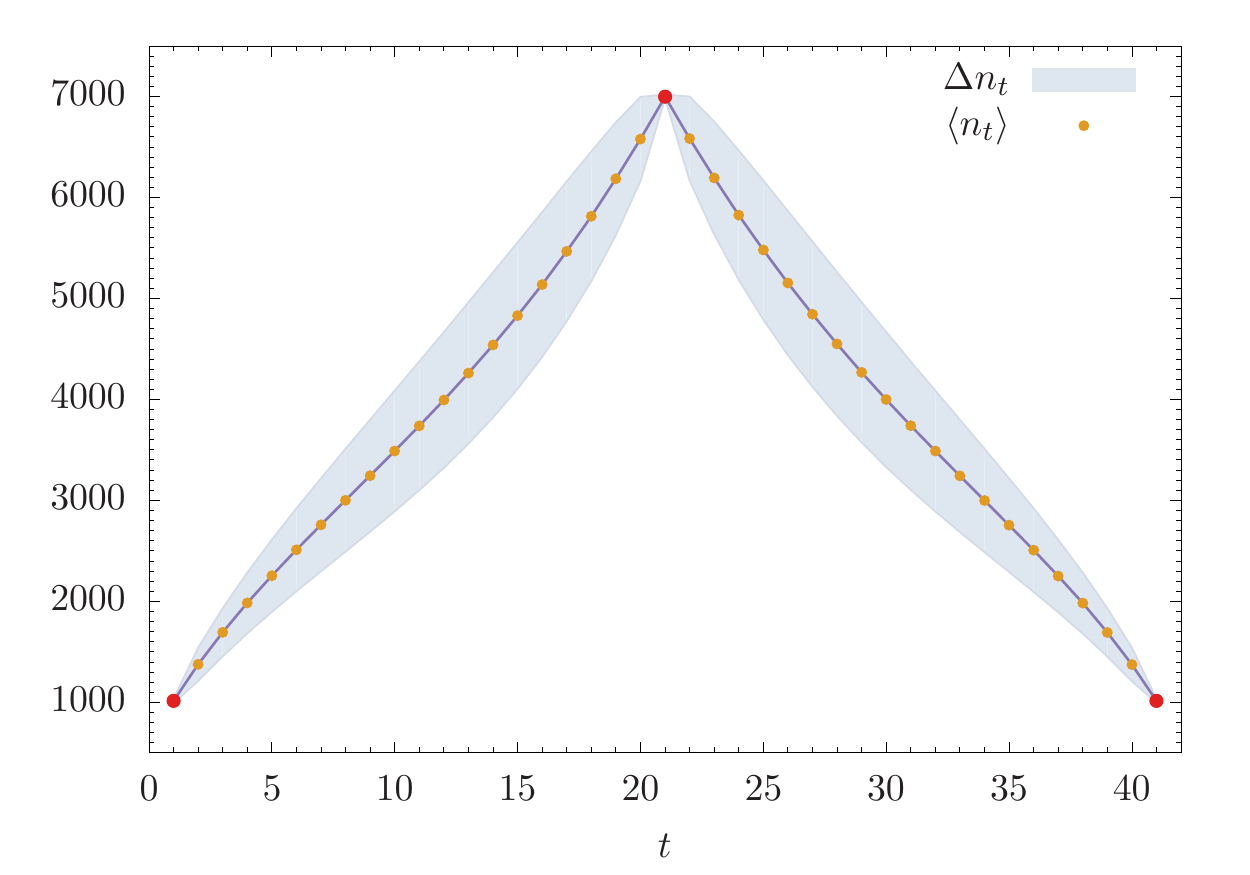}
\caption{\label{fig:pin} 
Volume profile $\langle n_t\rangle$ of CDT configurations with $t_{tot}\! =\! 40$ toroidal slices, with fixed spatial volumes
$\bar{n}_1\! =\! 1000$ and $\bar{n}_{21}\! =\! 7000$. The shaded region indicates
the size of the quantum fluctuations of $n_t$ around the mean.
Measurements taken at $(\kappa_0,\Delta)\! =\! (2.2,0.6)$ and $\kappa_4\! =\! 0.9225$. [Figure from \cite{torus1}.]
}
\end{figure}

These results were confirmed in a second, more detailed study, where the average volume profile was forced to be nonconstant,
to allow for measurements at different slice volumes within the same Monte Carlo simulation \cite{torus1}.
Instead of fixing the total volume, one (approximately) fixes the spatial volume at two times $t$ and $t'$ to
some target volumes $\bar{n}_t$ and $\bar{n}_{t'}$ by adding a term
\begin{equation}
S_{\rm fix}=\epsilon [(n_t-\bar{n}_t)^2+(n_{t'}-\bar{n}_{t'})^2],\;\;\; \epsilon >0,
\end{equation}
to the bare action. The numerical study used $t_{tot}\! =\! 40$, and fixed the volumes at times $t\! =\! 1$ and $t'\! =\! 21$
(Fig.\ \ref{fig:pin}). For this system, the inverse covariance matrix was again determined, confirming the 
functional form (\ref{storus}) of the effective action, with $\Gamma\! =\! 26.2\pm 0.1$ and $\gamma\!\approx\! 1.5$,
in reasonable agreement with the previous results \cite{impact}. The shape of the volume profile $\langle n_t \rangle$
of Fig.\ \ref{fig:pin} can be matched qualitatively well by classical solutions derived for the continuum analogue of the
action (\ref{storus}) \cite{torus1}.

The effective transfer matrix method introduced in Sec.\ \ref{subsec:eff} above has also been applied
to CDT with spatial tori, yielding compatible results for the volume behaviour. 
The matrix elements $\langle n|M|m\rangle$ have been measured in Monte Carlo simulations 
of the system with $t_{tot}\! =\! 2$, at the phase space point $(\kappa_0,\Delta)\! =\! (2.2,0.6)$ and 
for various fixed values of the volume $N_{41}\! =\! m+n$. The results are well approximated by an effective Lagrangian 
of the form 
\begin{equation}
L_{\rm eff}[n,m]=\frac{1}{\Gamma}\frac{(n-m)^2}{(n+m)}+\mu (n^{-\gamma}+m^{-\gamma})\! +\lambda (n+m).
\label{lagratorus}
\end{equation}
Alternatively, the same Lagrangian with a potential term $\mu (n+m)^{-\gamma}$ instead of
$\mu (n^{-\gamma}+m^{-\gamma})$ is an equally good fit to the data.
Following the strategy developed in \cite{bifurcation1}, it has been checked to what extent the volume profile $\langle n_t\rangle$
and the amplitude of the fluctuations $\Delta n_t$ obtained for the time extension $t_{tot}\! =\! 40$ (Fig.\ \ref{fig:pin}) can
be reconstructed from the matrix elements $\langle n|M|m\rangle$ of the effective transfer matrix alone \cite{torus1}.
It turns out that the volume profile can be reproduced well for a suitable choice of $\kappa_4$. The curve of the 
amplitude of the fluctuations looks qualitatively similar, but there is a noticeable, small discrepancy. This indicates that
excitations other than the three-volume induce correlations beyond pairs of neighbouring slices that are not captured
by the effective volume dynamics. 

\begin{figure}
\centering
\scalebox{.55}{\includegraphics{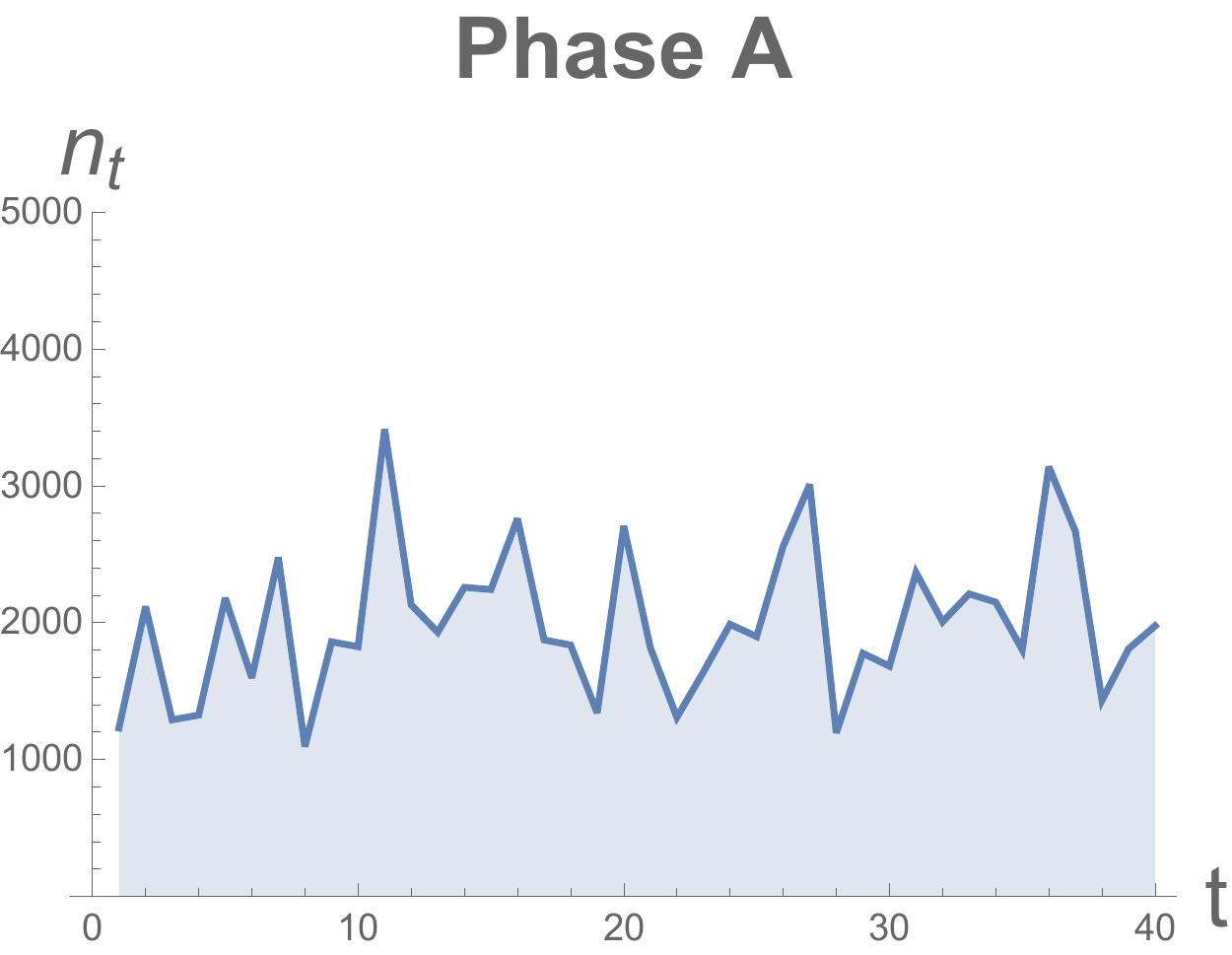}}
\scalebox{.55}{\includegraphics{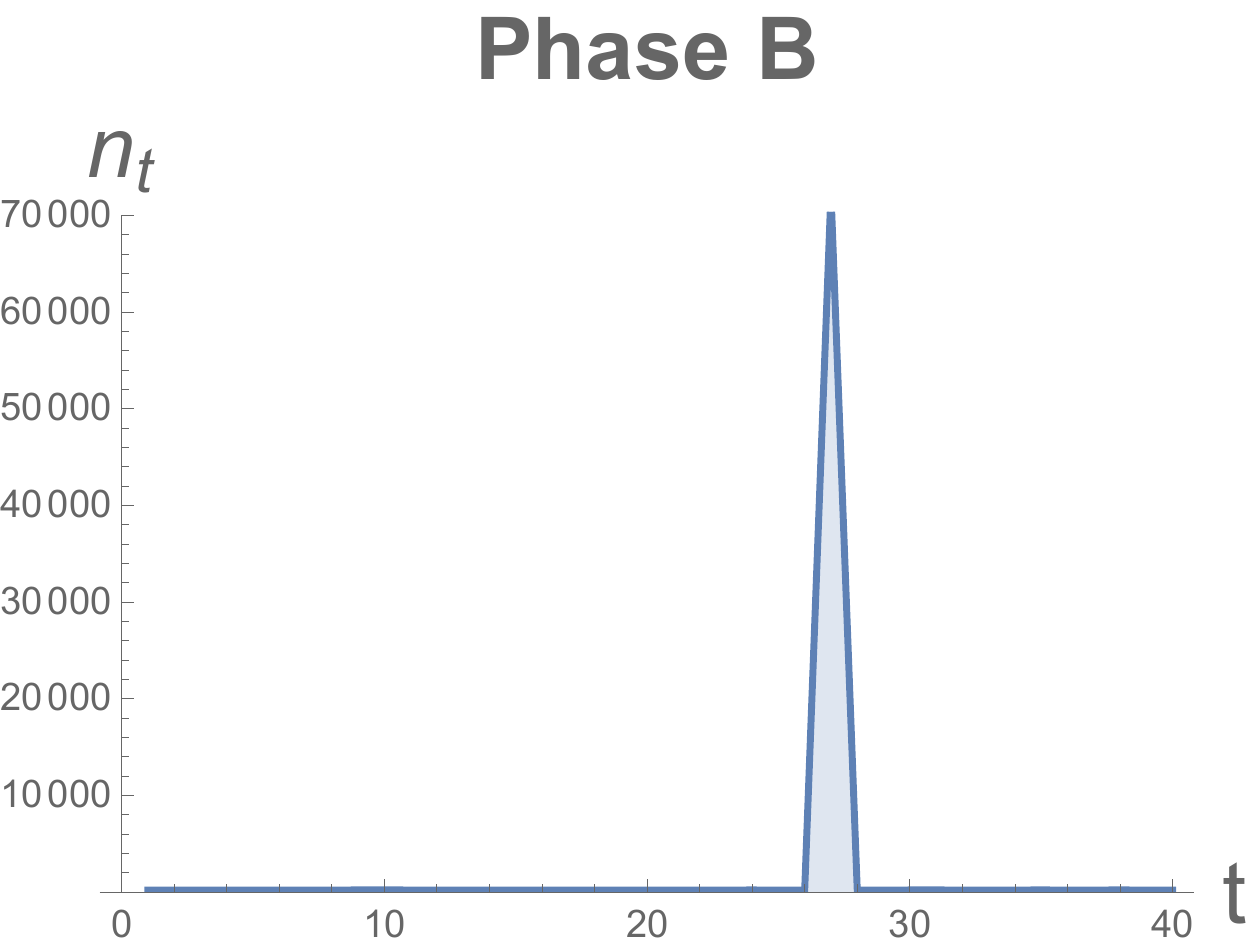}}
\caption{\label{fig:abtorus}  
Sample volume profiles $n_t$ of CDT configurations with $t_{tot}\! =\! 40$ toroidal slices and for
$N_{41}\! =\! 40k$ display characteristics familiar from phases $A$ and $B$ of
the spherical system. [Figures from \cite{torus2}.]}
\end{figure}

The phase structure of the toroidal system was investigated in \cite{torus2}, starting with a grid-like sampling of the 
$(\kappa_0,\Delta)$-space for fixed lattice volume $N_{41}\! =\! 80k$ and $t_{tot}\! =\! 40$.
For this purpose, one employs a set of order parameters,
\begin{equation}
N_0/N_4,\;\;\; N_{32}/N_{41},\;\sum_{t\in [1,t_{tot}]}\!\!\!\!\! (n_{t}-n_{t-1})^2,\;\;\; \max_{v\in T}\, O(v),
\label{neworder}
\end{equation}
which are variants of previously used parameters also applicable to constant volume profiles.
Scanning the phase space to establish in which regions
these parameters (or suitably rescaled versions) are large or small one finds an overall picture that is qualitatively
very similar to that for spherical slices, with typical characteristics of the $A$-, $B$- and $C$-phases. For 
example, as $\Delta$ is increased and $\kappa_0$ held fixed, $N_0/N_4$ increases and
the last two order parameters in (\ref{neworder}) decrease monotonically. One also observes volume profiles $n_t$
that are reminiscent of those of the unphysical phases $A$ and $B$ of the sphere case (Fig.\ \ref{fig:abtorus}). In the former, the entire
universe collapses in time, with only a single spatial slice whose volume is nonvanishing, and the latter is
associated with large, random volume differences between adjacent slices
(in both cases subject to the kinematical minimal-volume constraint for the spatial tori).

\begin{figure}
\centering
\includegraphics[width=0.65\textwidth]{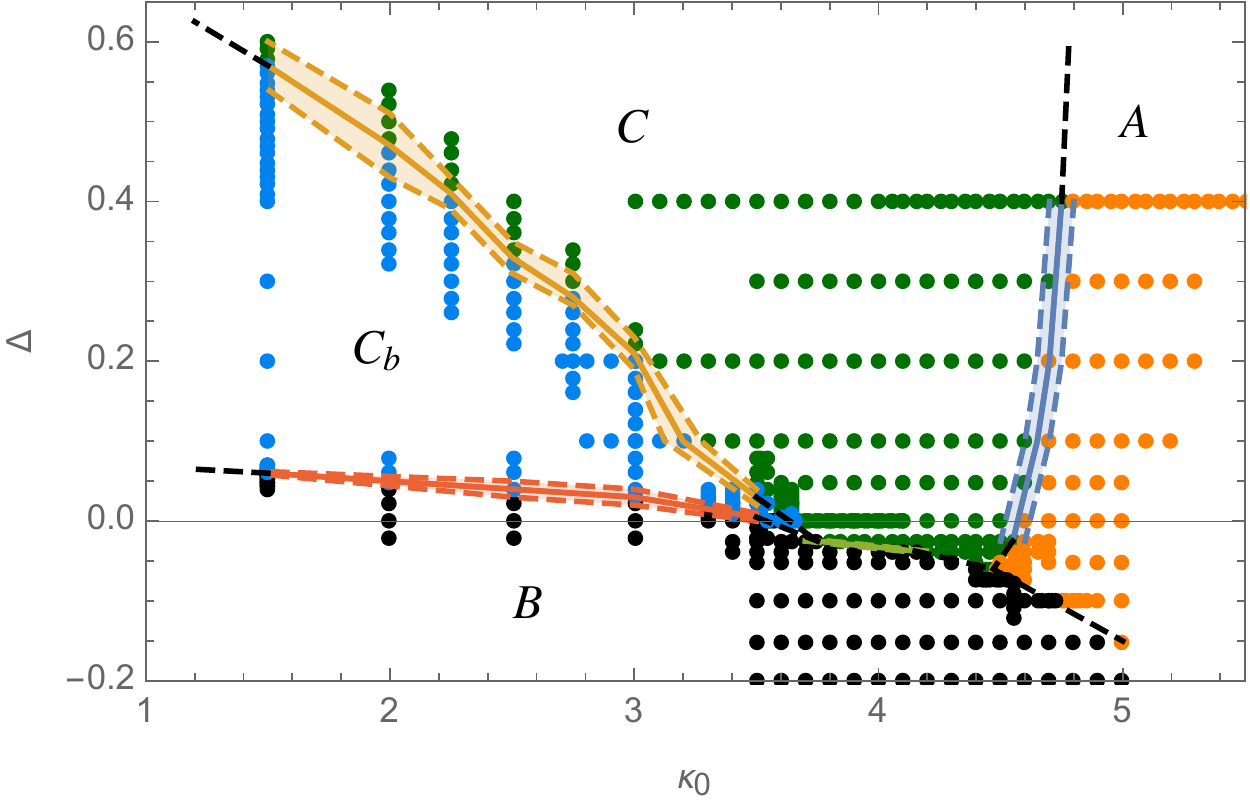}
\caption{\label{fig:phasediatorus} 
The phase diagram of CDT quantum gravity with $t_{tot}\! =\! 4$ toroidal slices, from measurements at 
volume $N_{41}\! =\! 160k$. Phase transitions are marked by solid lines, with shaded areas of the same colour indicating 
error bars. Dots mark measurement points, with different colours associated to the four different phases. The phase
denoted $C$ is the analogue of the de Sitter phase for spherical slices. Dashed black lines indicate extrapolations.
[Figure from \cite{torus2}.]   
}
\end{figure}

It is perhaps unsurprising that neither the bifurcation phase nor an analogue of the 
$C_b$-$C_{dS}$ phase transition are visible in these data; since the typical slice volumes are of the order of 2000,
and the volume profile is approximately constant in phase $C$, there is likely not enough three-volume 
to allow singular vertices to develop. This has motivated a further study of the torus model with
short time extension $t_{tot}\! =\! 4$, at total volume $N_{41}\! =\! 160k$ \cite{torus2}. In this set-up, the characteristic
structure of the bifurcation phase {\it is} visible, with vertices of very high coordination number ($\sim\! 60k$)
forming on every second spatial slice. In addition, extensive measurements of both the order parameters
(\ref{neworder}) and their susceptibilities have been made for these configurations, achieving a high resolution of the
location of the phase transitions. The resulting phase diagram is shown in Fig.\ \ref{fig:phasediatorus}. Its structure is clearly very 
similar to that for CDT quantum gravity for spherical slices (Fig.\ \ref{phasedia}), despite the fact that one
can still expect significant shifts in the location of the transition lines, due to strong finite-size effects for the torus 
measurements. One interesting difference with the spherical case is that the critical slowing-down observed near the
bottom right-hand corner of the phase diagram, where various transition lines meet, is much less pronounced.  
This will provide a good starting point for measurements of the order of the phase transitions, which 
in general are anticipated to be computationally expensive \cite{torus2}. 

\subsection{Renormalization group flow in CDT}
\label{sec:rengroup}

In the absence of an {\it a priori} defined length scale, it is not a given that renormalization group arguments can be 
formulated consistently in a background-independent theory of quantum gravity, where any notion of scale 
will typically have to be generated dynamically. However, a proof of principle has been given in \cite{renorm} that these 
difficulties can be overcome in CDT quantum gravity. One follows a strategy similar to what is done
in the lattice formulation of quantum fields on a fixed flat background, say, of scalar fields with a $\lambda\phi^4$-interaction 
(see, for example, \cite{momu}). The key idea is to identify {\it paths of constant physics}
in the space of bare coupling constants, follow them in the direction of decreasing lattice spacing or, equivalently, increasing
correlation length, and investigate whether they approach second-order transition points or lines where a continuum limit can be taken.
In the case of gravity, it would be highly interesting to verify the presence of UV fixed points of the renormalization group, as predicted
by asymptotic safety, and to
investigate the physical properties of the system in the approach to such a fixed point. 

Clearly, constant physics must be defined in terms of observables, 
which are much easier to identify in scalar field theory than in quantum gravity. 
A limiting factor in the renormalization group investigations at this stage is the scarcity of quantum-gravitational observables,
as well as the unclear phenomenological status of some of them, for example, a quantity like the spectral dimension.
The analysis of \cite{renorm} is based on observables with a clear-cut interpretation in terms of macroscopic physics,
at least inside phase $C$, where the macroscopic shape of spacetime can be matched to that of a de Sitter 
universe.\footnote{Note that \cite{renorm} interprets the measurements in the more general framework 
of Ho\v rava-Lifshitz gravity, allowing also for a ``deformed" de Sitter universe.}
They are the total four-volume of spacetime, the three-volume of the spatial universe at constant proper time, and 
the quantum fluctuations of the three-volume around the mean of the latter.

\begin{figure}
\centering
\includegraphics[width=0.48\textwidth]{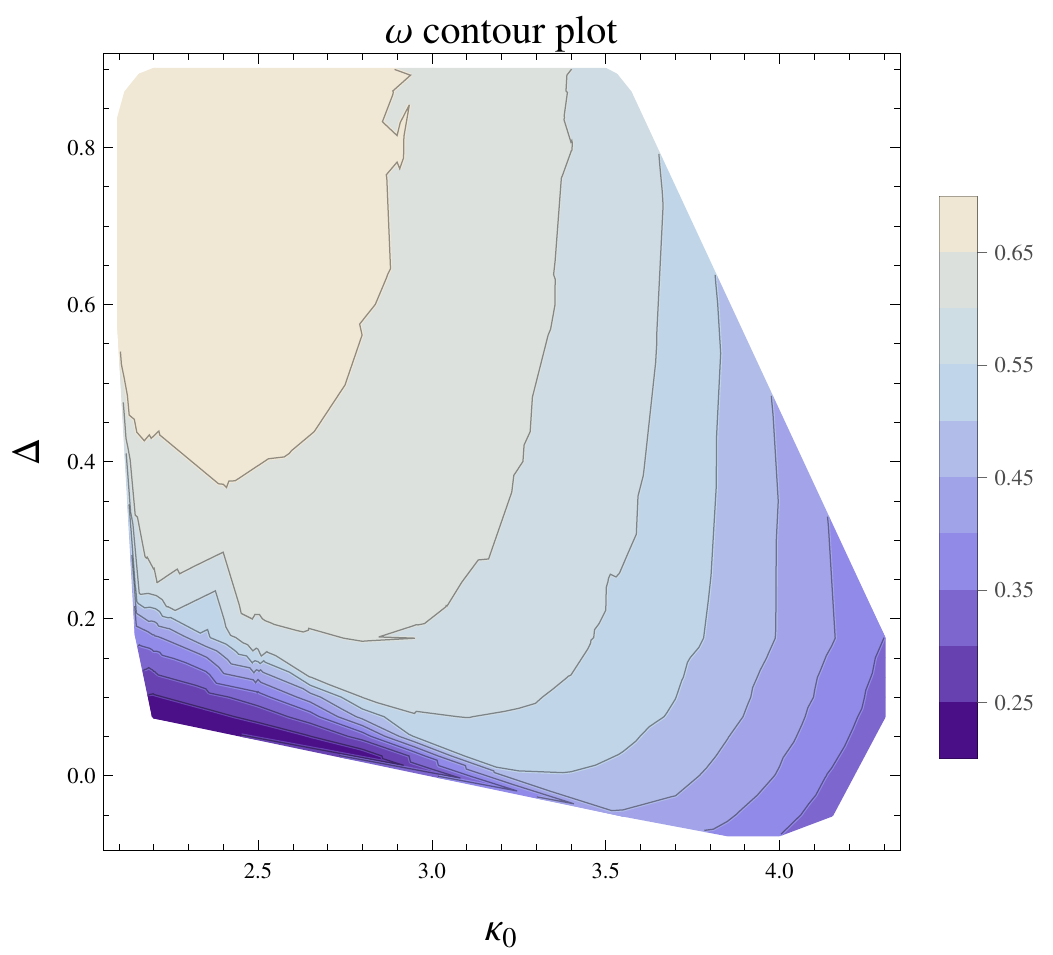}
\includegraphics[width=0.48\textwidth]{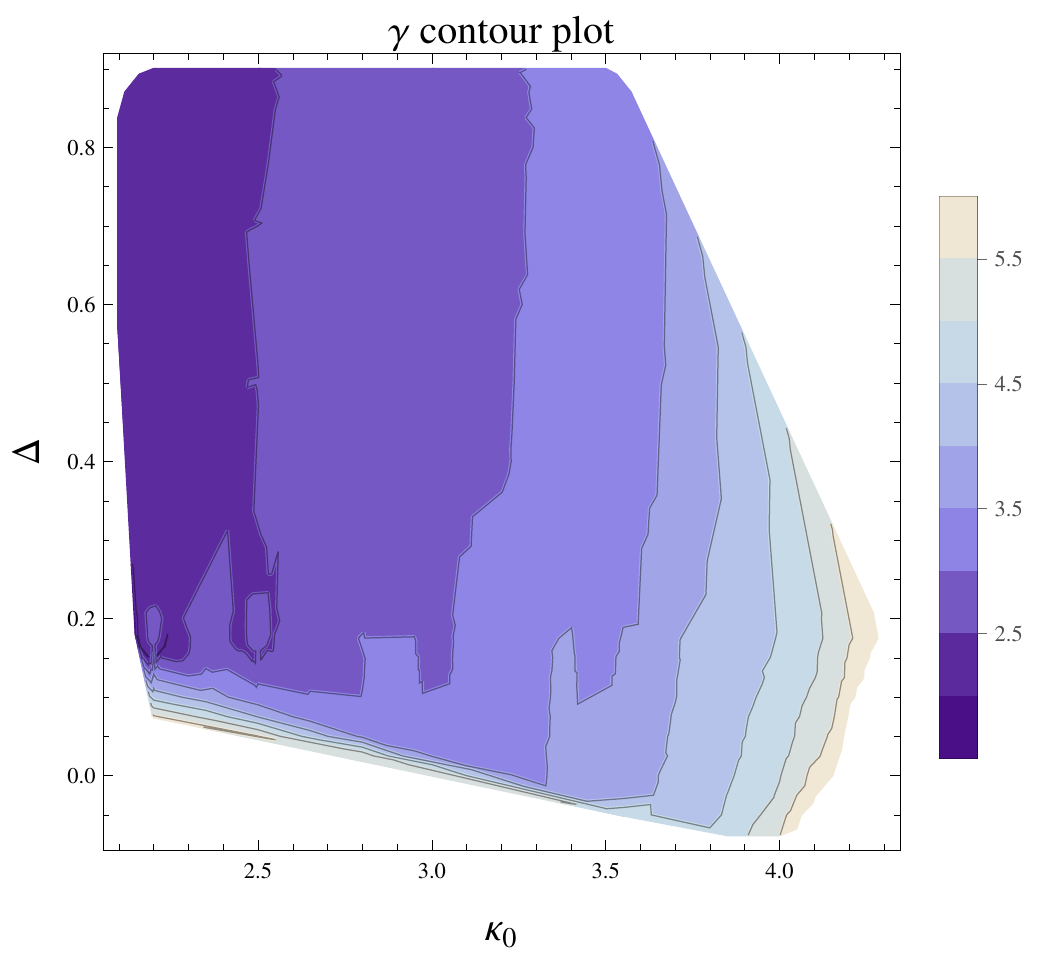}
\caption{\label{fig:flowlines} Lines of constant physics in the $\kappa_0$-$\Delta$ plane according to the renormalization
group analysis of \cite{renorm} are given by lines of constant $\omega$ (left) along which $\gamma$ (right) increases
$\propto\! N_4^{1/4}$ in a scaling limit $N_4\!\rightarrow\!\infty$. The parameters 
$\omega$ and $\gamma$ are extracted from the average volume profile and its fluctuation spectrum.}
\end{figure}

We currently do not have a good gravitational counterpart of a correlation length, which is the observable used
to set the scale in standard lattice field theory, in the sense of providing an identification of lattice units with 
dimensionful physical units. Instead, \cite{renorm} uses the quantum fluctuations of the spatial volume as a physical yardstick,
and identifies lines of constant physics as those with a constant ratio of the average size of these quantum fluctuations
and the average size of the spatial universe.\footnote{Ref.\ \cite{coop1} suggests the use of additional, alternative yardsticks,
which one would like to extract from the behaviour of the spectral dimension, see Sec.\ \ref{sec:specobs} below for further
comments.}
Fig.\ \ref{fig:flowlines} illustrates how flow lines of constant physics are extracted in phase $C$.
One further assumption underlying the analysis of \cite{renorm} is that    
the ratio of the proper length of a space-like and a time-like lattice unit does not depend on the values of
the bare coupling constants. This is arguably the most straightforward assumption, but certainly not the only one possible (see also
\cite{searchproc} for a related discussion).
It illustrates the subtleties of studying renormalization group flows in background-independent quantum gravity: 
in a situation where scales are generated dynamically, their possible dependence on the bare coupling constants 
must be analyzed with care, since it may affect the definition of what is meant by constant physics.  

Note that the study \cite{renorm} predated the discovery of the $C_b$-$C_{dS}$ phase transition, and 
therefore focused on the second-order $B$-$C_b$ transition. Its analysis
did not yield evidence of flow lines running into this transition line. However, this result should not be taken as conclusive,
since a critical slowing-down severely hampered the taking of data near the transition line and
made it very computationally intensive. Besides improving computational algorithms,
the scope for taking this research to the next level is clearly very rich, including other transition lines, alternative global
set-ups (torus topology, transfer matrix setting), alternative scaling assumptions and -- most importantly -- more
observables, together with a careful analysis of how they can yield reliable physical yardsticks. 
The larger value of the work of \cite{renorm}, which remains the only explicit nonperturbative renormalization group 
analysis in CDT quantum gravity, with an actual computation of renormalization group trajectories, is to provide a blueprint 
for a methodology that may well be relevant and applicable beyond this particular approach.

\section{Quantum-geometric observables} 
\label{observ:sec}

As we have been emphasizing throughout this review, identifying and measuring observables is key to understanding
the physical content of CDT quantum gravity and the existence and nature of its scaling limits in particular. 
Since the previous major review of the subject in \cite{physrep}, the already known observables have been studied in
greater detail and generality and new ones have emerged. This section summarizes the most important new developments.

\subsection{Spectral observables}
\label{sec:specobs}

Spectral observables refer to properties of the Laplace-Beltrami operator in the 
piecewise flat context of CDT, either on the full spacetime or on spatial slices of constant time. 
The most studied observable in this class is the spectral dimension $D_S$, whose original measurement in CDT quantum gravity
in \cite{spectral,reconstructing} exhibited the surprising phenomenon of {\it dynamical dimensional reduction}, 
namely, the smooth decrease of the
spectral dimension from its classical value of 4 on large scales to a smaller value compatible with 2 (within error bars) as the Planck
scale is approached. As has been described in detail elsewhere \cite{spectral,physrep}, this result was established by 
studying a diffusion process on the ensemble of CDT geometries, more precisely, by
measuring the expectation value of the average return probability $P_g(\sigma)$ of a random walker on a spacetime geometry $g$,
\begin{equation}
P_{g}(\sigma) = \frac{1}{V_g} \int  d^4\xi \sqrt{\det g(\xi)} \;
K_g(\xi,\xi;\sigma),
\label{probdiff}
\end{equation}
for a suitable range of diffusion times $\sigma$.
In eq.\ (\ref{probdiff}), $K_g(\xi,\xi_0;\sigma)$ denotes the probability density of diffusion from a point $\xi_0$ to a point $\xi$ in
diffusion time $\sigma$, and $V_g\! =\! \int d^4\xi \sqrt{\det g(\xi)}$ is the spacetime volume. The function $P_g(\sigma)$ is
also called the {\it heat trace} or {\it heat kernel trace}. 
For a smooth, $d$-dimensional Riemannian geometry $g$, the (nonnegative) eigenvalues $\lambda_i$ of the
Laplace-Beltrami operator $\Delta_g$ acting on scalar functions are related to the return probability by
\begin{equation}
P_g(\sigma)=\frac{1}{V_g}\sum_i {\rm e}^{-\lambda_i \sigma},
\label{retprob}
\end{equation}
where degenerate eigenvalues appear separately in the sum and for simplicity we have assumed that the
spectrum is discrete. At the same time, in the limit 
$\sigma\rightarrow 0$ there is an asymptotic expansion of $P_g(\sigma)$ of the form 
\begin{equation}
P_g(\sigma)\cong\frac{1}{(4\pi\sigma)^{d/2}\, V_g}\sum_{n=0}^\infty a_n\sigma^n,
\label{heatcoeff}
\end{equation}
where $a_0=V_g$ and the remaining $a_n$ are  
integrated curvature invariants of order $n$ \cite{vassilevich}. We can extract the topological dimension
$d$ of the Riemannian manifold by evaluating the logarithmic derivative of (\ref{heatcoeff}) in the limit
$\sigma\rightarrow 0$. More generally, we can define the {\it spectral dimension function}
\begin{equation}
D_S(\sigma):=-2\ \frac{d \log P_g(\sigma)}{d\log \sigma} 
\label{dspec}
\end{equation}
for arbitrary diffusion times $\sigma\geq 0$. One easily computes that on an infinite flat space, eq.\ (\ref{dspec}) evaluates to 
the topological dimension $D_S(\sigma)\! =\! d$, independent of $\sigma$. For general Riemannian spaces with 
nonvanishing curvature, it is clear from the expansion (\ref{heatcoeff}) that the spectral dimension $D_S(\sigma)$
away from $\sigma\! =\! 0$ will in general differ from $d$, such that
\begin{equation}
D_S(\sigma)=d +{\rm correction\; terms},
\label{dest}
\end{equation}
where the ``correction terms" can have either sign, depending on the values of the curvature invariants. 
The form of eq.\ (\ref{retprob}) illustrates that for small $\sigma$ the function $D_S(\sigma)$ probes fine-grained geometric properties
of the Riemannian space that are associated with the entire spectrum of the Laplace-Beltrami operator. By contrast, for large $\sigma$ 
only the low-lying part of the spectrum gives a significant contribution, which characterizes the large-scale geometry and topology of the space
in question. In particular, on a compact space the smallest nonvanishing eigenvalue of $\Delta_g$ determines the rate of the exponential
fall-off of the function $D_S(\sigma)$ for large $\sigma$. The linear scale associated with a given $\sigma$ is 
$\sqrt{\sigma}$, which is the typical linear distance travelled by a random walker away from its starting point after a diffusion time $\sigma$.

The relevance of this construction for quantum gravity and other applications of random geometry is the fact that
Laplace-type operators and diffusion processes can be formulated in a much more general context than that of smooth 
Riemannian manifolds, including for example fractals and graphs. The aim of the original treatment in CDT quantum gravity 
\cite{spectral,reconstructing} was to determine an effective spectral dimension of quantum spacetime {\it at short distances} 
from measuring the expectation
value $\langle P(\sigma)\rangle$ of the return probability and then using (\ref{dspec}),
in other words, a quantum analogue of the limiting prescription $\sigma\rightarrow 0$ of
$D_S(\sigma)$, which classically yielded the topological dimension $d$. 

\begin{figure}
\centering
\includegraphics[width=0.6\textwidth]{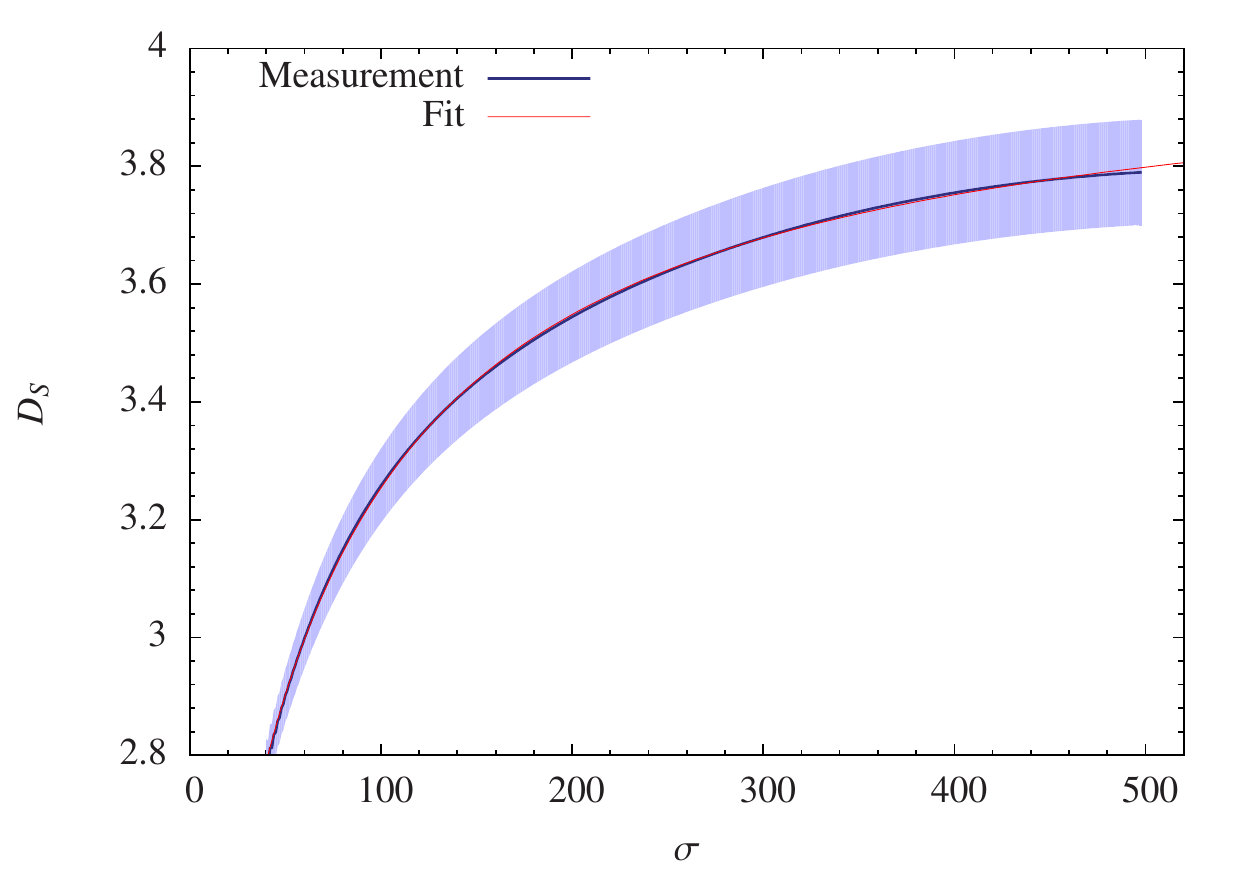}
\caption{\label{fig:39physrep} 
The spectral dimension $D_S$ of spacetime as a
function of the diffusion time $\sigma$, measured at
$(\kappa_0,\Delta)\! =\! (2.2, 0.6)$, for $t_{tot}\! =\! 80$ and a spacetime volume
$N_4\!=\! 181k$. The averaged measurements lie along the central curve,
together with a superimposed best fit
$D_S(\sigma)\! =\! 4.02- 119/(54+\sigma)$. The two outer
curves represent error bars. 
}
\end{figure}

Since it is generally {\it not} the case that a nonperturbative superposition of geo\-metries assembled from $d$-dimensional
building blocks will give rise to an effectively $d$-dimensional geometry on {\it any} scale beyond the cut-off, the result that
the short-scale spectral dimension measured in CDT is asymptotically compatible with 4 is highly nontrivial. 
On top of this came the surprising finding that at distances larger than the cut-off scale (and its associated lattice artefacts) but below 
the region where $D_S(\sigma)$ behaves classically, the spectral dimension undergoes a significant reduction, as already 
mentioned above.
A natural interpretation of this phenomenon is in terms of {\it quantum} corrections to the spectral dimension in a 
Planckian, genuinely nonperturbative regime.\footnote{Note that the Planck scale is not invoked in an ad hoc manner in CDT, but
can be extracted from the measured spectrum of the quantum fluctuations of the spatial volume aka the ``Friedmann factor".} 
Note that in the $\sigma$-interval where the spectral dimension could be measured and its thermodynamic limit 
be extracted with good confidence, $D_S(\sigma)$ takes on values $\lesssim 3.8$ (see Fig.\ \ref{fig:39physrep}) and therefore cannot be 
regarded as having reached its
classical regime. Nevertheless, the existence of a classical regime and the scale dependence of $D_S(\sigma)$ were established robustly with 
the help of heuristic three-parameter fits in the range of $\sigma$ where measurements are reliable. 

Since these early results, determining the short-scale spectral dimension has taken on a larger significance in nonperturbative 
quantum gravity, simply because it is a rare instance of a number that can be computed or measured in a large variety of 
candidate theories or, more generally, in nonclassical models of spacetime. This fact, and multiple indications -- beyond CDT -- that
there may be a ``true" or ``correct" value $D_S(\sigma\rightarrow 0)\! =\! 2$ (see, for example,
\cite{carlip}), make this reminiscent of computations of black hole entropy, whose value by contrast is a prediction 
of semi-classical gravity. On the one hand, the fact that 
``we finally can compute numbers" in quantum gravity, even if their phenomenological
implications are unclear, should certainly be welcomed. On the other hand, some degree of caution is advised, considering
that various derivations are performed in (sometimes highly) incomplete formulations of quantum gravity, rely heavily on additional 
ad hoc assumptions, employ semi-classical arguments of doubtful applicability, and/or are of a purely kinematical nature. In addition, some coincidences in the value of $D_S(\sigma\rightarrow 0)$
may be purely accidental, or are seen to occur only in particular (topological) dimensions (see \cite{benehenson} for a related
discussion).   

Recent developments regarding spectral observables in CDT include measurements of the spectral dimension away from the 
``canonical" phase space point $(\kappa_0,\Delta)\! =\! (2.2, 0.6)$, in phases 
$C_{dS}$ and $A$ \cite{evidence,searching}, the suggestion to investigate the function $D_S(\sigma)$ on all scales, generalising
a corresponding study of CDT in three dimensions \cite{benehenson}, and a comprehensive analysis of the spectral properties
of three-dimensional spatial slices in four-dimensional CDT quantum gravity \cite{clemente1}.

\begin{figure}
\centering
\includegraphics[width=0.75\textwidth]{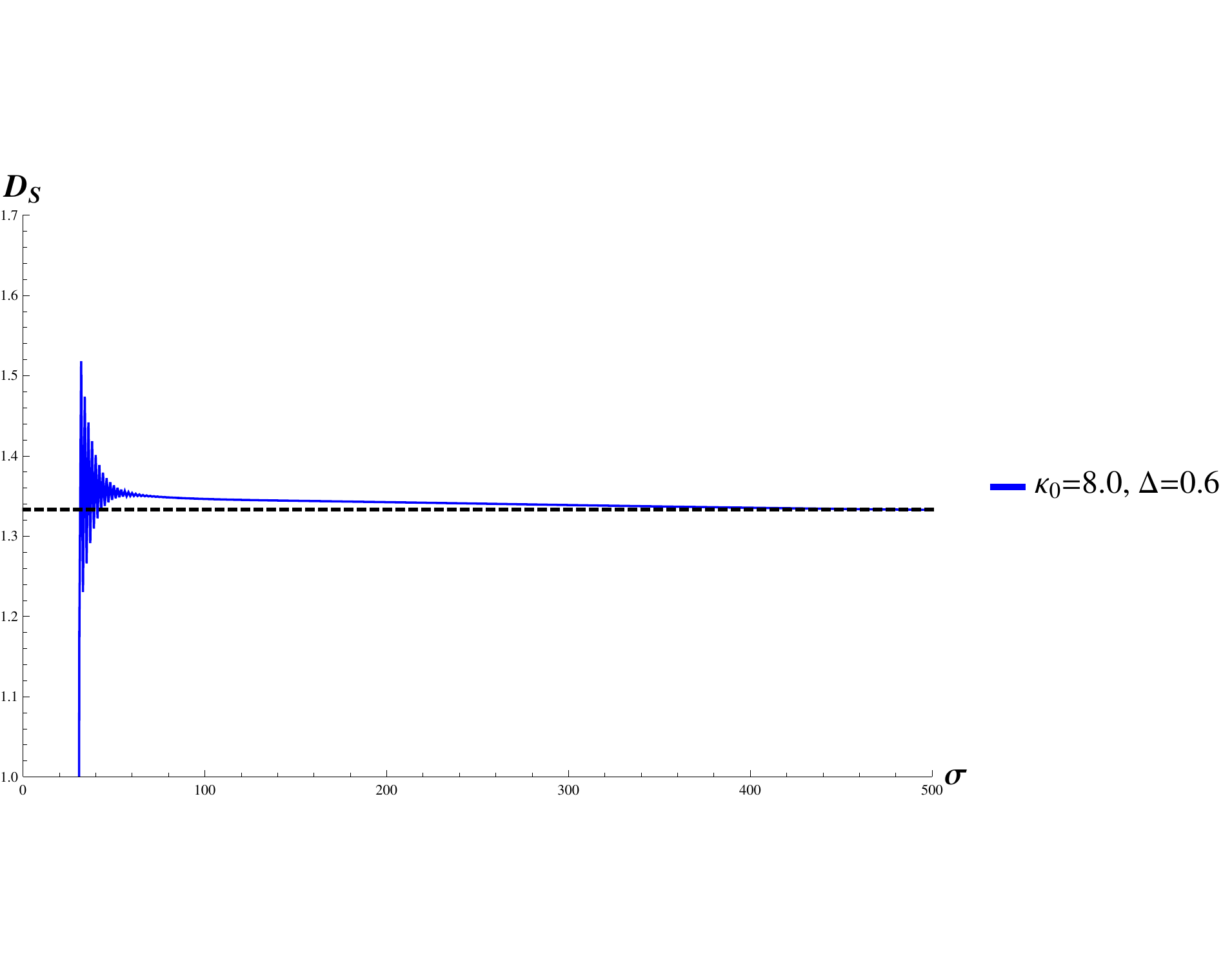}
\vspace{-1cm}
\caption{\label{fig:specdima} 
The spectral dimension $D_S$ of spacetime in phase $A$, measured at
$(\kappa_0,\Delta)\! =\! (8.0,0.6)$, for $t_{tot}\! =\! 80$ and $N_{41}\!=\! 160k$, exhibits a fractal, scale-inde\-pendent behaviour.
[Figure from \cite{evidence}.] 
}
\end{figure}

Ref.\ \cite{evidence} corroborates the original result $D_S\! \approx\! 2$
for the short-distance spectral dimension at $(\kappa_0,\Delta)\! =\! (2.2, 0.6)$, finding
$D_S(\sigma\!\rightarrow\! 0)\! =\! 1.97\pm 0.27$ 
in simulations at roughly twice the previous volume, $N_4\! =\! 367k$, with time extension
$t_{tot}\! =\! 80$, and otherwise following the procedure outlined in \cite{spectral}. The authors repeat the same measurement
at three other points in phase $C_{dS}$ that are closer to the first-order $A$-$C_{dS}$ transition line,
$(\kappa_0,\Delta)\! =\! (3.6, 0.6)$, $(4.4,0.6)$ and $(4.4,2.0)$, and for fixed $N_{41}$ of up to $300k$. 
Moving closer to this line, the curves for $D_S(\sigma)$ flatten considerably in the $\sigma$-range considered ($\sigma\leq 500$), 
making the fitting more challenging. Somewhat surprisingly, one finds $D_S(0)\!\approx\! 1.5$ at these points, 
within error bars, which seems at odds with the assumed universal character of this observable. 
On the other hand, one would expect short-distance properties to be closer to their universal values in the vicinity of
a higher-order transition, which the new phase space points examined in \cite{evidence} are not. Another possible
explanation for the discrepancy is that the systematic errors are underestimated and that the fitting functions used are less appropriate in
this region of phase space. Some evidence for this is the fact that the fitted curves for the spectral dimension functions 
corresponding to different phase space points do not match particularly well, even after a best rescaling of $\sigma$     
(Fig.\ 3 in \cite{evidence}). -- In addition to measurements in the de Sitter phase, \cite{evidence} also determined
the spectral dimension of CDT at the point $(\kappa_0,\Delta)\! =\! (8.0, 0.6)$ deep inside phase $A$, with strong
indications of ``branched-polymer" behaviour $D_S\!\approx\! 4/3$, independent of $\sigma$ in the range
$\sigma\in [60,500]$, see Fig.\ \ref{fig:specdima}.

A related line of argument is pursued in \cite{searching}, where in addition to the phase space points 
$(\kappa_0,\Delta)\! =\! (2.2, 0.6)$, $(3.6,0.6)$ and $(4.4,0.6)$, a sequence of points {\it along} the first-order
$A$-$C_{dS}$ transition line is considered. The aim is to investigate the behaviour of two different notions of
an (effective) lattice spacing as a function on phase space. Comparing measurements at different points in phase 
space is of course not meaningful a priori unless they lie on the same renormalization group trajectory of
constant physics, as described in Sec.\ \ref{sec:rengroup} above (see also \cite{coop3} for related comments).\footnote{Moving along
such a trajectory towards a UV fixed point, the lattice spacing $a$ should vanish.}
The paper \cite{searching} illustrates some of the issues that arise in the absence of a full-fledged renormalization group analysis. 
Its first proposition is to extract what the authors call an ``absolute lattice spacing"
from measuring the quantum fluctuations around de Sitter space, as introduced in \cite{desitter1,desitter2}. 
A difficulty here is the fact that the effective action for the spatial volume at the 
$A$-$C_{dS}$ transition is very different from that inside phase $C_{dS}$, because the kinematic term vanishes,
as described in Sec.\ \ref{subsec:eff}. To what extent a de Sitter description still captures this situation adequately
is unclear. In addition, \cite{searching} shows that different ways of accounting for the asymmetry between the 
numbers of (4,1)- and (3,2)-simplices, which depends on the location in phase space, lead to significant differences 
for the measurements of the lattice spacing for points away from the $A$-$C_{dS}$ transition. 

The second proposition of \cite{searching} is to extract a ``relative lattice spacing" $a_{rel}$ by fitting spectral dimension
curves at different points in phase space, using a fit function
\begin{equation}
D_S(\sigma)=a-b/(c+\sigma/a_{rel}^2),
\label{dsrescale}
\end{equation}
where $a$, $b$ and $c$ are constants and $a_{rel}$ is adjusted to achieve a best overlap between different curves. 
Here, the difficulties already faced in \cite{evidence} come to a head when trying to
measure the spectral dimension very close to the $A$-$C$ transition. The curves become extremely flat, with
$D_S$ barely exceeding 1.6 at the upper end $\sigma\! =\! 500$ of the $\sigma$-range considered. Consequently,
the curves have to undergo a very large rescaling compared to the curve at the reference point $(\kappa_0,\Delta)\! =\! (2.2, 0.6)$,
leading to extremely small ranges in the rescaled $\sigma$ available for fitting. One has to set $a\!\approx\! 4$ by 
hand in eq.\ (\ref{dsrescale})
to be able to make any comparison between the curves. --
The systematic and theoretical uncertainties just described make it difficult to judge whether
the qualitative increase in the lattice spacing for growing coupling $\kappa_0$ (at fixed $\Delta\! =\! 0.6$) observed in 
\cite{evidence,searching} is a genuine physical effect, which would imply that these simulations probe closer into the
Planckian regime. A resolution of these issues will most likely require a larger set of observables and a better understanding of
the behaviour of CDT quantum gravity under renormalization.

An idea first suggested in the context of three-dimensional CDT \cite{benehenson} is to consider the 
entire spectral dimension function $D_S(\sigma)$, not just for small $\sigma$, to learn more about quantum geometry and
its classical limit. An obvious difficulty with this proposal is that, even classically, away from the asymptotic regimes 
$\sigma\! \rightarrow \! 0$ and
$\sigma\! \rightarrow \!\infty$, $D_S(\sigma)$ contains information about the curvature, shape and topology of a space in 
a rather summary and integral manner, such that thinking of $D_S$ as an effective dimension is in most instances not particularly 
meaningful.
Evaluating $D_S$ on ensembles of triangulated manifolds adds finite-size effects, discretization 
artefacts\footnote{see \cite{thuerigen}
for examples of discretization effects on some regular lattices} and genuine quantum effects
to the mix, in a way that in general will be difficult to disentangle, certainly at the lattice volumes currently within reach. 
Similarly, the suggestion of extracting absolute physical scales from the behaviour of $D_S(\sigma)$ for the purpose 
of defining renormalization group flows \cite{coop1} seems difficult to realize: the relatively robust short-scale dimension
does not refer to any sharply defined transition point to classicality, while the universal nature of other
features of the spectral dimension function at intermediate scales (like its maximum) are not well understood either.

Noteworthy in the context of spectral observables is a recent attempt to tap into the potentially rich information contained 
in the eigenvalues and eigenvectors of the Laplace-Beltrami operator on spatial slices with topology $S^3$ in CDT \cite{clemente1}.
Which properties of the spectrum lend themselves to ensemble averaging and are ultimately related to
geometric observables in the continuum is largely unexplored territory. 
The fractal and largely nonclassical geometric nature of the slices of constant proper time has been  
exhibited previously \cite{reconstructing,geometry}.
The authors of \cite{clemente1} 
solve the eigensystem for the Laplacian associated with the four-valent graphs dual to spatial triangulations
for sample points in all four phases of CDT, searching for typical features of the spectrum that characterize and distinguish between the
different phases, and for new order parameters to help analyze the nature of the phase transition lines.
The simulations were performed at fixed volumes $N_{41}\! \leq \! 80k$ and for $t_{tot}\! =\! 80$,
and a range of discrete spatial volumes $V_S$ of up to about 3000 in terms of the number of tetrahedra.
\begin{figure}
\vspace{-0.5cm}
\centering
\includegraphics[width=0.65\textwidth]{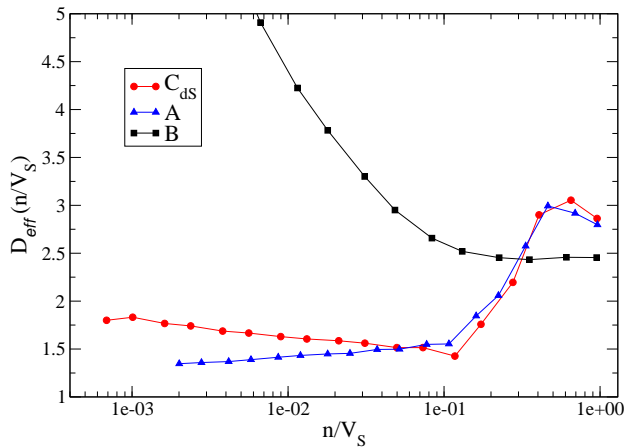}
\caption{\label{fig:16fromclemente1} 
Effective spectral dimension $D_{\it eff}$ of spatial slices at $(\kappa_0,\Delta)\! =\! (5.0, 0.6)$ in phase $A$,
$(2.2, -0.2)$ in phase $B$ and $(2.2, 0.6)$ in phase $C_{dS}$, measured 
for $t_{tot}\! =\! 80$, spacetime volume $N_{41}\!=\! 80k$ (phases $A$ and $C_{dS}$) and $N_{41}\! =\! 16k$ (phase $B$),
and in small bins of $n/V_S$ \cite{clemente1}. 
In the kinematical limit $n/V_S\!\rightarrow\! 1$ the effective dimensions are close to 3, primarily 
determined by the discrete lattice set-up. [Figure from \cite{clemente1}.]}
\end{figure}

Since the eigenvalues $\lambda_n$, $n\! =\! 0,1,2,\dots$, can be ordered according to size, with $\lambda_0\! =\! 0$ and
$\lambda_1$ denoting the
smallest nonvanishing eigenvalue, one can for example monitor the distribution of a particular eigenvalue with label $n_0$,
where the outcome will in general depend on both the location in phase space and the spatial volume of the slice.
When measuring the density $\rho(\lambda)$ of the first 100 eigenvalues for different spatial volumes, \cite{clemente1} found  
a completely different behaviour in phases $B$ and $C_{dS}$. While in phase $B$ there is a gap $\Delta\lambda\! =\! \lambda_1
\!\approx\! 0.1$ that stays practically constant when the slice volume is varied, there is no such gap in the de Sitter phase,
where instead $\lambda_1$ converges to zero in the thermodynamic limit.\footnote{Both of these have been confirmed in 
a more detailed follow-up study \cite{clemente2}. Interestingly, the expectation values 
$\langle \lambda_n \rangle$ of low-lying eigenvalues in phase $B$ seem to approach a discrete set of distinct values in the infinite-volume 
limit.}
The former is characteristic for a highly connected space or graph with large Hausdorff dimension and small diameter, and
is consistent with what is already known about the geometry of phase $B$. 
By contrast, in phase $C_{dS}$ one finds a behaviour compatible with
Weyl's law for the asymptotic scaling of $n(\lambda)$ -- the number of eigenvalues below the value $\lambda$ -- for large
$\lambda$, namely,
\begin{equation}
n(\lambda\rightarrow\infty) \propto V \lambda^{d/2}/(2\pi)^d,
\label{weyl}
\end{equation}
valid on a $d$-dimensional smooth space of volume $V$. In analogy with the earlier described method of extracting
a spectral dimension from the return probability of a random walker on a piecewise flat space by generalising the
smooth-space relation (\ref{dspec}), the relation (\ref{weyl}) can be used to extract an effective spectral dimension
\begin{equation}
D_{\it eff}:=2\ \frac{d \log n/V_S}{d\log \lambda_n} 
\label{dseff}
\end{equation}
for nonsmooth
spaces. In this context, considering the large-$\lambda$ limit is the analogue of considering the small-$\sigma$ regime of the random walker. 
Beyond very short distances, which are dominated by lattice artefacts, the spectral dimension determined in this way is
$D_{\it eff}\!\approx\! 1.5$ in phase $C_{dS}$, in agreement with earlier measurements using the return 
probability \cite{reconstructing,Gorlichthesis}. As illustrated by Fig.\ \ref{fig:16fromclemente1}, the same 
approximately constant value is found in phase $A$, whereas in phase $B$
one finds a strongly increasing $D_{\it eff}$ for decreasing $n/V_S$, compatible with an infinite spectral dimension.

In line with the properties of the bifurcation phase discussed earlier, the spectral analysis in this phase reveals characteristic
phase-$B$ and phase-$C_{dS}$ behaviour on alternating spatial slices. For the slices of $B$-type this implies the existence
of a spectral gap $\lambda_1$ and a general shift to larger values for eigenvalues $\lambda_n$ of the same order $n$.
The jumps between large and small values are particularly apparent for the low-lying part of the spectrum, and
become somewhat less pronounced as $n$ increases. The suggestion in \cite{clemente1}, further elaborated in \cite{clemente2}, is
to approach the $C_b$-$C_{dS}$ phase transition from inside the bifurcation phase $C_b$, and use the discrete set of
eigenvalues $\langle \lambda_n\rangle$ associated with the slices of type $B$, which one expects to vanish as
the transition line is approached, as order parameters. In the exploratory study \cite{clemente2}, the coupling $\kappa_0$ is kept
fixed as $\Delta$ is increased towards the phase transition. The critical values found for $\Delta$ are consistent with those
found for other order parameters described in Sec.\ \ref{subsec:eff} above, but the size of the simulations does not yet allow
for a reliable measurement of critical exponents, or for a verification that the spectral gaps do indeed vanish at the transition.  

To summarize, although there clearly is a lot of (fine-grained) information contained in the Laplacian spectrum of ensembles of causal
dynamical triangulations, it remains a challenge to extract from it observables -- beyond the relatively well-understood short-scale
spectral dimension -- which describe genuine, universal 
quantum properties of spacetime, and upon coarse-graining relate to (quasi-)local semi-classical geometric features.

\subsection{Curvature observables}
\label{sec:curv}

Given that curvature is such a central notion in classical general relativity, one may wonder what role 
it has to play in nonperturbative quantum gravity.
There are multiple obstacles in representing curvature in the quantum theory, which have hampered 
developments so far. However, there has been a recent and promising breakthrough in defining a notion of Ricci curvature in CDT,
which we will describe later in this section.
One reason why the need for {\it quantum curvature} may not appear as pressing is the existence of a straightforward prescription for 
the curvature scalar $R(x)$ in Regge calculus in terms of deficit angles \cite{regge}, which is also used in representing
the Einstein-Hilbert action (\ref{actlor}) in CDT quantum gravity, see \cite{physrep} for a derivation. Of course, one needs to keep in
mind that the curvature scalar, like any other local curvature invariant, is not a good observable unless it is integrated over spacetime, say.

However, even the integrated form of the curvature scalar turns out to be not well suited as a quantum observable. 
Not unexpectedly for the counterpart of a second-order differential operator in the continuum, the total Regge scalar curvature
diverges in the continuum limit, but it is not clear how to renormalize it appropriately. Put differently,
the deficit angle prescription is associated with the cut-off scale and there is no obvious way of scaling it up, in the sense 
of associating ``effective" deficit angles to coarse-grained regions of spacetime.
Another shortcoming is the fact that the curvature scalar captures only a small part of the information 
contained in the full Riemann tensor. It is certainly possible to construct finite-difference expressions that represent
the Riemann tensor on piecewise flat manifolds, but they inevitably would be unwieldy and highly nonunique, 
and -- to the extent that
they are still based on deficit angles representing the Gaussian curvature of two-dimensional subspaces -- would suffer from
the same problems as the scalar curvature. 

To ameliorate the singular nature of curvature in the quantum realm and realize gauge invariance at the same time,
an old idea coming from gauge field theory 
is to capture curvature information in terms of nonlocal Wilson loop observables, obtained by integrating the associated gauge connection 
along one-dimensional closed curves in spacetime.
Similarly, in gravity one can consider the {\it gravitational Wilson loop} $W_\gamma$ associated with a loop $\gamma$ 
on a spacetime manifold $M$ with metric $g_{\mu\nu}$ and
associated Levi-Civita connection $\Gamma^\lambda_{\mu\nu}$, as the trace of the path-ordered exponential of $\Gamma$
along $\gamma$,
\begin{equation}
W_\gamma [\Gamma]:={\rm Tr}\, {\cal P}\, {\rm exp}\Big( -\oint_\gamma \Gamma\Big).
\label{wilsonloop}
\end{equation} 
The path ordering, indicated by $\cal P$, is needed because of the nonabelian nature of the connection. The expression under the
trace in (\ref{wilsonloop}) is a $SO(3,1)$-valued functional, a so-called ``holonomy", 
acting on tangent vectors at the base point $p\in M$ where the loop $\gamma$ begins and ends. 
(The analogous quantity in the Wick-rotated CDT setting will be $SO(4)$-valued.)
Unlike in gauge theory, Wilson loops in pure gravity are {\it not} observables, because they
refer to a closed curve $\gamma$ in spacetime and therefore are not diffeomorphism-invariant. 
There are different ways of constructing observables from Wilson loops,
for example by ``marking" the location of the loop in terms of matter degrees of freedom or by performing averages over loops or
subsets of loops that share certain invariant geometric features regarding their length and shape. 

These considerations are relevant for CDT, where parallel transport and Wilson loops can be defined in a
straightforward way, as described in \cite{wilson}. 
Because of the piecewise flat nature of the geometry this turns out to be far easier than on a smooth 
curved manifold. A natural prescription is to use straight path segments between the 
centres of pairs of adjacent four-simplices and consider the holonomies of piecewise straight loops made from
such segments. To compute the path-ordered
exponentials requires the introduction of a coordinate system\footnote{Barycentric coordinates are a convenient choice \cite{wilson}.} 
in each simplex at an intermediate stage of the calculation, which drops out again upon taking the trace. 

On a smooth four-dimensional Riemannian manifold,
the gravitational Wilson loop of an infinitesimal square loop $\gamma_{[\mu\nu]}$ of geodesic edge length $\varepsilon$ 
in the $\mu$-$\nu$ plane depends to lowest nontrivial order on the entries $R^.{}_{.\mu\nu}$ of the 
Riemann curvature tensor according to
\begin{equation}
W_{\gamma_{[\mu\nu]}} = 4+\varepsilon^4 R^\kappa{}_{\lambda\mu\nu}  R^\lambda{}_{\kappa\mu\nu}+{\cal O}(\varepsilon^5),
\label{riemwilson}
\end{equation}
where the 4 comes from the trace of the unit matrix in the defining representation of $SO(4)$ (see, for example, \cite{diak}).
On a general curved manifold, 
there is no obvious way to relate the Wilson loop of a {\it noninfinitesimal} loop to 
some (integrated or coarse-grained) form of local curvature in a similar way.\footnote{A gravitational version of the nonabelian
Stokes' theorem is unlikely to be useful because curvature appears only inside a complicated, nonlocal expression
involving area ordering \cite{ournewpaper}.} However, it is a priori conceivable that in nonperturbative quantum gravity 
like CDT, Wilson loops which probe the quantum geometry on scales sufficiently large compared to the Planck scale but are still
infinitesimal from a classical point of view
display a semiclassical behaviour, where quantum fluctuations mostly ``average out", leading to
expectation values for the Wilson loops close to (4 times) the identity, with small deviations that characterize some effective curvature scale.

\begin{figure}[t]
\centerline{\scalebox{0.6}{\rotatebox{0}{\includegraphics{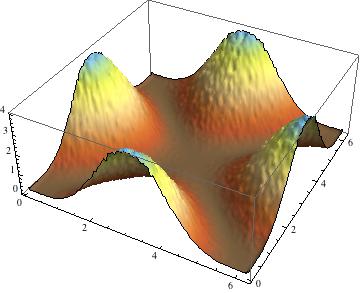}}}}
\caption{
Histogram of the invariant angles $(\theta_1,\theta_2)$, extracted from Monte Carlo measurements
of Wilson loops in CDT \cite{wilson}, matching the functional form (\ref{angles}) up to an overall normalization.}
\label{fig:wilsondist}
\end{figure}

This hypothesis was investigated in simulations of four-dimensional CDT with spherical slices
at $(\kappa_0,\Delta)\! =\! (2.2,0.6)$, for $t_{tot}\! =\! 80$ and $N_4\! =\! 20k$ and loops of winding
number 1 with respect to the compactified time direction \cite{wilson}. To obtain a genuine Wilson loop observable,
the underlying loop $\gamma$ was identified with the spacetime trajectory of a massive particle moving forward in time along one of the 
piecewise straight paths mentioned earlier. The total action of the system consists of the usual Einstein-Hilbert piece
and an extra term $m_0 \ell (\gamma)$ proportional to the discrete length $\ell$ of the loop, and its partition function is a double sum over 
triangulated geometries {\it and} nonselfintersecting paths $\gamma$. The observable extracted from the Wilson loop measurements
was the expectation value of the probability distribution $P(\theta_1,\theta_2)$ of two invariant angles $\theta_i\!\in\! [0,\pi]$ that
characterize elements of $SO(4)$ up to conjugation. It is straightforward to compute $P(\theta_1,\theta_2)$ analytically for the case that
the holonomies are uniformly distributed with respect to the Haar measure on the group manifold of $SO(4)$. The normalized
distribution is given by
\begin{equation}
P(\theta_1,\theta_2) = \frac{1}{\pi^2} \,
\sin^2 \bigg( \frac{\theta_1 +\theta_2}{2}\bigg) \sin^2  \bigg(\frac{\theta_1 -\theta_2}{2}\bigg).
\label{angles}
\end{equation}
It turns out that for a wide range of bare masses $m_0$, the distributions measured in the CDT simulations
match this functional form perfectly 
(see Fig.\ \ref{fig:wilsondist}), from which several conclusions may be drawn. On the one hand, there are no obvious discretization
effects connected to the fact that (for the four-dimensional equilateral simplices used) all deficit angles come in integer units of
$\arccos (1/4)$, the angle between two three-dimensional faces sharing a two-dimensional triangle. 
On the other hand, there is no sign that for the ensemble of loops under consideration the Wilson loops cluster around
the identity $(\theta_1,\theta_2)\! =\! (0,0)$, which would be a potential indicator of an averaging-out of small-scale fluctuations 
of the curvature. It could mean that the loops considered here are too large -- their minimal discrete length is $\ell\! =\! 360$, due to
the way the dual paths move between adjacent spatial layers. This is a hypothesis that could be tested relatively easily. 
Alternatively, there may be more subtle ways of extracting
interesting geometric information from Wilson loops, perhaps from higher-order correlations. It is also possible that Wilson loops
are simply not suited as a measure of coarse-grained local curvature in quantum gravity, because they are still too singular.

\begin{figure}[t]
\centering
\includegraphics[width=0.55\textwidth]{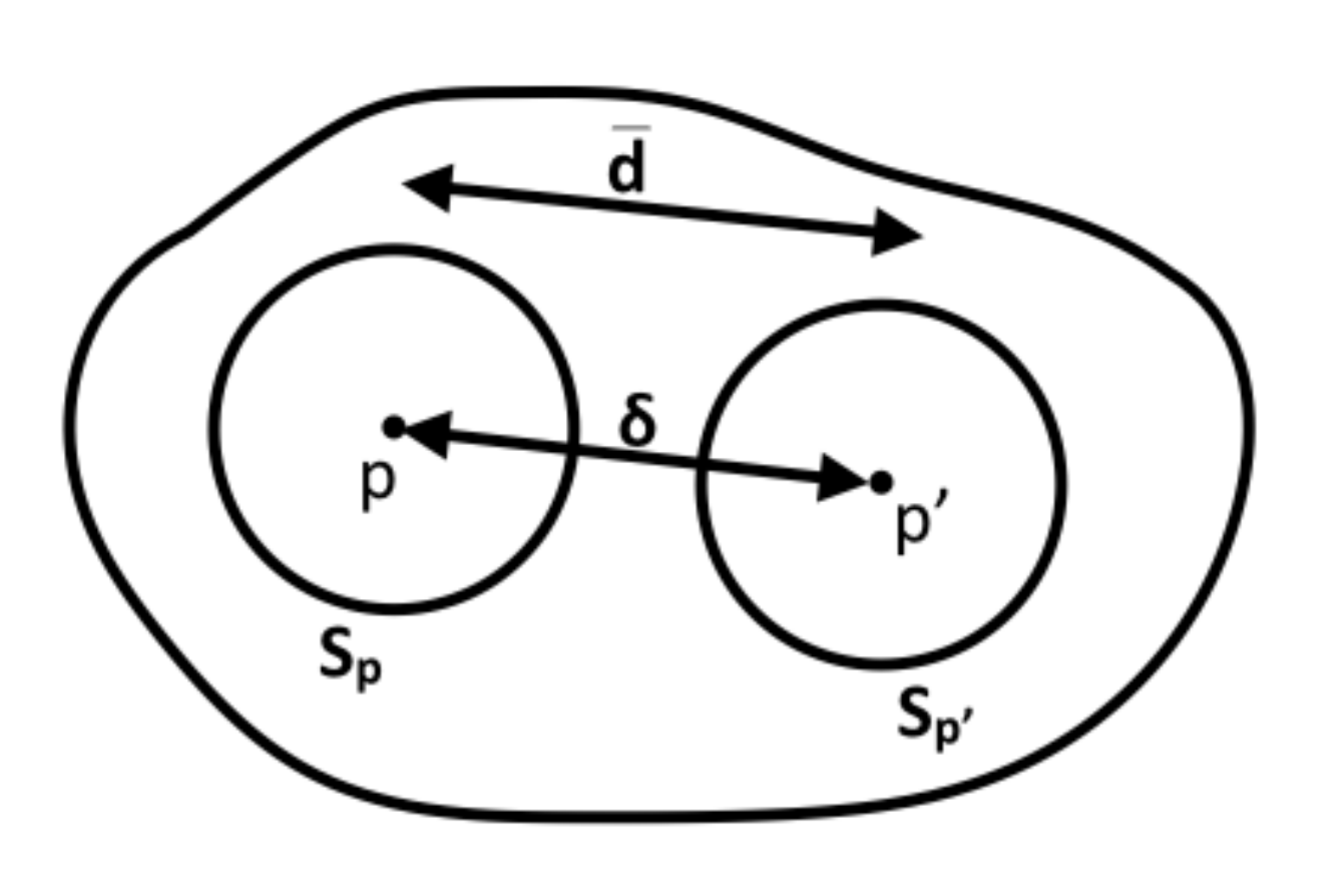}
\caption{\label{fig:spheredist} To obtain the local Ricci curvature $Ric(v,v)$ associated with a vector $v$
on a smooth Riemannian space, one compares the
distance $\bar d$ of two small nearby spheres $S_p$ and $S_{p'}$ with the distance $\delta\! =\! |v|$ of their centres $p$ and $p'$,
measured along the unique geodesic from $p$ to $p'$ in the direction of $v$.}
\end{figure}

The Wilson loop investigation just described provides another motivation for finding more useful curvature observables with better
averaging properties. The recent suggestion of defining a scalable notion of {\it quantum Ricci curvature} \cite{qrc1} without referring to
deficit angles is a promising step in this direction, which has already been tested for dynamical triangulations 
in lower dimensions \cite{qrc2}. The key idea, inspired by a classical characterization of curvature on smooth $D$-dimensional
Riemannian spaces and
later generalized to general metric spaces \cite{ollivier}, is to compare the distance $\bar{d}$ of two $(D\! -\! 1)$-spheres with the distance
$\delta$ between their centres (see Fig.\ \ref{fig:spheredist}). More specifically, 
on a Riemannian space with positive
(negative) Ricci curvature, the distance $\bar{d}(S_p,S_{p'})$ of two small nearby spheres $S_p$ and $S_{p'}$ 
is smaller (bigger) than the distance $\delta$. 

The quantum Ricci curvature defined in \cite{qrc1} and evaluated on a variety of piecewise flat spaces, including regular lattices and
Delaunay triangulations, uses a variant where the sphere radii are both set equal to the centre distance $\delta$. 
In addition, the sphere distance is defined as the average distance of $S_p^\delta$ and $S_{p'}^\delta$,
\begin{equation}
\bar{d}(S_p^{\delta},S_{p'}^{\delta}):=\frac{1}{vol(S_p^{\delta})}\frac{1}{vol(S_{p'}^{\delta})}
\int_{S_p^{\delta}}d^{D-1}q\; \sqrt{h} \int_{S_{p'}^{\delta}}d^{D-1}q'\; \sqrt{h'}\ d(q,q'),
\label{sdist}
\end{equation}   
or a suitable analogue of this prescription on the metric space in question. In (\ref{sdist}),
$h$ and $h'$ are the determinants of the metrics induced on the two $(D-1)$-dimensional ``spheres"\footnote{Note that
when using this construction for macroscopic scales $\delta$, the sphere $S^\delta_p$ is defined as the set of all points
at geodesic distance $\delta$ from the point $p$, and will in general not have the topology of an $S^{D-1}$-sphere.}, which are also used 
to compute the sphere volumes $vol(S)$, and $d(q,q')$ denotes the geodesic distance between the points $q$ and $q'$. 

\begin{figure}[t]
\centering
\includegraphics[width=0.6\textwidth]{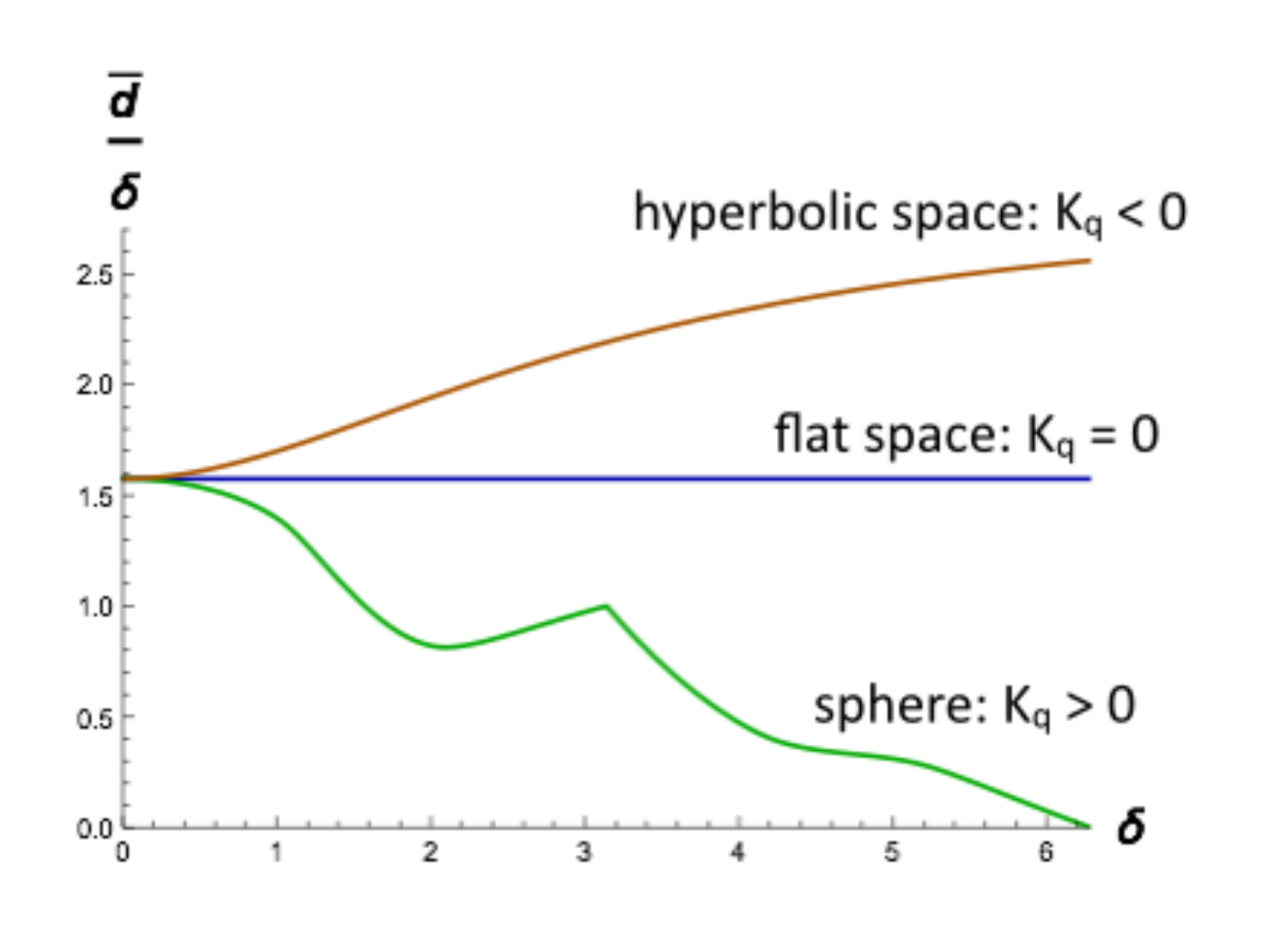}
\caption{\label{fig:constcurv} The normalized average sphere distance $\bar{d}/\delta$ as a function of $\delta\! \in\! [0,2\pi]$
for three constantly curved model spaces in two dimensions. The analogous curves in higher dimensions look qualitatively similar.
The curvature radius has been set to 1 wherever applicable.}
\end{figure}

The presence of a single linear distance scale $\delta$ then allows us to define a scalable notion of {\it quantum Ricci curvature $K_q$ at
the scale} $\delta$ from the quotient 
\begin{equation}
\frac{\bar{d}(S_p^{\delta},S_{p'}^{\delta})}{\delta}=c_q (1 - K_q(p,p')),\;\;\; \delta =d(p,p'),
\label{qric}
\end{equation}
where $c_q$ is a positive constant with $0\! <\! c_q\! <\! 3$, which depends on the metric space under consideration,
and $K_q$ captures any nontrivial dependence on $\delta$. 
The beauty of this prescription is that it relies only on being able to measure geodesic distances and volumes, both of
which are easily accessible in (C)DT, and that it captures directional, tensorial information. 
Part of the strategy when measuring quantum Ricci curvature in CDT quantum gravity will be to try and understand
whether in any range of scales its behaviour resembles that of a classical curved space. An important point of comparison
are the curves for the normalized average sphere distance $\bar{d}/\delta$ for constantly curved smooth spaces (with positive, negative and
vanishing curvature) in any dimension, which can be computed analytically (c.f. Fig.\ \ref{fig:constcurv}). 
A first indication of the robustness of the quantum Ricci curvature comes from evaluating it on the ensemble of Euclidean 
dynamical triangulations in two dimensions \cite{qrc2}, which is known to have a highly nonclassical and fractal geometry.
The measured expectation value of its curvature can be mapped best to that of a five-dimensional continuum
sphere, where it should be recalled that Euclidean quantum gravity in two dimensions has a Hausdorff dimension of four (which
at least comes close). Measurements of the quantum Ricci curvature in CDT in four dimensions are under way \cite{preparation}, 
and will for the first time give an indication to what extent the dynamically generated de Sitter universe -- identified on the basis of 
its large-scale dimension and volume profile -- resembles a de Sitter space also in terms of its quasi-local curvature properties.

\section{Prospects for cosmology}
\label{lessons:sec}

On the basis of the quantum observables known and measured in CDT quantum gravity so far, what is the status of 
extracting physical predictions, without invoking additional ad hoc assumptions? 
At this stage, no statements of substance have been made about specific, quantitative predictions of new physics due to
genuine quantum gravity effects. However, it should be noted that 
{\it CDT predicts a positive (renormalized) cosmological constant $\Lambda$}.
This feature of dyna\-mically tri\-angulated models has to do with the convergence requirement of the nonperturbative 
path integral and the renormalization behaviour of the cosmological constant in the continuum limit.
CDT does not predict any particular value for the physical $\Lambda$ per se, only that
it should be positive, which is of course in agreement with our current understanding of the universe based on $\Lambda$CDM 
cosmology.  

The most obvious place to look for physical implications is early universe cosmology. In this context, it is encouraging that 
CDT has been shown to reproduce several aspects of the classical theory, which is {\it not} the norm in nonperturbative quantum 
gravity, as pointed out earlier. The matching with a semiclassical behaviour of the de Sitter volume profile including quantum 
fluctuations for a universe of only about 20 Planck lengths across \cite{desitter1,desitter2} is remarkable, and an 
indication that -- at least for selected observables -- the scale gap from Planckian to semiclassical is perhaps
less daunting than it appears na\"ively. Nevertheless, the observables currently accessible in CDT quantum gravity are still 
too few and too coarse to allow for a more detailed comparison. 

A next step in the development of the quantum theory is therefore to ascertain whether and 
on what scale the dynamically generated de Sitter universe can be said to resemble
a smooth, constantly curved de Sitter space in terms of its local geometric properties, beyond displaying the correct large-scale
dimensionality and volume profile. Conversely, what would be desirable on the cosmology side is the formulation of
explicitly diffeomorphism-invariant and {\it background-independent} observables. 
Because of the vastly different languages used in the full quantum theory (nonperturbative,
nonsmooth, no fixed background, manifest diffeomorphism invariance) and in standard cosmology (perturbative, smooth, 
cosmological background metric, gauge-fixed coordinates), such observables constitute the natural common ground
where the two formulations can be compared.

Let us emphasize that all CDT results discussed so far were extracted {\it without} making assumptions about the
homogeneity or isotropy of spacetime at the outset. For instance, the expectation value of the volume 
profile for spacetimes of topology $S^1\!\times\! S^3$ 
was obtained by integrating over all geometric modes in the path integral, 
without resorting to any a priori symmetry reduction or a symmetry-related truncation of the degrees of freedom.
Remarkably, {\it it then turned out that in an extended region of phase space the outcome can be interpreted in terms
of an isotropic and homogeneous solution of the Einstein equations.}

\begin{figure}
\centering
\includegraphics[width=0.85\textwidth]{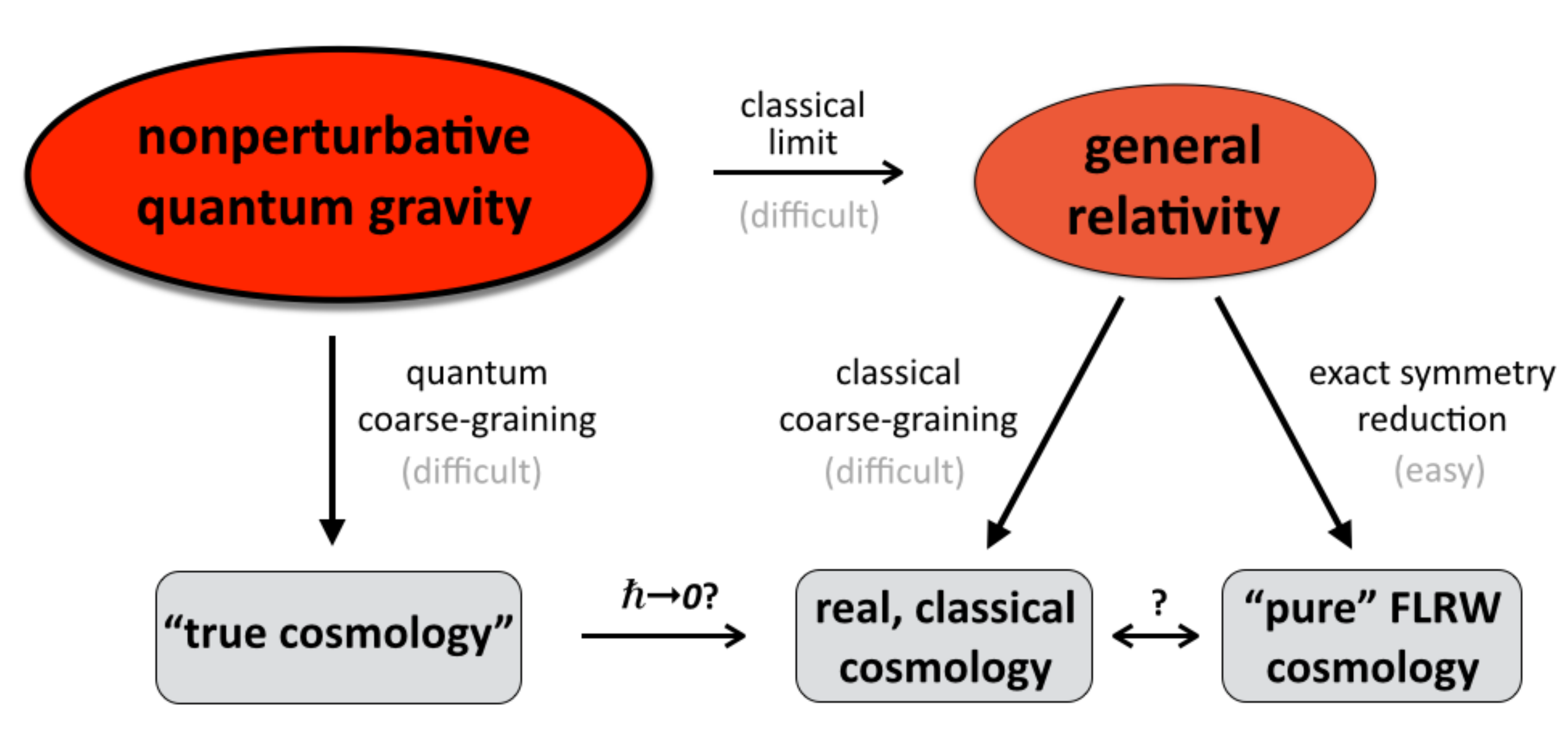}
\caption{\label{fig:QGGR} Relations in theory space: different ways to derive cosmology from nonperturbative 
quantum gravity and general relativity.}
\end{figure}
Of course, given a fundamental, nonperturbative theory of quantum gravity, all quantum and classical aspects of gravity and
cosmology must ultimately follow from it, including the quantum behaviour of the very early universe and finding
the nonperturbative quantum ground state of the universe, if it exists. It is an interesting question whether there are observable aspects of
large-scale cosmology today that can {\it only} be understood in terms of a full-fledged theory of quantum gravity, and not 
classical general relativity (or perhaps quantum cosmology) alone. 
To exhibit phenomenological consequences in cosmology or elsewhere is of course one of the raisons d'\^etre of quantum gravity.

Fig.\ \ref{fig:QGGR} illustrates various ways of arriving at cosmology from either quantum or classical gravity, assuming that
general relativity is the correct theory classically. 
By ``true cosmology" we mean all large-scale phenomena derived directly from nonperturbative quantum gravity
by integrating out or coarse-graining its microscopic degrees of freedom. 
It should be contrasted with standard quantum cosmology, which denotes a quantum-mechanical treatment of
only those degrees of freedom that are left after a classical symmetry reduction\footnote{In the gravitational sector of 
Friedmann-Lema\^itre-Robertson-Walker (FLRW) cosmologies, this is just the scale factor $a(t)$.},
that is, a quan\-tum version of the ``pure FLRW cosmology" in the bottom right-hand corner of Fig.\ \ref{fig:QGGR}. 
Readers interested in further considerations on the relation between CDT and cosmology should consult \cite{cosmo},
which also compares the robustness of the classical FLRW paradigm with the fact that in candidate theories of
nonperturbative quantum gravity the emergence of a scale factor exhibiting FLRW behaviour 
is definitely not automatic, {\it unless}
one adopts causality conditions \`a la CDT, as explained in Sec.\ \ref{time:sec} above. Whether and how 
the two ``effective" scale factors and their dynamics 
are related requires further research. Another interesting question is to what extent quantum coarse-graining,
especially for a quantity like the Ricci curvature \cite{qrc1}, can shed light on the little-understood coarse-graining properties 
of the classical Einstein or Friedmann equations\footnote{The so-called 
averaging problem of cosmology refers to the difficulty of quantifying the effect of inhomogeneities on smaller scales
on the dynamics of the universe on larger scales, beyond a linear regime, see \cite{challenges,inhomogeneous} for recent reviews.}
and vice versa. 

Generally speaking, looking for observables that can quantify the presence of (approximate) global symmetries 
at a given scale of a quantum geometry is an important and interesting topic, especially after one has established 
the presence of an extended geometry, as in the $C$-phases of CDT.  
It has been suggested to use the variance of the return probability $P(\sigma)$ (discussed in Sec.\ \ref{sec:specobs})
and the variance of the
normalized volume of geodesic balls of radius $r$ for a given triangulation $T$ as measures for the homogeneity 
of the geometry of $T$ \cite{coop2}. However, a small-scale simulation for CDT in 2+1 dimensions of the expectation values of these
quantities did not produce a conclusive result on whether they can be regarded as meaningful observables in
a continuum limit. -- Of course, homogeneity can in principle be tested for arbitrary local quantities whose averages give rise to
observables, leading us back to a recurring theme of this review, namely, the need for more observables (which by the way is
also an issue familiar to cosmologists). 

\section{Outlook}
\label{summary:sec}

CDT quantum gravity has come a considerable way since its inception as a nonperturbative Lorentzian path integral for 
gravity \cite{cdt2d,cdt1,cdt2}. While the first big review in \cite{physrep} gave a thorough account of the ingredients and
construction of the model, and described some of CDT's trademark results, the present review documents the next, more
mature stage of the theory as a serious contender for nonperturbative quantum gravity. 
As emphasized in Sec.\ \ref{portrait:sec}, a characteristic feature of this approach is to 
take numerical tools seriously as a legitimate way of advancing our quantitative understanding of ``bulk quantum gravity" in a 
Planckian regime, to which we currently do not have analytical access, either by conventional or exotic means.  
A partial exception are functional renormalization group techniques in an asymptotic safety 
scenario \cite{ReuSaubook,percaccibook}, although also 
this ansatz relies crucially on numerical methods. This observation and the methodological complementarity of the
two approaches suggests that developing them in tandem may be a fruitful research strategy. 

Structurally, the parallels between CDT lattice gravity and lattice gauge theory are coming into sharper focus, 
despite the obvious differences in terms of
their dynamical degrees of freedom and the nature of their gauge symmetries. One beautiful and possibly key aspect of CDT
is its manifest diffeomorphism invariance, as spelled out in Sec.\ \ref{geom:sec}, which may have been somewhat underappreciated
in the past. Although realized in a different way, it can be thought of as the gravitational counterpart of the exact gauge invariance of 
Wilson's time-honoured lattice formulation of QCD \cite{kgwilson}, which 
is also preserved, despite the lattice discretization. Apart from the enormous technical con\-venience of not
having to deal with coordinate redundancy, this also implies
that the physical and geometric content of CDT are not obscured by gauge issues.

In complete analogy with lattice gauge field theory, 
computational resources are a limiting factor, and finding ever better ways of addressing standard issues of lattice Monte Carlo 
simulations (such as finite-volume effects, lattice artefacts, statistical errors, numerical efficiency and critical slowing down) will 
be a continuing challenge. However, because of the dynamical nature of gravity -- translated into the 
dynamical nature of the triangulated lattices of CDT -- and the absence of a 
fixed background reference frame, many of these issues come with a new and interesting twist. 
The ensuing subtleties were illustrated well by our discussion in Sec.\ \ref{sec:rengroup} about
defining renormalization group trajectories of constant physics without a predefined notion of scale. 
Progress in CDT provides a blueprint for how these difficulties, which are not present in quantum field theory
on a fixed background, may be addressed and overcome.

There is a clear road ahead for CDT quantum gravity, with plenty of interesting research problems to tackle, 
many of which have already been mentioned throughout this review.
They are still mostly of a fundamental nature, but include technical challenges of implementability and measurability.
Analyzing the observables described in Sec.\ \ref{observ:sec} in detail and coming up with new ones is a rich 
playground for creative minds and will play an important role in addressing key open issues.  
These include (i) getting a better handle on the renormalization group analysis, thus paving the way toward verifying the asymptotic
safety conjecture; (ii) obtaining a more detailed understanding of the properties of quantum spacetime near the Planck scale, especially in
the vicinity of the second-order phase transitions, and fleshing out the nature of any UV completion;
(iii) investigating the physical nature of the new bifurcation phase and the mechanism behind the $C_b$-$C_{dS}$ phase
transition; (iv) getting a better theoretical understanding of how to couple matter and relate it in a meaningful way to
semiclassical observables; 
(v) further strengthening the evidence for a well-defined classical limit and quantifying quantum deviations from it; and
(vi) creating a bridge toward cosmology and coming up with signatures of quantum gravity phenomenology. 

It is fortunate that CDT quantum gravity comes with a theoretical framework and computational tools that allow us to
address these questions directly.
It would be even more remarkable if in a few years' time from now we would find ourselves in a situation where CDT is no longer 
regarded as a candidate theory of quantum gravity, but ``merely" as a powerful method that enables us to understand 
and quantify quantum gravity nonperturbatively, in the same way as lattice gauge theory has given us unique access to the
nonperturbative properties of quantum chromodynamics. 

\vspace{0.2cm}

\subsection*{Acknowledgments} 
It is a pleasure to thank J.\ Ambj\o rn, A.\ G\"orlich and J.\ Jurkiewicz for long-standing collaboration and for
their feedback on the manuscript. --
This research was supported in part by Perimeter Institute for Theoretical Physics. 
Research at Perimeter Institute is supported by the Government of Canada through the Department of Innovation, Science and Economic Development and by the Province of Ontario through the Ministry of Research and Innovation.

\vspace{0.3cm}


\end{document}